%% file: main.tex
\documentclass[nofootinbib, superscriptaddress, longbibliography, 10pt]{revtex4-1}

\include{header}

\begin{document}

\raggedbottom
\clubpenalty = 10000
\widowpenalty = 10000
\displaywidowpenalty = 10000
\brokenpenalty = 10000


\title{
Poisson Clustering and Seed Effect of PBHs: Implications for Microlensing,\\
Gravitational Waves, Little Red Dots and Ultra-Faint Dwarf Galaxies
}

\author{Bernard Carr}
\email{b.j.carr@qmul.ac.uk}
\affiliation{School of Physics and Astronomy,
	Queen Mary University of London,
	Mile End Road,
	London E1 4NS,
	United Kingdom
	}

\author{S{\'e}bastien Clesse}
\email{sebastien.clesse@ulb.be}
\affiliation{
	Service de Physique Th{\'e}orique, Universit\'e Libre de Bruxelles (ULB), 
	Brussels Laboratory of the Universe (BLU-ULB),
	Boulevard du Triomphe,
	CP225, 1050, Brussels,
	Belgium
	}

\author{Juan~Garc{\'i}a-Bellido}
\email{juan.garciabellido@uam.es}
\affiliation{
	Instituto de F{\'i}sica Te{\'o}rica UAM/CSIC,
	Universidad Aut{\'o}noma de Madrid,
	Cantoblanco,
	28049 Madrid,
	Spain
	}

\author{Michael~R.~S.~Hawkins}
\email{mrsh@roe.ac.uk}
\affiliation{
	Royal Observatory, Edinburgh,
	Blackford Hill,
	Edinburgh EH9 3HJ,
	Scotland
	}

\author{Florian~K{\"u}hnel}
\email{florian.kuehnel@physik.uni-muenchen.de}
\affiliation{
	Arnold Sommerfeld Center,
	Ludwig-Maximilians-Universit{\"a}t,
	Theresienstr.~37,
	80333 M{\"u}nchen,
	Germany}
\affiliation{
	Fakult{\"a}t Physik,
	Technische Universit{\"a}t Dortmund,
	August-Schmidt-Str.~4,
	44221 Dortmund,
	Germany}

\date{\formatdate{\day}{\month}{\year}, \currenttime}

\begin{abstract}
The existence of primordial black holes (PBHs) could have significant effects on the formation of cosmic structures through both the seed effect and the Poisson effect. For an extended PBH mass function, this should lead to dark virialised bound clusters of black holes with modest mass, together with central black holes of larger mass. We consider the combination of three effects: the inevitable gravitational binding of small-scale inhomogeneities due to Poisson and seed fluctuations; the dilution through dynamical heating of all clusters smaller than around $10^{6}\.\Msun$;\. the survival of clusters in the outer Galactic halo (i.e.~towards the Magellanic Clouds) but their destruction in the central halo region through tidal disruption and collisions. We examine the implications of this for the constraints on the fraction of dark matter in PBHs from microlensing surveys and gravitational-wave events. Although the clustering of PBHs with the expected mass and radius is unlikely to relax the microlensing limits, the decline of the Galactic rotation curve might do so, allowing much of the dark matter to be in solar-mass PBHs formed at the quantum chromodynamics transition. We also emphasise that binary merger rates must be revised: binaries formed in PBH clusters can dominate the merger rate, which may be compatible with the LIGO--Virgo--KAGRA gravitational-wave limits if PBHs provide the dark matter. More positively, Ultra-Faint Dwarf Galaxies and Little Red Dots might arise through a combination of the seed and Poisson effects. Our conclusions could be modified for PBHs produced from primordial fluctuations with exponential non-Gaussian tails, since the clustering would then exceed the Poissonian prediction, opening up new possibilities to explain dark matter with solar-mass PBHs.
\end{abstract} 

\AddToShipoutPictureFG*{
	\AtPageLowerLeft{
	\put(
		\LenToUnit{\dimexpr 1.1in+\oddsidemargin+\textwidth\relax},
		\LenToUnit{\dimexpr 43mm+\voffset\relax}){
			\makebox(0,0)[br]{
			{\fontsize{10}{11}\selectfont
			\renewcommand{\arraystretch}{1}
			\begin{tabular}{@{}l@{}}
				\itshape \ldots you can look into the seeds of time\\
				\itshape And say which grain will grow and which will not\ldots\\[4pt]
				\multicolumn{1}{@{}r@{}}{
				{\color{grey}\normalfont\fontsize{9}{9}\selectfont
				\textsc{---\.William Shakespeare}, \textit{Macbeth}, I.3}}
			\end{tabular}}
			}
		}
	}
}

\maketitle

\enlargethispage{2\baselineskip}
\tableofcontents
\newpage

\section{Introduction}
\label{sec:Introduction}

\noindent The possibility that primordial black holes (PBHs) might make up the dark matter goes back 50 years and is a natural consequence of the realisation that Big Bang nucleosynthesis (BBN) implies that the dark matter could not be baryonic \cite{Cyburt:2003fe}. This is because, unlike astrophysical black holes, PBHs would originate in the radiation-dominated period before BBN. Until recently, most dark matter theorists have favoured hypothetical non-baryonic elementary particles but there is still no evidence for this from either direct searches or collider experiments. On the other hand, PBHs are also non-baryonic and share many features of particle `cold dark matter' (CDM). 

Historically, three events have stimulated interest in PBHs. The first was the 1996 detection by the MACHO collaboration~\citep{1996ApJ...461...84A} of microlensing of stars in the Magellanic Clouds by objects in the Galactic halo. Although they found a large excess of stellar-mass lenses, they ultimately concluded that the population of unidentified lenses was insufficient to make up the expected dark matter content of the halo~\citep{MACHO:2000qbb}. However, we will argue below that this conclusion may be invalidated in the light of modern measurements of the kinematics of the Galaxy and the effects of an extended PBH mass function. The second event was the 2016 detection of gravitational waves from merging black holes with somewhat larger masses than expected and subsequently there have been around three hundred similar events. Although the mainstream view is that these are associated with stellar black holes, there are reasons to believe that at least some{\,\textemdash\,}notably those in the low and high mass gaps where stellar black holes should not arise{\,---\,}may be PBHs. The third event was the recent discovery by the {\it James Webb Space Telescope} (JWST) that supermassive black holes formed at the centres of the first galaxies much earlier than expected in the standard CDM picture without PBHs~\cite{2025ApJ...987..185Z, Zhang:2025oyl}, leading to the suggestion that massive PBHs may play a r{\^o}le in seeding them. 

There are reasons to believe that some of these effects could be explained by PBHs, although this does not require them to provide {\it all} the dark matter. Subsequently, there have been many other arguments for the existence of PBHs, these being associated with other microlensing effects and various dynamical and accretion effects. These arguments have been collated in our recent review~\cite{Carr:2023tpt}, advocating what we term a `positivist' approach to the subject. Although not all researchers are convinced by this evidence, there has certainly been a huge surge in publications about PBHs in the last decade. 

The arguments advocated in our positivist approach suggest that {\it most} of the dark matter is contained in PBHs with masses around one solar mass, although some fraction should also be in the sub-solar and supermassive ranges. Indeed, we have claimed that this is compatible with a conventional model for the thermal history of the Universe~\cite{2018JCAP...08..041B, Carr:2019kxo}. In particular, we argue that a series of sound-speed dips in the period $10^{-10}$ to $10\.$s can produce a PBH mass function with bumps in the range $10^{-6}$ to $10^{6}\.\Msun$. The most striking dip is associated with the QCD transition at $10^{-5}\.$s and a mass of around $1\.\Msun$. On the other hand, this proposal is challenged~\cite{Andres-Carcasona:2024wqk} by current constraints, at least for a monochromatic PBH mass function, these being associated with the same effects which purportedly provide the evidence. So other PBH enthusiasts argue that they should be in the asteroidal or planetary mass range. This view has the advantage that it is not excluded by current constraints, although there is no positive evidence for it and recent works, based on star destruction by PBHs in Ultra-Faint Dwarf Galaxies (UFDGs), disfavour this possibility~\cite{Esser:2025pnt}. Whether the dark matter could be in asteroidal-mass or solar-mass PBHs is one of the most important current issues in PBH research. For a recent review discussing both the evidence for PBHs and the constraints on them in various mass ranges, see Ref.~\cite{Carr:2026hot}.

A major difference between dark matter particles and PBHs comes from the larger mass of the latter. This implies that PBHs can be a significant source of density fluctuations on some scales via two effects: 
	(1) the Poisson effect, associated with the $\sqrt{N}$ fluctuation in the number of black holes;
	(2) the seed effect, associated with the Coulomb attraction of a {\it single} black hole. 
The Poisson effect was first pointed out by~\citet{Meszaros:1975ef}, although he overestimated the size of the effect~\cite{1977A&A....56..377C}, and in the following years several other authors investigated this~\cite{1983ApJ...268....1C, 1983ApJ...275..405F, Chisholm:2005vm, Chisholm:2011kn}. More recently, there has been a proliferation of papers on this topic, as cited later. The Coulomb effect, which might be regarded as a special case of the Poisson effect with $N = 1$, has attracted less attention but has a long history, with~\citet{1966RSPSA.290..177H} first invoking it to explain the formation of galaxies in the context of the {\it Steady State} theory.

These effects have two important consequences. On the one hand, they may weaken some of the constraints in the solar-mass range which are currently assumed to exclude our favoured scenario~\cite{Belotsky_2019}. On the other hand, they may lead to the formation of bound cosmological structures (e.g.~dwarf galaxies) at a much earlier epoch than expected in the standard CDM model~\cite{khlopov2014primordialblackholeclusters}. This may be supported by the recent discovery of ``little red dots'' (LRDs) at high redshift~\cite{Zhang:2025tgm, Matteri:2025klg, Matteri:2025vnv}. Poisson clusters may also be identified with UFDGs, since these are dominated by dark matter and have the expected mass and radius, and PBH clustering has important implications for gravitational-wave observations since it affects the PBH merger rate. The Poisson effect was emphasised in our positivist review and we explore it in more detail here but the seed effect was not considered in our review. Clustering could be even more important if there are extra clustering effects associated with non-Poisson ones and we consider this further in the present paper.

The relative importance of the seed and Poisson effects depends on the PBH mass function. This was discussed in Ref.~\cite{Carr:2018rid} for a monochromatic mass function or a simple power-law extended mass function but the situation is more complicated for the multimodal mass functions expected in some scenarios. For example, in the {\it Thermal History} model, one may have solar-mass PBHs providing most of the dark matter, while $10^{6}\.\Msun$ PBHs with a small density provide seeds for the supermassive black holes in galactic nuclei. At sufficiently early times the seed effect dominates, so this may be relevant for LRDs, while the Poisson effect dominates at later times and may be relevant for UFDGs. However, in reality one may have a complicated mixture of both effects at all redshifts, resulting in bound clusters of PBHs with modest mass together with central black holes of much larger mass. This can only be studied properly by numerical calculations but we provide a preliminary analytical discussion. The implications of an extended PBH mass function have also been studied in Ref.~\cite{Magaraggia:2026ngp}.

One goal of this paper is to clarify the process of PBH clustering through Poisson fluctuations by combining different aspects that are often considered separately:
	(1) the linear evolution of these fluctuations until they form non-linear bound structures; 
	(2) their subsequent dynamical heating and dilution in larger clusters to reach masses and sizes possibly similar to those of UFDGs; 
	(3) gas accretion onto PBHs that naturally leads to ultra-diffuse clusters; (4) the disruption of some of these clusters by the Galactic tidal field, dynamical friction or collisions; 
	(5) their impact on the thickness of the Galactic disc.
Clustering depends on the whole PBH mass function and not only on the component contributing most of the density.

For a wide mass distribution, the merger rates are dominated by binaries formed in clusters at late times. This is because early binaries are perturbed by nearby PBHs from the low-mass tail or inside the clusters. We study the impact of a light PBH population on the binary disruption in PBH clusters. The merger rates also depend on the mass segregation in such Poisson clusters, which should boost the rate for $\Ocal( 10 )\.\Msun$ PBHs and reduce it for solar-mass ones. Deriving new PBH merger rates from our results on the Poisson clustering is another important goal of the paper. We show that the recent LIGO--Virgo--KAGRA (LVK) observations \cite{LIGOScientific:2026ctl} do not exclude solar-mass PBHs from being the dark matter. We estimate the impact of these clusters on microlensing limits, which could be sizeable if their mass were comparable to the dark matter content within the cone probed by microlensing surveys towards the Magellanic Clouds. Although this is unlikely to apply, we suggest some other possibilities for relaxing the microlensing constraints. Our analysis supports a unified scenario with solar-mass PBHs, where many of the observed binary coalescences and microlensing events are due to PBHs, and UFDGs and LRDs are associated with some combination of the seed and Poisson effects. The rough agreement between the PBH cluster masses and sizes required to explain these different observations should motivate further investigation.

The plan of this paper is as follows. Section~\ref{sec:Formation-of-PBH-Clusters-from-Poisson-Fluctuations} describes the formation of PBH clusters from the Poisson and seed effects, extending the analysis in our positivist review, while Section~\ref{sec:Dynamical-Constraints-on-PBH-Clusters} describes the dynamical constraints on the mass and radius of the clusters. Section~\ref{sec:Ultra--Faint-Dwarf-Galaxies,-Little-Red-Dots,-Blue-Monsters} considers how PBH clusters could explain features of UFDGs and some recent results from JWST, while Section~\ref{sec:PBH-Clustering-from-Non--Gaussian-Effects} discusses non-Gaussian clustering beyond the Poisson effect. Sections~\ref{sec:Effect-of-Clusters-on-Microlensing} and~\ref{Effect-of-Clusters-on-Gravitational-Waves-from-PBH-Mergers} study the implications of PBH clusters for lensing and gravitational wave constraints. Section~\ref{sec:Thermal-History-Model-and-Multimodal-Mass-Functions} reviews relevant aspects of PBH formation in the {\it Thermal History} model. Finally, Section~\ref{sec:Conclusion} draws some general conclusions.

\section{Formation of PBH Clusters from Seed and Poisson Effects}
\label{sec:Formation-of-PBH-Clusters-from-Poisson-Fluctuations}

\noindent Some of the discussion in this section appeared in our earlier review \cite{Carr:2023tpt} but it has been updated to cover an extended PBH mass function and we correct some minor errors in our earlier analysis. \citet{Carr:2018rid} have clarified the relationship between these two effects and we start by presenting their heuristic analysis before discussing the seed and Poisson effects in more detail. If the PBHs have a single mass $m$, the initial fluctuation in the matter density on a scale $M$ is~\cite{Carr:2018rid}
\begin{equation}
	\delta_{\mspace{1mu}\irm}
		\approx
			\begin{cases}
				m / M
					& ( {\text{seed}} )
				\\[1mm]
				( \fPBH\,m / M )^{1/2}
					& ( {\rm Poisson} )
				\, ,
			\end{cases}
			\label{eq:initial}
\end{equation}
where $\fPBH$ is the fraction of the dark matter in the PBHs, assumed to be constant during the clustering process. If $\fPBH \sim 1$, the Poisson effect dominates for all $M$; if $\fPBH \ll 1$, the Poisson effect dominates for $M > m / \fPBH$ and the seed effect for $M < m / \fPBH$. Indeed, the mass bound by a single seed can never exceed $m / \fPBH$ because of competition from other seeds. There is always a mass $M_{\text{CDM}}$ below which the PBH fluctuations dominate the CDM fluctuations, so this produces extra power on small scales. However, the CDM fluctuations fall off more slowly than both the Poisson and seed fluctuations with increasing $M$ and so generally dominate for sufficiently large $M$. Therefore structure formation is as in the standard model on sufficiently large scales but it can be radically different on small scales. Although Poisson clustering is less important for asteroidal PBHs than for solar-mass ones, since $\delta_{\mspace{1mu}\irm}$ decreases as $m$ becomes smaller, it could still be relevant for some constraints and observational signatures.

Both types of PBH fluctuations are frozen during the radiation-dominated era~\cite{1974A&A....37..225M} but grow as $( 1 + z )^{-1}$ from the start of the matter-dominated era, which corresponds to a redshift $z_{\text{eq}} \approx 3400$, so long as $\delta$ remains small. If one uses this linear growth formula, an overdense region has $\delta \approx 1.06$ at turn-around and 1.686 at virialisation~\cite{Gunn:1972sv}, so the mass virialising at redshift $z_{\crm}$ would be
\vs{-1mm}
\begin{equation}
	M_{\crm}
		\approx
			\begin{cases}
				2.0 \times 10^{3}\;m\.( 1 + z_{\crm} )^{-1}
					& ( {\text{seed}} )
					\\[0.5mm]
				4.1 \times 10^{6}\.\fPBH\,m\.
				( 1 + z_{\crm} )^{-2}
					& ( {\rm Poisson} )
				\, .
			\end{cases}
			\label{eq:bind}
\end{equation}
This is more accurate than the expression in Ref.~\cite{Carr:2018rid} but still neglects a numerical factor which is included in a more precise analysis below and introduces a further $10\%$ correction. The actual overdensity at maximum expansion is $9\mspace{1mu}\pi^{2}/16 \approx 5.5$ and this increases by a further factor of $32$ at virialisation. (This is because the radius decreases by 2 and the background density by 4 during virialisation.) However, there is an important difference between the two effects. In the Poisson case, most of the Universe goes into bound regions of mass $M_{\crm}$ at redshift $z_{\crm}$. In the seed case, the fraction of the Universe in bound regions at redshift $z_{\crm}$ is approximately $\fPBH\.( z_{\text{eq}} + 1) / ( z_{\crm} + 1 )$, which is initially small for $\fPBH \ll 1$ and only reaches $1$ at $z_{\crm} + 1 \approx \fPBH\.( z_{\text{eq}} + 1 )$. Before this, the seed-bound regions should not be described as PBH clusters since they only contain a single black hole. After this, competition between the seeds will limit the mass of each seed-bound region to at most $M \sim m / \fPBH$. This is just the value of $M$ above which the Poisson effect dominates. Each PBH is surrounded by a local bound region of this mass, so one has a combination of the seed and Poisson effects.

We mainly emphasise the Poisson effect in this paper but the seed effect may also be important at redshifts above $\fPBH\.z_{\text{eq}}$ and this may be relevant in the discussion of UFDGs and LRDs. We note that the Poisson effect really leads to clusters with a range of masses at a given redshift. So even if the PBHs have a monochromatic mass function, the clusters would have a range of masses. 
We derive the associated mass function later.

Reference~\cite{Carr:2018rid} also considers the case of a power-law mass function, in which the relationship between the Poisson and seed effects is more complicated. In this work we consider an even more complicated situation with a multimodal mass function and this has not been considered in the literature before. In this case, the first bound regions (seeded by the larger PBHs) may still contain smaller black holes and thus be described as clusters. 

The primordial adiabatic perturbations, such as those arising in the standard $\Lambda$CDM model (i.e.~with particle dark matter), also generate PBH clusters. Therefore perturbations below the threshold for black hole formation can also lead to clustering of previously formed PBHs. This can be calculated using the primordial spectrum and the appropriate matter transfer function. However, this is smaller than the Poisson contribution to clustering on sufficiently small scales.

\subsection{Cluster Formation for Monochromatic and Extended PBH Mass Functions}
\vs{-1mm}
For both the Poisson and seed effects, the bound structures would comprise both PBH clusters and baryons (i.e.~gas), together with any other dark matter component; the gas could sink to the centre of the bound structure to form stars. For sufficiently massive and abundant PBHs, the bound PBH clusters could be very massive and this has several consequences:
	(1) some of the PBH constraints{\,---\,}including the microlensing limits{\,---\,}may be weakened;
	(2) clustering influences the PBH merger rate and thereby provides a connection between the dark matter and gravitational-wave observations;
	(3) clustering may explain the minimum size and large mass-to-light ratio of UFDGs, with an extra population of dark haloes being detectable with the next generation of instruments. 

We now go beyond our treatment in Ref.~\cite{Carr:2023tpt} by considering the case of an extended PBH mass function. Since the variance scales like $m\mspace{1.5mu}\fPBH$ in the monochromatic case, we can generalise to any PBH mass function by replacing $m$ with
\begin{equation}
	\langle\mspace{1mu}m\mspace{1mu}f \rangle 
		\equiv
			\int{\d\mspace{-1mu}\ln m}\;m\mspace{1mu}f( m )
			\, ,
\end{equation}
where $f( m ) \equiv ( 1 / \rho^{}_{\text{PBH}} )\.\d\mspace{1mu}\rho^{}_{\text{PBH}} / \d\mspace{-1mu}\ln m$ is the normalised PBH density distribution (i.e.~the fraction of the PBH density in black holes of mass $m$). The fraction of the dark matter in PBHs of all masses is still denoted by $\fPBH$ and we must distinguish between these ``f'' factors since they appear with different dependencies below. The initial fractional density fluctuation due to Poisson fluctuations in the number of PBHs in a volume containing a mass $M$ is then
\begin{equation}
	\delta_{\mspace{2mu}\irm}
		= 
			\mspace{-2mu}
			\left(
				\frac{ \fPBH\mspace{1mu}\langle\mspace{1mu}m\mspace{1mu}f \rangle }{ M }
			\right)^{\!1/2}
\end{equation}
and in the monochromatic case this just corresponds to Eq.~\eqref{eq:initial}. The formation of Poisson-induced PBH clusters has been studied using $N$-body simulations~\cite{Inman:2019wvr, Raidal:2018bbj} and analytically using the Press--Schechter formalism and the theory of spherical collapse~\cite{Afshordi:2003zb}. For example, Figure 1 of Ref.~\cite{Inman:2019wvr} shows the density distribution at redshift $z = 100$ obtained for monochromatic PBHs with $m = 30\.\Msun$ and $\fPBH = 10^{-5}$ and $10^{-1}$. These results also apply for any combination with $m\mspace{1.5mu}\fPBH/\Msun = 0.0003$ and $3$. The difference with the standard cosmological scenario, in which the Universe is still homogeneous at such a high redshift, is striking.

As also pointed out in Ref.~\cite{Kashlinsky:2016sdv}, the contribution from Poisson fluctuations to the matter power spectrum becomes dominant on small scales and for stellar-mass PBHs leads to the high-redshift gravitational binding of almost all small-scale perturbations into clusters with masses up to $10^{6}$\,--\,$10^{7}\.\Msun$. The Poisson contribution to the matter power spectrum is independent of wavenumber $k$ (white noise) and its amplitude at redshift $z$ is well approximated by~\cite{Afshordi:2003zb}\vs{-1mm}
\begin{equation}
\label{eq:power}
	P_{\Prm} 
		\simeq
			1.03 \times 10^{-3}\,
			\fPBH 
			\mspace{-1.5mu}
			\left(
				\frac{ \langle\mspace{1mu}m\mspace{1mu}f \rangle }
				{ 3\mspace{1mu}\Msun }
			\right)\mspace{-3.5mu}
			\left[
				\frac{ g( z )}{ 1 + z_{\text{eq}} }
			\right]^{2}
			{\rm Mpc}^{3}
			\, ,
\end{equation}
where $g( z )$ is the growth factor for isocurvature fluctuations; this increases approximately linearly with the scale factor in the matter-dominated era, when the clusters form, so $g( z ) \approx ( 1 + z_{\text{eq}} )/( 1 + z )$ then. More precisely, one has \cite{Inman:2019wvr}
\vs{-1mm}
\begin{equation}
	g( z )
		=
			{_2}F_{1}\mspace{-2.5mu}
			\left(
				\frac{ 1 - \sqrt{1 + 24\.\gamma\vphantom{1^{1}}} }{ 4 },\mspace{1mu}
				\frac{ 1 + \sqrt{1 + 24\.\gamma\vphantom{1^{1}}} }{ 4 },\mspace{1mu}
				1,\mspace{1mu}
				-\.\frac{ 1 + z_{\text{eq}} }{ 1 + z }
			\right)
			,
\end{equation}
where ${_2}F_{1}$ is the hypergeometric function and $\gamma$ is the fraction of dark and baryonic matter that clusters. This is $5/6$ if the baryons are unclustered and $1$ if they are clustered, corresponding to the square-root factor being $\sqrt{21}$ or 5, respectively. In the latter case $g( z ) = 1 + 3\.( 1 + z_{\text{eq}} ) / 2\.( 1 + z )$, which gives the 9/4 enhancement of Ref.~\cite{Afshordi:2003zb}. The growth factor $g( z )$ tends to unity when $z \gg z_{\text{eq}}$, but $g( z_{\text{eq}} )$ is $2.5$ for $\gamma = 1$ and $2.2$ for $\gamma = 5/6$. The expression for $g( z )$ neglects the impact of dark energy on the evolution of Poisson fluctuations but we mostly focus on cluster formation redshifts when dark energy remains subdominant. 

\begin{figure*}[t]
	\vs{-8mm}
	\hs{7mm}\includegraphics[width=0.90\textwidth]{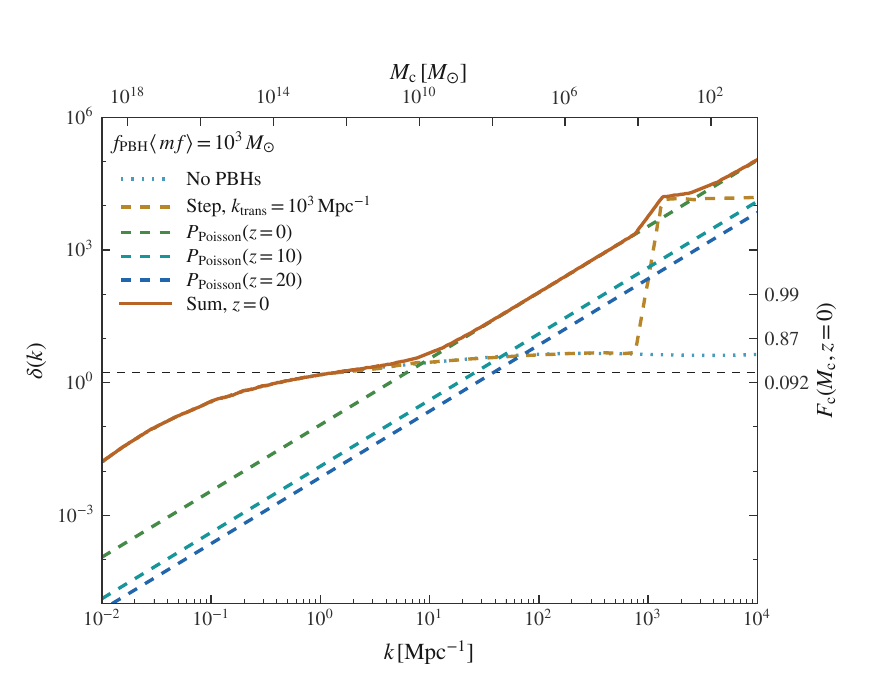}
	\caption{
		Dimensionless linear density contrast $\delta( k )$ today in the standard $\Lambda$CDM model without PBHs (dotted blue line) and with them for $\fPBH\mspace{1mu}\langle\mspace{1mu}m\mspace{1mu}f \rangle = 10^{3}\.\Msun$, including the inevitable Poisson term in the matter power spectrum (dashed lines) estimated at three different redshifts ($20$, $10$ and $0$). For comparison, we also show the possible power spectrum enhancement leading to PBH formation for a transition scale at $k_{\text{trans}} = 10^{3}\.{\rm Mpc^{-1}}$ (dashed orange line). The horizontal dashed line represents the critical threshold for halo formation $\delta_{\crm} \simeq 1.686$. The upper $x$-axis is the halo mass implied by Eq.~\eqref{eq:Mlambda}. The right $y$-axis shows the halo collapsed fraction today $F_{\crm}( M_{\crm},\.z = 0)$; this is almost $1$ for $k \gtrsim 100\.{\rm Mpc}^{-1}$, meaning that all PBHs are regrouped in clusters with masses below $10^{8}\.\Msun$.
		}
	\label{fig:deltas}
\end{figure*}

For example, if we consider the standard $\Lambda$CDM model with particle dark matter but also PBHs with $m = 10^{3}\.\Msun$ and $\fPBH = 1$, the Poisson term dominates for comoving wave-modes $k \gtrsim 100\.{\rm Mpc}^{-1}$ at $z \approx 20$. The current (dimensionless) density perturbation $\delta( k )$, related to the power spectrum through $\delta( k )^{2} \equiv P( k )\.k^{3} / ( 2\mspace{1mu}\pi^{2} )$, is shown in Fig.~\ref{fig:deltas} (neglecting non-linear effects). For comparison, we also show the expected perturbation for the standard (particle) $\Lambda$CDM scenario with a primordial spectrum with a sharp enhancement at $k_{\text{trans}} = 10^{3}\.{\rm Mpc^{-1}}$. This generates PBHs at small scales, without violating the constraints from cosmic microwave background (CMB) distortions. It modifies the spectrum below the scale at which the Poisson effect becomes dominant. Even for lower values of $m$ or $\fPBH$, one may have to account for the effect of the enhanced primordial power spectrum on PBH cluster formation, an effect that is sometimes overlooked.

One can associate a cluster mass scale with each comoving radius $R$:
\vs{-1mm}
\begin{align}
\begin{split}
\label{eq:Mlambda}
	M_{\crm}( R )
		=
			\frac{ 4\mspace{1mu}\pi }{ 3 }\. 
			\rho_{\mrm}^{(0)}\mspace{1mu}R^{3}\.
		\simeq
			1.6 \times 10^{11}\mspace{-1mu}
			\left(
				\frac{ \Omega_{\mrm}\.h^{2} }{ 0.142 }
			\right)\mspace{-4.5mu}
			\left(
				\frac{ R }{ {\rm Mpc} }
			\right)^{\!3}
			\Msun
			\, ,
\end{split}
\end{align}
where $\rho_{\mrm}^{(0)}$ is the current matter density (dark plus baryonic) and we have assumed the standard value for $\Omega_{\mrm}\.h^{2}$. If the initial overdensity is assumed to grow as $( 1 + z )^{-1}$, or more precisely as $g( z )$, according to the linear theory in the matter era, then its value when the PBH cluster binds and virialises is $\delta_{\crm} \simeq 3\.(12 \pi)^{2/3}\mspace{-1.5mu}/20 \simeq 1.686$. (Recall this is not the actual overdensity.) Assuming $\gamma = 1$, this implies a virialisation redshift 
\begin{equation}
	z_{\crm} + 1
		\simeq
			3
			\left(
				\frac{ \langle\mspace{1mu}m\mspace{1mu}f \rangle
				\mspace{1.5mu}\fPBH }{ \.\Msun }
			\right)^{\!1/2}
			\left(
				\frac{ M_{\crm} }
				{ 10^{6}\mspace{1mu}\Msun }
			\right)^{-1/2}
\end{equation}
in the matter-dominated era [cf.~Eq.~\eqref{eq:bind}]. For example, if $\langle\mspace{1mu}m\mspace{1mu}f \rangle \mspace{1.5mu}\fPBH = 3\.\Msun$, one gets $z_{\crm} + 1 \approx 11$ for $M_{\crm} = 10^{5}\.\Msun$ and for $\langle\mspace{1mu}m\mspace{1mu}f \rangle \mspace{1.5mu}\fPBH = 10^{3}\.\Msun$, one gets $z_{\crm} + 1 \approx 63$ for $M_{\crm} = 10^{6}\.\Msun$. When a bound cluster virialises, the theory of spherical collapse predicts that its density is $18\.\pi^{2} \approx 178$ times the background density, giving a cluster radius at formation
\begin{equation}
	r_{\crm}
		\simeq 
			1600\,{\rm pc} 
			\left(
				\frac{ \langle\mspace{1mu}m\mspace{1mu}f \rangle
				\mspace{1.5mu}\fPBH }{ \.\Msun }
			\right)^{\!-1/2}
			\left(
				\frac{ M_{\crm} }
				{ 10^{6}\mspace{1mu}\Msun }
			\right)^{\!5/6}
			\, ,
			\label{eq:rM}
\end{equation}
which is shown by the diagonal straight lines in Fig.~\ref{fig:relax}. For example, if $\langle\mspace{1mu}m\mspace{1mu}f \rangle \mspace{1.5mu}\fPBH = 3\.\Msun$, one has $r_{\crm} \approx 135\,{\rm pc}$ for $M_{\crm} = 10^{5}\.\Msun$; if $\langle\mspace{1mu}m\mspace{1mu}f \rangle\.\fPBH = 10^{3}\.\Msun$, one has $r_{\crm} \approx 50\,{\rm pc}$ for $M_{\crm} = 10^{6}\.\Msun$.

In the seed scenario with a monochromatic mass function, which only applies if $\fPBH \ll 1$ and for objects forming at redshifts above $\fPBH\.z_{\text{eq}}$, the radius is 
\begin{equation}
	r_{\crm}
		\simeq 		
			160\,{\rm pc}
			\left(
				\frac{ m }{ 10^{4}\.\Msun }
			\right)^{\!-1}
			\left(
				\frac{ M_{\crm} }
				{10^{6}\mspace{1mu}\Msun }
			\right)^{\! 4/3} \quad \text{for}
			\quad
	M_{\crm}
		<
			\fPBH^{-1}\.m
			\, .
			\label{eq:seed}
\end{equation}
Although these bound objects should not be described as ``clusters'', since they just contain a single black hole at their centre, there is a smooth transition between Eqs.~\eqref{eq:seed} and \eqref{eq:rM} at the mass $m/\fPBH$. An interesting implication of Eq.~\eqref{eq:seed} is that the density profile around the black hole always has the form $\rho \propto r^{-9/4}$~\cite{Gunn:1972sv}.
For a bimodal mass function, the situation is more complicated. For example, consider a scenario with most of the dark matter in PBHs with $m_{1}$ and $f_{1} \approx 1$ but another population with $m_{2}$ and $f_{2} \ll 1$. Then Eq.~\eqref{eq:bind} implies that the seed fluctuation dominates (i.e.~generates the largest bound regions) for $m_{2} > 10^{3}\,m_{1}( 1 + z_{\crm} )^{-1}$. This condition is always satisfied if $m_{2} > 10^{3}\.m_{1}$. However, only a small fraction of the Universe is in the seed-bound regions, where Eq.~\eqref{eq:seed} applies. Most of it is in the Poisson-bound regions, where Eq.~\eqref{eq:rM} applies. So we need to decide which observations relate to each of these. For example, LRDs may be associated with the seed effect and UFDGs with the Poisson effect.

For extended mass functions, such as the \textit{Thermal History} model in which most of the dark matter is composed of solar-mass PBHs, there exists a regime in which the seed and Poisson effects lead to the same cluster mass--radius relation. Indeed, when the fraction of the Universe collapsing into Poisson clusters of mass $M_{\crm}$ approaches unity, the less numerous clusters induced by the seed effect with the same mass $M_{\crm}$ cease to grow. Their virialisation redshift and radius are therefore identical to those of the corresponding Poisson clusters. The subsequent dynamical evolution, however, could in principle differ because of the presence of a massive seed at the cluster centre. As discussed below, the seed mass, $ m_{\text{seed}} = M_{\crm}\.( 1 + z_{\crm} )/( 1 + z_{\text{eq}} )$, is typically much smaller than the cluster mass, implying that its impact on the subsequent dynamical heating is in fact negligible. This regime does not apply when the collapse fraction into Poisson clusters is significantly below unity, as occurs for sufficiently large cluster masses or for $\fPBH\mspace{1mu}\langle\mspace{1mu}m\mspace{1mu}f \rangle \ll \Msun$, as discussed in more detail below. 

Finally, for extended mass functions, it is interesting to compare the mass and size of PBH clusters from the seed and Poisson effects to the ones resulting from the growth of the primordial {\it adiabatic} fluctuations in the standard CDM scenario. If these fluctuations are amplified on cluster scales, as expected if they form massive PBHs, then they grow linearly with the scale factor in the matter era and logarithmically in the radiation era~\cite{Meszaros:1974tb}. This effect is often overlooked but could play an important role in PBH clustering, especially for PBHs formed from non-Gaussian tails. For a region of mass $M_{\crm}$, the primordial density fluctuation $\delta_{\text{prim}}$ grows as
\begin{equation}
	\delta( z )
		\approx
			\delta_{\text{prim}}\.\ln\!
			\left(
				\frac{ z_{\text{H}}( M_{\crm} ) +1 }{ z_{\text{eq}} + 1 }
			\right)
			g( z )
			\, ,
\end{equation}
where $z_{\text{H}}( M_{\crm} )$ is the horizon-entry redshift of the comoving mode associated with the cluster mass. With this approximation, the cluster forms at a redshift given by
\begin{equation}
	1 + z_{\crm}
		\approx
			\delta_{\text{prim}}\.
			\big[
				4 \times 10^{4} - 10^{3}\.\ln( M_{\crm} / \Msun )
			\big]
			\. .
\end{equation}
One can then obtain the mass--radius relation using the same method as for the Poisson and seeded clusters. As explained in Ref.~\cite{Inman:2019wvr}, there is no logarithmic term for the PBH Poisson effect since this is an isocurvature fluctuation. For the \textit{Thermal History} model, the amplitude of the primordial density fluctuations is approximately $\delta_{\text{prim}} \approx 0.02$, not far from the value which produces more compact clusters.

\begin{figure*}[t]
	\vs{-3mm}
	\includegraphics[width=0.95\textwidth]{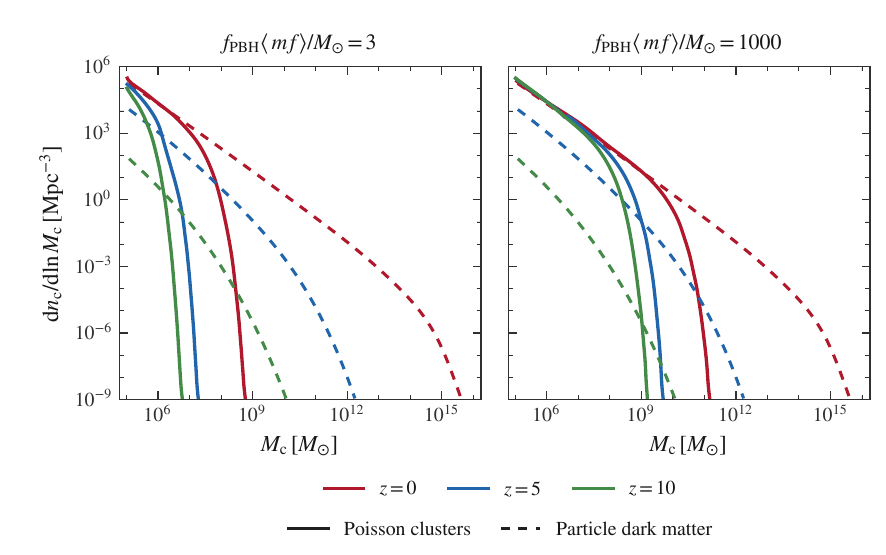}
	\caption{Mass function for Poisson clusters (solid lines) at redshift $z = 0$ (red), $z = 5$ (blue) and $z = 10$ (green) for $\fPBH\mspace{1mu}\langle\mspace{1mu}m\mspace{1mu}f \rangle = 3\.\Msun$ (left panel) and $1000\.\Msun$ (right panel), compared to the standard halo mass function for non-interacting particle dark matter (dotted lines).}
	\label{fig:hmf}
\end{figure*}

According to the Press--Schechter formalism~\cite{Bond:1990iw, Bower:1991kf, Lacey:1993iv, Lacey:1994su}, the fraction of mass collapsed into haloes with a mass above $M_{\crm}$ due to the Poisson fluctuations alone is
\begin{equation} 
	F_{\crm}( M_{\crm},\mspace{1mu}z )
		=
			{\text{erfc}}\!
			\left[
				\frac{ \delta_{\crm} }
				{ \sqrt{2}\.
				\sigma( M_{\crm},\mspace{1mu}z) }
			\right]
			,
			\label{eq:F}
\end{equation}
where
\begin{equation}
	\sigma^{2}( z )
		=
			\int\!\d\mspace{-1mu}\ln k\;
			\frac{ k^{3} }{ 2\mspace{1mu}\pi^{2} }\.
			P_{\Prm}( k,\mspace{1mu}z )\mspace{1mu}\big| W( k ) \big|^{2}
			\, .
\end{equation}
In order to associate a characteristic wavenumber with the scale $R$, we assume $R = \pi / k =\lambda / 2$ and use a Fourier-space top-hat function for the window function $W( k\mspace{1mu} )$ with width corresponding to unit logarithmic cluster mass bins. Figure~\ref{fig:deltas} shows the collapsed fraction $F_{\crm}( M_{\crm},\mspace{1mu}z )$ for $\fPBH\mspace{1mu}\langle\mspace{1mu}m\mspace{1mu}f \rangle = 10^{3}\.\Msun$. The associated halo mass function is
\begin{equation}
	\frac{ \d\mspace{1mu}n_{\crm} }{ \d\mspace{-1mu}\ln M_{\crm}} 
		= 
			\frac{ \rho^{}_{\text{CDM},\mspace{1mu}0} }{ M_{\crm} }\.
			F_{\crm}( M_{\crm},\mspace{1mu}z )
			\, .
\end{equation}
Since ${\text{erfc}}( x ) \simeq {\text{exp}}\big( {-\.x^{2}} \big) / \big( \sqrt{\pi}\.x \big)$ for $x \gg 1$, this gives $\drm\mspace{1mu}n_{\crm} / \drm\mspace{1mu}M_{\crm} \propto M_{\crm}^{-2}$ and an exponential cut-off when the collapse fraction is much smaller than one. Figure~\ref{fig:hmf} represents the halo mass function for Poisson clusters at redshifts $0$, $5$ and $10$ for $\fPBH\mspace{1mu}\langle\mspace{1mu}m\mspace{1mu}f \rangle = 3\.\Msun$ and $1000\.\Msun$. This is compared to the standard Press--Schechter halo mass function in the particle dark matter case. As expected, the difference appears at high redshifts when Poisson clusters form before particle dark matter haloes, inducing a strong enhancement of the halo mass function for low cluster masses.

\subsection{Relaxation, Dynamical Heating and Evaporation of PBH Clusters}
\label{sec:Relaxation,-Dynamical-Heating-and-Evaporation-of-PBH-Clusters}

When Poisson-induced PBH clusters have formed, it takes some time for the PBH velocities and the cluster size to relax towards the values expected from the virial theorem. In the monochromatic case, the relaxation time for a halo of $N = \fPBH\.M_{\crm} / m$ objects is 
\begin{align}
\label{eq:reltime}
	t_{\text{rel}}
		\simeq
			\left[
				\frac{ 0.14\.N }{ \ln( 0.14\.N ) }
			\right]\mspace{-1mu}
			\sqrt{\frac{ r_{\crm}^{3} }
			{ G M_{\crm} }}
		= 
			2.1 \times 10^{6}\.{\rm yr}\,
			\frac{ \sqrt{M_{\crm}/\Msun}\.\fPBH} 
			{ (m/\Msun)\.
			\ln\!
			\big[
				0.14\.\fPBH \.M_{\crm} / m
			\big] }\mspace{-1mu}
			\left(
				\frac{ r_{\crm}}{\rm pc}
			\right)^{\!3/2}
			\, ,
\end{align}
where the $\fPBH$ factor indicates that the PBHs may provide only part of the dark matter and $r_{\crm}$ is the edge radius for a uniform-density cluster. After relaxation, the clusters continue to expand due to dynamical heating. This process is taken into account, for instance, to derive limits on $\fPBH$ from the minimum size of UFDGs or their globular clusters~\cite{Brandt:2016aco, Green:2016xgy}. For an extended mass function, the cluster radius increases according to 
\begin{equation}
	\frac{ \drm\mspace{1mu}r_{\crm} }{ \d\mspace{1mu}t }
		=
			\frac{ 4\.\sqrt{2}\pi\.G\fPBH\.
			\langle\mspace{1mu}m\mspace{1mu}f \rangle\mspace{1mu}
			\ln\mspace{-2mu}
			\big[
				\fPBH\mspace{1mu}
				M_{\crm} / 2\. \langle\mspace{1mu}m\mspace{1mu}f \rangle
			\big] }
			{ 2\.\beta\.v_{\text{vir}}\.r_{\crm} } 
			\, ,
			\label{eq:growth}
\end{equation}
where $v_{\text{vir}} = \sqrt{G M_{\crm} / (2\,r_{\crm})}$ is the halo virial velocity and $\beta$ parametrises the efficiency of secular expansion. This depends on the cluster mass profile and $\beta = 10$ is usually assumed~\cite{Brandt:2016aco, Green:2016xgy}.

The derivation of Eq.~\eqref{eq:growth} may be generalised to characterise the efficiency of secular cluster expansion. The value $\beta \simeq 10$ applies for a stellar population heated by massive objects in a dark matter halo and its application to a self-gravitating PBH population requires additional assumptions~\cite{Brandt:2016aco, Green:2016xgy}. This is because encounters redistribute energy within the system, while sustained expansion depends on energy generation and transport. An extended mass spectrum and subsequent mass segregation can modify these processes by producing a compact population of massive PBHs embedded in a more diffuse halo. The simulations of Ref.~\cite{Siles:2024yym} do indeed exhibit approximately stationary core radii alongside increasing half-mass radii, illustrating that different parts of a cluster need not expand at the same rate.

A further mechanism affecting the energy supply is the presence of a retained intermediate-mass black hole (IMBH) at the centre of the cluster. The simulations of \citet{Barcelo-Sastre:2026nbs} find that such an object can strongly suppress dynamical heating. This motivates treating the core and outer halo separately when assessing departures from the simplest expansion prescription~\cite{Belotsky_2019}. More precisely, if there is an additional gravitational potential, due to either a central IMBH or a compact core of heavier black holes induced by rapid mass segregation, then $\beta $ should be considered as an effective parameter taking into account both the core and the cluster. As long as $ \fPBH\.\langle\mspace{1mu}m\mspace{1mu}f \rangle\mspace{1mu}/\Msun \sim 1$ and $M_{\crm}$ much exceeds the core mass, $M_{\text{core}}$, the impact of this on the dynamical heating of the cluster is unimportant. However, it can be important if one has an extended mass function with $ \fPBH\.\langle\mspace{1mu}m\mspace{1mu}f \rangle\mspace{1mu}/\Msun \gg 1$
even above the mass $M_{\text{peak}}$ at which the PBH density peaks. For example, this is expected in the \textit{Thermal History} model. In this case, PBHs much larger than $M_{\text{peak}}$ will rapidly fall towards the centre, producing a compact core and possibly a central IMBH. They would then no longer contribute to the cluster dynamical heating, which would be mostly due to the PBHs with mass around $M_{\text{peak}}$. Therefore $ \langle\mspace{1mu}m\mspace{1mu}f \rangle\mspace{1mu}$ should be replaced by $f( M_{\text{peak}} )\.M_{\text{peak}}$ in the heating equation, which is equivalent to
\begin{equation}
	\beta^{\text{eff}} =
			\frac{ \langle\mspace{1mu}m\mspace{1mu}f \rangle\mspace{1mu} }
			{ f( M_{\text{peak}} )\.M_{\text{peak}} }\mspace{-2mu}
			\left[
				\frac{ \alpha\.M_{\crm} }{ 2\mspace{1mu}M_{\text{core}} } + \beta
			\right]
		\gg
			1
			\, ,
			\label{eq:effbeta}
\end{equation}
where the second parameter $\alpha$ would be $0.4$ for a central star cluster~\cite{Brandt:2016aco} but is more general here. We consider benchmark values of 10 and 1000 for $\beta^{\text{eff}}$ in the following discussion, with the latter representing weaker net expansion, and henceforth drop the label ``eff''. For fixed cluster mass, radius and PBH mass function, the expansion time $t_{\text{exp}} \equiv r_{\crm} / ( \d\mspace{1mu}r_{\crm} / \d\mspace{1mu}t )$ is proportional to $\beta$ and therefore 100 times longer for $\beta = 1000$. However, the comparison $t_{\text{exp}} \sim t_{\text{rel}}$ applies only to the original $\beta \sim 10$ estimate; the relaxation time in Eq.~\eqref{eq:reltime} has no explicit $\beta$ dependence for fixed cluster properties. 

\begin{figure*}[t]
	\includegraphics[width=0.8\textwidth]{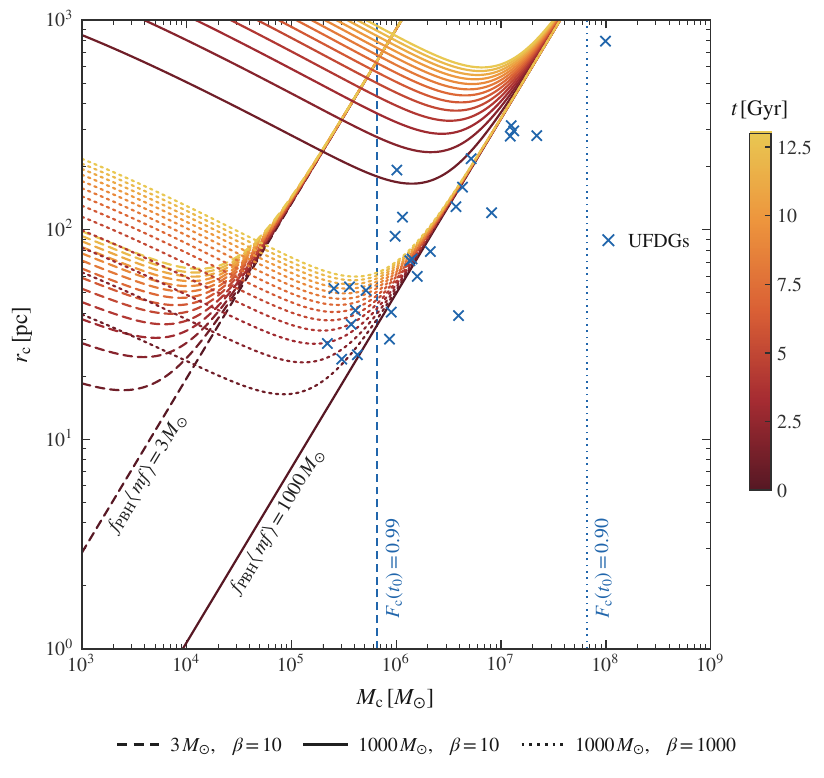}
	\caption{
		Evolution of the radius $r_{\crm}$ and mass $M_{\crm}$ from~\eqref{eq:dyntime} for $\beta = 10$ and $\beta = 1000$. The colour scale gives the elapsed time in Gyr. The initial radii come from Eq.~\eqref{eq:rM}; increasing $\beta$ reduces subsequent expansion and shifts the turnover towards smaller masses and radii. The crosses show UFDG size and mass estimates (see Sec.~\ref{sec:Ultra--Faint-Dwarf-Galaxies-and-Accretion-in-PBH-Clusters}). The vertical blue lines denote $F_{\crm}( t_{0} ) = 0.99$ and $0.9$.
		}
\label{fig:relax}
\end{figure*}

This analysis applies for any mass function but it neglects the small changes of the logarithmic term in Eq.~\eqref{eq:growth}. These effects could become important for compact clusters, for instance in PBH scenarios with a strong initial clustering. For clusters from Poisson fluctuations and the type of mass distributions considered here, mass segregation is expected for clusters that are dynamically heated and diluted inside a larger PBH cluster, but it is likely less efficient for larger and less compact clusters that have a relaxation time much larger than the age of the Universe. We have also neglected the cooling due to the friction induced by the gravitational interaction with any other CDM component with a fractional density $( 1 - \fPBH )$. The arguments of Ref.~\cite{Brandt:2016aco}, applied to a PBH cluster rather than a star cluster, imply that the cooling term should dominate over the dynamical heating term if $\fPBH \ll 1$, preventing clusters from expanding dynamically. The dynamical friction timescale, however, may remain larger than the age of the Universe, leading to stable clusters. By integrating Eq.~\eqref{eq:growth}, one can express the current cluster radius as
\vs{-1mm}
\begin{equation}
	r_{\crm}( t )
		\approx
			\left\{
				\left(
					\frac{ r_{\crm,\mspace{1mu}\irm} }{ \rm pc }
				\right)^{\!3/2} + 1260 \.\beta^{-1}
				\left(
					\frac{ M_{\crm} }{\.\Msun }
				\right)^{\!-1/2}
				\fPBH
				\left(
					\frac{ \langle\mspace{1mu}m\mspace{1mu}f \rangle }{\.\Msun }
				\right)
				\ln\!
				\left[
					\frac{ \fPBH\.M_{\crm} }
					{ 2\.\langle\mspace{1mu}m\mspace{1mu}f \rangle }
				\right]
				\frac{ t }{ \rm Gyr }
			\right\}^{2/3}
			{\rm pc}
			\, ,
			\label{eq:dyntime}
\end{equation}
where $r_{\crm,\mspace{1mu}\irm}$ is the initial cluster radius, given by Eq.~\eqref{eq:rM}. Figure~\ref{fig:relax} compares the evolution for $\beta = 10$ and $\beta = 1000$. Increasing $\beta$ by a factor of $100$ decreases $r_{\crm}( t )$ by a factor $100^{-2/3} \approx 0.0464$. However, this only applies when the $\alpha$ term in Eq.~\eqref{eq:effbeta} dominates. When the $\beta$ term dominates, the two curves approach the same relation~\eqref{eq:rM} and the initial Poisson-collapse relation is not modified.

Clusters may no longer exist if they are destroyed by other dynamical processes. One relevant effect is the decrease of mass due to evaporation, which means that the evolutionary arrow is inclined to the left. Eventually, a PBH cluster may evaporate completely and it is sometimes claimed that this happens on a timescale $t_{\text{evap}} \sim 100\,t_{\text{rel}}$. If one can neglect the initial radius, the timescales $t_{\text{rel}}$ and $t_{\text{evap}}$ have the same dependence on $M_{\crm}$, $\fPBH$ and $m$ but with different numerical factors. They have sometimes been confused. For example, in the context of the microlensing limits, the evaporation time was considered in Ref.~\cite{Petac:2022rio} in order to find the minimum cluster mass, whereas the dynamical heating time should have been used. 
\newpage

The above analysis of cluster evaporation is simplistic since it treats the cluster as a homogeneous sphere, whereas it is expected to have a core and some radial density profile outside that (cf.~the Galactic halo). In fact, a more detailed investigation~\cite{Siles:2024yym} indicates that PBH clusters do not evaporate completely; they only lose $20\%$ of their mass, with their compact cores maintaining their initial radius. The cluster stability would be further enhanced by the presence of a massive central black hole; this might be expected for an extended PBH mass function for which more massive holes would sink to the centre through dynamical friction. In this case, one must distinguish between the cluster core, which does not change its radius, and the expanding outer regions.

As discussed below, clusters larger than this minimum mass may still lose their integrity as a result of collisions within a larger cluster. However, this can lead to their disruption only if the velocity dispersion $\sigma_{\crm}$ within clusters increases with their mass. So long as the Poisson effect dominates, Eq.~\eqref{eq:rM} implies ${\sigma_{\crm} \sim M_{\crm}^{1/12}}$, which increases only very weakly, so it is marginal whether the hierarchy of clustering can be destroyed. However, on scales larger than about $10^{8}\.\Msun$, the cold dark matter fluctuations take over (see Fig.~\ref{fig:deltas}) and in this case $\sigma_{\crm} \sim M_{\crm}^{1/3}$. This means that collisional disruption may occur, at least in the central region of the larger bound region. In particular, we show below that collisions in our own halo may destroy clusters within a sufficiently small Galactocentric radius.

Figure~\ref{fig:relax} illustrates the initial relation~\eqref{eq:rM} and subsequent expansion~\eqref{eq:dyntime} for both $\beta = 10$ and $\beta = 1000$. However, the previous discussion assumes that dynamical heating dominates over cooling. In the opposite case, i.e.~if only a small dark matter fraction is made of solar-mass PBHs, one expects a different and more complex picture that can be compared to the time evolution of globular clusters. First, PBHs would sink towards the cluster centre while making the dark matter halo expand dynamically. But this phase must stop at some point and the PBH distribution must relax due to their regular interactions. Then, the PBH core would finally expand due to dynamical heating, leading to a further expansion of the particle dark matter halo.

This discussion assumes that the evolution of a cluster can be considered in isolation. In a hierarchical Poisson-clustering scenario, however, sub-clusters can be incorporated into larger haloes and lose their identity as they expand. In this context a characteristic survival scale is obtained by comparing the initial-radius and expansion contributions in Eq.~\eqref{eq:dyntime}. At fixed age and $\fPBH$, and neglecting the variation of the Coulomb logarithm, this gives $M_{\text{min}} \propto \fPBH\mspace{1mu}\langle\mspace{1mu}m\mspace{1mu}f\mspace{1mu}\rangle\,\beta^{-4/7}$ and $r_{\text{min}} \propto \big[ \fPBH\mspace{1mu}\langle\mspace{1mu}m\mspace{1mu}f\mspace{1mu}\rangle \big]^{1/3}\.\beta^{-10/21}$. Thus, changing from $\beta = 10$ to $\beta = 1000$ lowers these characteristic scales by factors of $14$ and $9$, respectively. Later we will compare the expected mass and radius of PBH clusters with those of UFDGs, the latter being represented by crosses in Fig.~\ref{fig:relax}. Suffice it to say at this point that a fit requires $\fPBH\mspace{1mu}\langle\mspace{1mu}m\mspace{1mu}f \rangle \simeq 10^{3}\.\Msun$ and $\beta = 1000$. The first high value is required to ensure the clusters are sufficiently massive; the second to ensure that their radius is sufficiently small (i.e.~due to the reduced expansion). The vertical lines show the thresholds for $F_{\crm}( M_{\crm},\,z_{0} ) = 0.9$ and $0.99$.

\section{Dynamical Constraints on PBH Clusters}
\label{sec:Dynamical-Constraints-on-PBH-Clusters}

\begin{figure*}[t]
	\vs{-3mm}
	\includegraphics[width=0.75\textwidth]{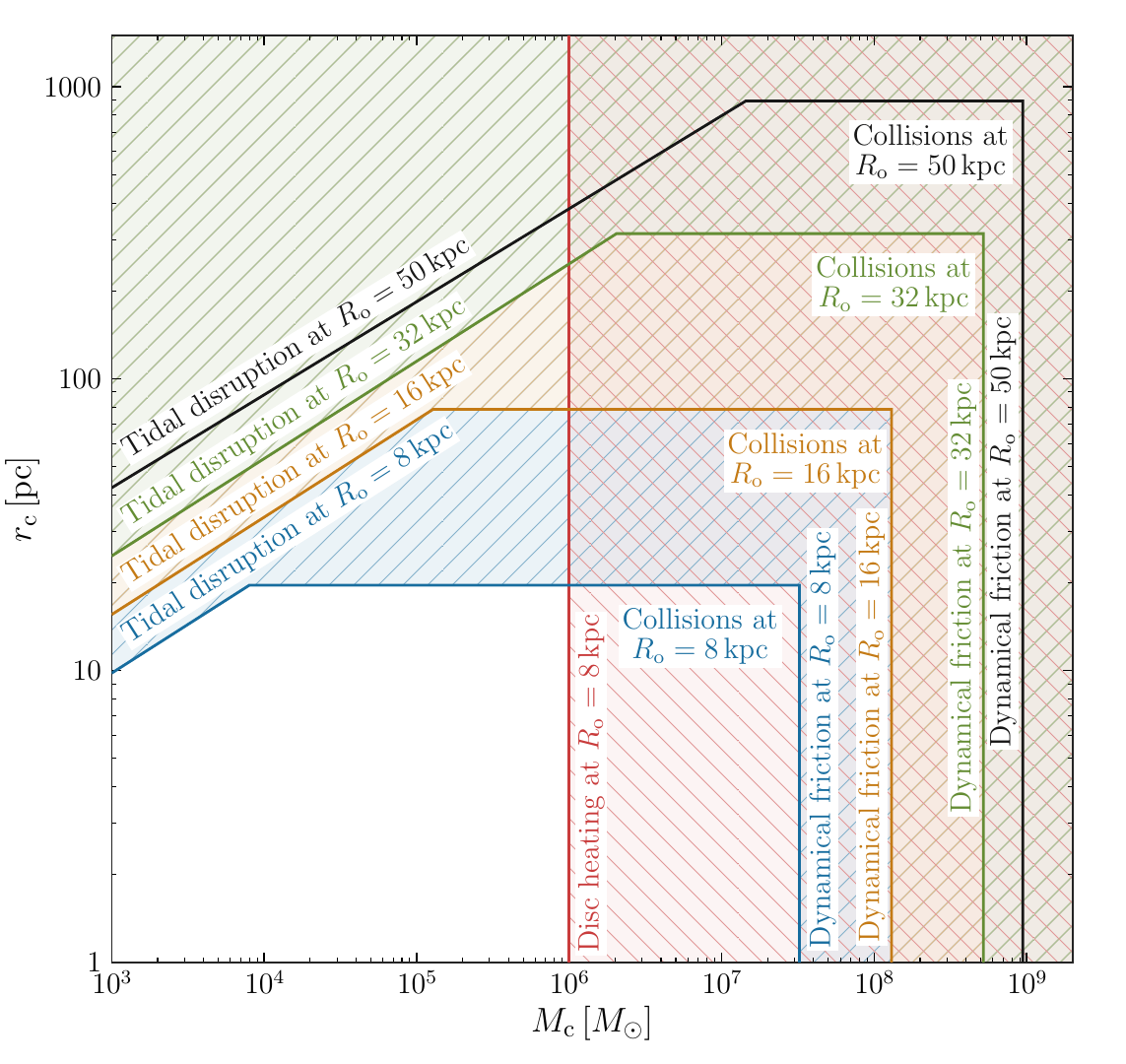}
	\caption{
		Regions in the $( M_{\crm},\mspace{1mu}r_{\crm} )$ plane for PBH clusters with $f=1$ that are excluded by various dynamical effects at Galactocentric distances of $8$\.kpc (blue region), $16$\.kpc (red region), $32$\.kpc (green region) and $50$\.kpc (black region). The effects are: 
	(1) Galactic disc heating at $8$\.kpc (middle vertical line); 
	(2) disruption by the Galactic tidal field (diagonal left lines); 
	(3) cluster collisions (middle horizontal lines); 
	(4) dynamical friction (vertical right lines). 
For smaller values of $f$, the collisional boundaries move up and the disc heating boundary moves to the right but the tidal and dynamical friction boundaries are unaffected. 
		}
	\label{fig:rcvsmc}
\end{figure*}

\noindent In this section we summarise the dynamical constraints on the mass and radius of PBH clusters, with the results shown in Fig.~\ref{fig:rcvsmc}. These constraints have been the focus of several previous papers~\cite{1987ApJ...316...23C, Carr:1997cn} and were reviewed in Ref.~\cite{Carr:2023tpt}. Here we update the earlier studies and correct some mistakes in our review. The discussion below is somewhat simplified, though more detailed than in Ref.~\cite{Carr:2023tpt}, and more precise equations are derived in the Appendix of Ref.~\cite{1987ApJ...316...23C}. We also include the dependence on $f_{\hrm}$, the fraction of the halo mass in PBHs, which was assumed to be $1$ in Ref.~\cite{1987ApJ...316...23C}. It should be stressed that every dynamical constraint is a potential signature, so the emphasis here is more on the positive evidence for clusters.

Although dynamical constraints preclude the possibility that galactic haloes comprise supermassive black holes (SMBHs), these constraints may be removed or weakened if haloes are made of supermassive clusters of black holes of more modest mass. Indeed, many years ago~\citet{1987ApJ...316...23C} suggested that the dark matter could consist of clusters of compact objects (MACHOs), each cluster having a mass of around $10^{6}\.\Msun$, this being the Jeans mass at decoupling. The dynamical limits may then be circumvented if the clusters are disrupted by collisions before the dynamical effects become operative. Although they did not discuss the nature of the MACHOs or assume any particular mass for them, it is natural to assume that they are PBHs if they provide the dark matter because other MACHO candidates would have to derive from baryons and could only comprise 20\% of the dark matter. The constraints shown in Fig.~\ref{fig:rcvsmc} relate to the Milky Way halo but the arguments could also be applied to other galaxies. 

\subsection{Collisional and Tidal Disruption}

Collisions between clusters will lead to their disruption if the velocity dispersion of the Galactic halo objects $\sigma_{\hrm}$ exceeds their internal dispersion $\sigma_{\crm}$. Otherwise collisions would lead to their merging. Outside the halo core ($R > R_{\crm}$), where $\sigma_{\hrm}$ is roughly constant, disruption requires
\be
	r_{\crm}
		> 
			\frac{ 2\.G M_{\crm} }{ 5\.\sigma_{\hrm}^{2} }
		\approx
			0.05
			\left(
				\frac{ M_{\crm} }{ 10^{6}\.\Msun }
			\right)
			{\rm pc}
			\, 
\ee
and this is generally satisfied. However, this condition is not likely to be satisfied before the formation of our galactic halo, so clusters should merely merge before then. There is some ambiguity in the precise definition of the cluster radius (outer radius, half-mass radius, virial radius); for simplicity we treat the cluster as a uniform sphere with a sharp cut-off at radius $r_{\crm}$. The value of $R_{\crm}$ is uncertain but it is probably in the range $2$\,--\,$8$\.kpc. The timescale for collisional disruption is constant within the core and increases further outward, so clusters will only be disrupted if the timescale at $R_{\crm}$ is less than the age of the Galaxy ($t_{\Grm} \approx 13\.$Gyr). This requires
\be
	r_{\crm}
		>
			\frac{ 3 }{ 5 }\.\sqrt{\frac{ \pi }{ 2 }}\.
			\frac{ R_{\crm}^{2} }
			{ f_{\hrm}\.\sigma_{\hrm}\.t_{\Grm} }
		\approx 
			0.9\,f_{\hrm}^{-1}\mspace{-2mu}
			\left(
				\frac{ R_{\crm} }{ 2\.{\rm kpc} }
			\right)^{\!2}
			{\rm pc}
\ee
and clusters will then be disrupted within a Galactocentric distance 
\be
	R_{\text{dis}}
		= 
			\left(
				\frac{ 5 }{ 3 }\.\sqrt{\frac{ 2 }{ \pi }}\.
				f_{\hrm}\.\sigma_{\hrm}\.r_{\crm}\.t_{\Grm}
			\right)^{\!1/2}
		\approx
			2.1\.\sqrt{f_{\hrm}}\.
			\sqrt{\frac{ r_{\crm} }{ {\rm pc} }}\;
			{\rm kpc}
			\, . 
			\label{eq:disrupt}
\ee
Requiring that the clusters survive at a Galactocentric distance $R_{\orm}$ implies an upper limit 
\be
	r_{\crm}
		<
			15 \.f_{\hrm}^{-1}\mspace{-2mu}
			\left(
				\frac{ R_{\orm} }{ 8\.{\rm kpc} }
			\right)^{\!2}
			{\rm pc}
			\, .
\ee
We show this line in Fig.~\ref{fig:rcvsmc} for $R_{\orm} = 8\.$kpc (our own Galactocentric distance) and $R_{\orm} = 16\.$kpc. This limit is independent of $m$ or the PBH mass function but it is interesting that the upper limit at $8\.$kpc corresponds to the minimum size of PBH clusters if $\fPBH\,m /\.\Msun \sim \Ocal( 1 )$. In this case, one expects PBH clusters to be disrupted by collisions in the inner part of the Milky Way halo but not at the larger Galactocentric distances typical of UFDGs.

Clusters will also be destroyed by the halo's tidal field at a Galactocentric radius $R_{\orm}$ if
\vs{-1mm}
\begin{align}
\begin{split}
	r_{\crm}
		&>
			\mspace{-1mu}
			\big(
				M_{\crm}/ M_{\hrm}( R_{\orm} )
			\big)^{1/3} 
			\left(
				3 - \frac{ \d\mspace{-1mu}\ln M_{\hrm} }{ \d\mspace{-1mu}\ln R }
			\right)^{\!-1/3}
			R_{\orm}
			\\[1.5mm]
		&\approx
			100\.
			\big(
				M_{\crm} / 10^{6}\.\Msun
			\big)^{1/3}
			\left(
				\frac{ R_{\orm} }{ 8\.{\rm kpc} }
			\right)^{\!2/3}
			{\rm pc}
			\, ,
	\label{eq:tidal}
\end{split}
\end{align}
where $M_{\hrm}( R_{\orm} )$ is the halo mass within $R_{\orm}$; this is assumed to scale as $R_{\orm}$ outside the core, so that the term in parentheses in the first line is just 2. We show this limit in Fig.~\ref{fig:rcvsmc} for $R_{\orm} = 8\.$kpc and $R_{\orm} = 16\.$kpc. Again, the predictions for PBH clusters cross these regions, implying that they should have been destroyed by the tidal field in the inner part of the halo. Furthermore, the condition for tidal and collisional disruption is even more likely to be satisfied as a result of expansion due to dynamical heating. Therefore tidal disruption adds to dynamical expansion to dilute small compact clusters into any larger host cluster up to a mass scale that corresponds to UFDGs. We note that clusters would not be destroyed by the tidal field of any larger clusters which contain them. This is because, in this case, one just replaces $M_{\hrm}( R_{\orm} )$ and $R_{\orm}$ in Eq.~\eqref{eq:tidal} by the mass and radius of the larger cluster; since $r_{\crm} \propto M_{\crm}^{5/6}$ from Eq.~\eqref{eq:rM}, this condition is necessarily satisfied.

\subsection{Dynamical Friction} 

Encounters between either SMBHs or clusters in our halo and lower-mass objects (such as the spheroid stars or halo objects) will transfer energy from these massive objects to the lower-mass population, causing the massive objects to drift towards the Galactic centre. We assume that the spheroid has a density profile~\cite{1985ApJ...299..633L}
\vs{-1mm}
\be
	\rho_{\srm}( R )
		=
			\rho_{1}
			\begin{cases}
				\big(
					R_{1} / R
				\big)^{9/5}
					& \big( R < R_{1} \big)
					\\[1mm]
				\big(
					R_{1} / R
				\big)^{3}
					& ( R > R_{1} ) 
				\, ,
			\end{cases}
\ee
where $R_{1} = 800\.$pc and $\rho_{1} = 1.8\.\Msun\.$pc$^{-3}$, and that the halo has an isothermal density profile
\vs{-1mm}
\be
	\rho_{\hrm}( R )
		=
			\begin{cases}
				\rho_{\crm}
					& ( R < R_{\crm} )
					\\[1mm]
				\rho_{\orm}\.( R_{\orm} / R )^{2}
					& ( R > R_{\crm} ) 
				\, ,
			\end{cases}
\label{eq:rhohalo}
\ee
where $R_{\orm} = 8\.$kpc is our own Galactocentric distance, $\rho_{\orm} = 0.01\.\Msun$\.pc$^{-3}$ is the local halo density, $R_{\crm}$ is the halo core radius and $\rho_{\crm} = \rho_{\orm}\.( R_{\orm} / R_{\crm} )^{2} \approx 0.16\.( R_{\crm}/2\.{\rm kpc} )^{-2}\.\Msun$\.pc$^{-3}$ is the density within the halo core. The radius beyond which the halo density dominates the spheroid density is $R_{2} \approx 1.8\.( R_{\crm}/2\.{\rm kpc} )^{2/3}\.$kpc and this is always less than $R_{\crm}$. However, the mass within the radius $R$ will be dominated by the spheroid up to a larger radius $R_{3}$, given explicitly in Ref.~\cite{Carr:1997cn}. This is well beyond $R_{2}$ and generally larger than $R_{\crm}$.

The clusters or SMBHs will reach the nucleus within the age of the Galaxy ($t_{\Grm} \approx 13\.$Gyr) if they initially lie within a Galactocentric radius 
\be
	R_{\text{df}}
		\approx
			\begin{cases}
				1.3\.{\rm kpc}
				\left(
					\frac{ M_{\crm} }{ 10^{6}\mspace{1mu}\Msun }
				\right)^{\!10/21}
					&
					\big(
						R_{\text{df}}
							<
								R_{1}
						\;\Rightarrow\;
						M_{\crm}
							<
								M_{1}
							=
								4 \times 10^{5}\.\Msun
					\big)
					\\[2mm]
				2.2\.{\rm kpc} 
				\left(
					\frac{ M_{\crm} }{ 10^{6}\mspace{1mu}\Msun }
				\right)^{\!2/3} f_1(R_{\text{df}})
					&
					\big(
						R_{1}
							<
								R_{\text{df}}
							<
								R_{2}
						\;\Rightarrow\;
						M_{1}
							<
								M_{\crm}
							<
								M_{2}
							=
								10^{6}\mspace{1.5mu}( R_{\crm}/2\.{\rm kpc} )\.\Msun
					\big) 
					\\[2mm]
				1.4\.{\rm kpc}
				\left(
					\frac{ M_{\crm} }{ 10^{6}\mspace{1mu}\Msun }
				\right)^{\!1/2} f_2(R_{\text{df}})
					&
					\big(
						R_{\text{df}}
							>
								R_{3}
						\Rightarrow
						M_{\crm}
							>
								M_{3}
							=
								10^{6}\mspace{1.5mu}( R_{3} / 1.4\,{\rm kpc} )^{2}\.
								\Msun
					\big) 
				\, .
			\end{cases}
			\label{eq:DF}
\ee
This is generally less than the halo core radius, although the latter is very uncertain. Here $R_{3}$ corresponds to the transition to the outer-halo approximation for dynamical friction. The first two expressions were derived in Ref.~\cite{1987ApJ...316...23C}, while the third allows for the dynamical friction effects of the halo objects themselves if they have an extended mass function and is taken from Ref.~\cite{Carr:1997cn}. The functions $f_{1}$ and $f_{2}$ depend weakly on the ratio $R_{\text{df}}/R_{1}$. We omit the expression for $R_{2} < R_{\text{df}} < R_{3}$, because it is complicated and leads to a discontinuity at the transition values of $R$. For a given Galactocentric distance, this implies that there can be no SMBHs or clusters with mass exceeding the value given implicitly by Eq.~\eqref{eq:DF}. This corresponds to the vertical lines on the right of Fig.~\ref{fig:rcvsmc}. Combined with the tidal disruption lines on the left and the horizontal collisional disruption lines, this specifies a permitted ($M_{\crm},\mspace{1mu}r_{\crm}$) wedge for each value of $R$.

However, there is another interesting constraint. In the SMBH case, the mass dragged into the Galactic nucleus for $R_{\text{df}} < R_{\crm}$ is 
\be
	M_{\Nrm}
		=
			\frac{ 4\mspace{1mu}\pi }{ 3 }\.f_{\hrm}\.
			\rho_{\crm}\.R_{\text{df}}^{3}
		\approx
			\begin{cases}	
				7 \times 10^{8}\.f_{\hrm}\mspace{-2mu}
				\left(
					\frac{ M_{\crm} }{ 10^{6}\.\Msun }
				\right)^{\!10/7}\!
				\left(
					\frac{ R_{\crm} }{ 2\.{\rm kpc} }
				\right)^{\!-2}\Msun
					& \big( R_{\text{df}} < R_{1} \big)
				\\[2mm]
				2 \times 10^{9}\.f_{\hrm}\mspace{-2mu}
				\left(
					\frac{ M_{\crm} }{ 10^{6}\.\Msun }
				\right)^{\!2}\!
				\left(
					\frac{ R_{\crm} }{ 2\.{\rm kpc} }
				\right)^{\!-2}\Msun
					& \big( R_{1} < R_{\text{df}} < R_{2} \big)
				\\[2mm]
				8 \times 10^{9}\.f_{\hrm}\mspace{-2mu}
				\left(
					\frac{ M_{\crm} }{ 10^{6}\.\Msun }
				\right)^{\!1/2}\!\.\Msun
					& \big( R_{\text{df}} \gg R_{2} \big)
					\, .
			\end{cases}
\ee
$M_{\Nrm}$ will exceed the observational upper limit of $4 \times 10^{6}\.\Msun$~\cite{1998ApJ...509..678G} unless one imposes an upper limit on the SMBH mass:
\be
	M_{\crm}
		<
			\begin{cases}
				3 \times 10^{4}\.( R_{\crm} / 2\.{\rm kpc} )^{1.4}\.
				f_{\hrm}^{-0.7}\.\Msun
					& \big( M_{\crm} < M_{1} \big)
					\\[1.5mm]
				4 \times 10^{4} \.( R_{\crm} / 2\.{\rm kpc} )\.
				f_{\hrm}^{-1/2}\.\Msun
					& \big( M_{1} < M_{\crm} < M_{2} \big)
					\\[1.5mm]
				0.3\.f_{\hrm}^{-2}\.\Msun
					& \big( M_{\crm} \gg M_{2} \big)
					\, .
				\end{cases}
\ee
However, this conclusion can be avoided for halo clusters since they may be disrupted by collisions before dynamical friction can drag them into the nucleus. Indeed, clusters will be destroyed within the Galactocentric radius~\eqref{eq:DF} at which dynamical friction would drag them into the nucleus if 
\be
	r_{\crm}
		> 
			\begin{cases}

				0.4\,f_{\hrm}^{-1}\mspace{-3mu}
				\left(
					\frac{ M_{\crm} }{ 10^{6}\.\Msun }
				\right)^{20/21}
				{\rm pc}
					& \big( M_{\crm} < M_{1} \big)
					\\[2mm]
				0.4\,f_{\hrm}^{-1}\mspace{-3mu}
				\left(
					\frac{ M_{\crm} }{ 10^{6}\.\Msun }
				\right)^{\!4/3}
				{\rm pc}
					&
					\big( M_{1} < M_{\crm} < M_{2} \big)
					\\[2mm]
				0.8\,f_{\hrm}^{-1}\mspace{-3mu}
				\left(
					\frac{ M_{\crm} }{ 10^{6}\.\Msun }
				\right)
				{\rm pc}
					&
					\big( M_{\crm} \gg M_{2} \big)
					\, .
			\end{cases}
\ee
This condition is necessary in order to obviate the upper limit on the mass which would apply for SMBHs. For any Galactocentric radius, this corresponds to the intersection of the collisional disruption and dynamical friction lines.
\vs{-2mm}

\subsection{Dynamical Heating of Milky Way Disc}
\label{sec:Dynamical-Heating-of-Milky-Way-Disc}

As halo objects traverse the Galactic disc, they will impart energy to the stars there. This will lead to a gradual puffing up of the disc, with older stars being heated more than younger ones. This problem was first studied by~\citet{1984ASIC..117..351L} in the context of a very general analysis of disc heating and then by~\citet{1985ApJ...299..633L}, who argued that SMBHs could generate the observed disc thickening if they provide the dark matter ($\fPBH \sim 1$). In particular, they claimed that this could explain: 
	(1) why the velocity dispersion of disc stars $\sigma$ scales with age as $t^{1/2}$;
	(2) the relative velocity dispersions in the radial, azimuthal and vertical directions; and
	(3) the existence of a high-energy tail of stars with high velocities (cf.~\citet{1985A&A...149..408I}). The velocity dispersion of the stars evolves as
\vs{-2mm}
\be
	\sigma( t )
		=
			\sqrt{
				\sigma_{\orm}^{2}
				+ \alpha\.G^{2}\.n\mspace{1mu}M^{2}\.t\./v_{\crm}
			}
			\, ,
\ee
where $v_{\crm}$ is the circular velocity and $\alpha$ is a function of $v_{\crm} / \sigma_{\hrm}$ and $M / M_{\hrm}$. At a time $t_{\srm}$, corresponding to the age of the disc, this yields a limit
\vs{-2mm}
\be
	\rho_{\hrm}\.M
		<
			\sigma_{\rm obs}^{2} v_{\crm} /
			\big(
				{\alpha\.G^{2}\.t_{\srm}}
			\big)
			\, .
\ee
Combining this with the local halo density, $\rho_{\hrm} = n\.M \approx 0.01\.\Msun\.{\rm pc}^{-3}$, then gives 
\be
	M
		<
			2 \times 10^{6}\.
			\big(
				\sigma_{\rm obs} / 60\.{\rm km\,s}^{-1}
			\big)^{2}\.
			\big(
				t_{\srm} /13\.{\rm Gyr}
			\big)^{-1}\.\Msun
			\, .
\ee
\citet{1988LNP...306..100W} claimed that black hole heating could also explain the dependence of the velocity dispersion upon Galactocentric distance. If some other mechanism explains disc heating, these arguments still give an upper limit on the fraction of dark matter in SMBHs:
\be
	f_{\hrm}( M )
		<
			{\text{min}}
			\Big[
				1,\mspace{1mu} 
				\big(
					M / 10^{6}\.\Msun
				\big)^{-1}
			\Big]
			\, .
\ee
For example, this would apply if the heating were due to giant molecular clouds, which is the more mainstream explanation.

Although disc heating can no longer be regarded as providing evidence for {\it supermassive} PBHs, because this is now excluded by other constraints, Carr \& Lacey pointed out that it may still provide evidence for clusters of much smaller PBHs~\cite{1987ApJ...316...23C}. Indeed, since PBH clusters with masses close to the value of $10^{6}\.\Msun$ required for disc heating are {\it expected} to form, this might be regarded as a prediction if PBHs provide the dark matter. The stellar velocity change in close encounters is smaller than in the SMBH case because of the finite size of the clusters but most of the disc heating comes from distant encounters anyway. However, scattering off the clusters can only produce the high-velocity disc stars with $v > 100\,$km\,s$^{-1}$ if $r_{\crm} < 0.5\,( M_{\crm} / 10^{6}\.\Msun )\.{\rm pc}$ and the associated constraint is shown by the red region in Fig.~\ref{fig:rcvsmc}. Note that disc heating should be modified within the radius $R_{\text{dis}}$ given by Eq.~\eqref{eq:disrupt}. Indeed, since $\sigma^{2} \propto \rho_{\hrm}\.t_{\text{dis}}$ and $t_{\text{dis}} \propto \rho_{\hrm}^{-1}$ there, the predicted disc velocity dispersion should be constant within this radius.

\section{Ultra-Faint Dwarf Galaxies, Little Red Dots, Blue Monsters}
\label{sec:Ultra--Faint-Dwarf-Galaxies,-Little-Red-Dots,-Blue-Monsters}

\begin{figure}
	\vs{-5mm}
	\includegraphics[width=0.65\textwidth]{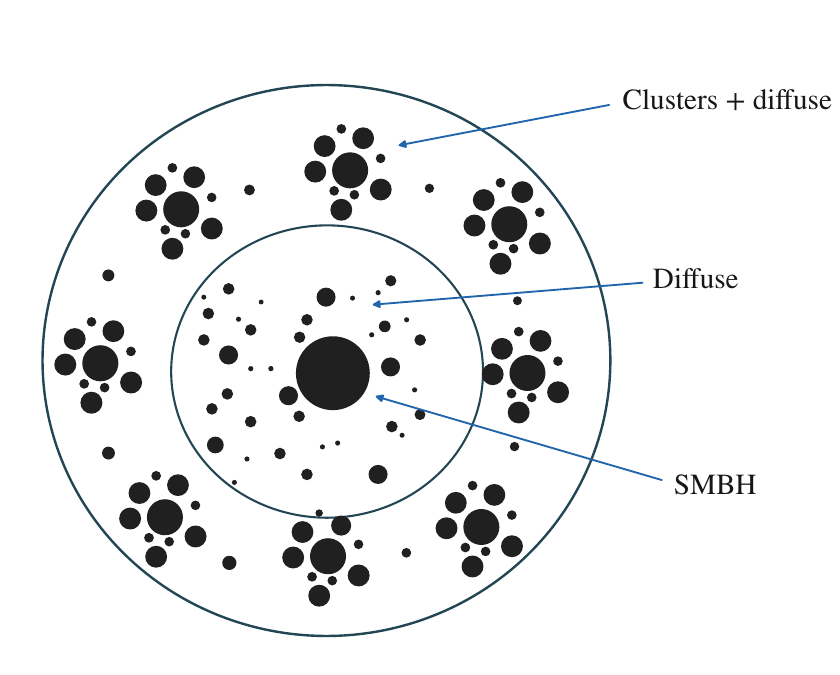}
	\caption{
		Schematic illustration of the expected Galactic distribution of clustered and diffuse PBHs.
		}
	\label{fig:clusters}
\end{figure}

\noindent The dynamical evolution of PBH clusters in galaxies depends on their Galactocentric radius. Some of the cluster mass will be lost at each radius, giving rise to a smoothly distributed PBH contribution. However, the fraction of mass lost will increase with decreasing radius, implying an increase in the smooth-to-clustered ratio. Indeed, within some critical Galactocentric radius, the clusters should be completely disrupted. If the PBHs have an extended mass function, one would also expect the more massive ones to migrate to the centre of the galaxy as a result of dynamical friction. They could then merge to form a larger central black hole, it being observed that all galactic nuclei harbour such a black hole. The situation is depicted qualitatively in Fig.~\ref{fig:clusters}, although more careful numerical calculations are clearly required to investigate this scenario in detail. We consider three applications of this proposal{\,---\,}UFDGs, LRDs and what are termed ``Blue Monsters''{\,---\,}and also study the implications of accretion.

\subsection{Ultra-Faint Dwarf Galaxies and Accretion in PBH Clusters}
\label{sec:Ultra--Faint-Dwarf-Galaxies-and-Accretion-in-PBH-Clusters}

We first discuss whether UFDGs could be associated with PBH clusters. The disc-heating argument in Sec.~\ref{sec:Dynamical-Heating-of-Milky-Way-Disc} can also be applied to constrain the PBH abundance in UFDGs~\cite{2020MNRAS.492.5247S}. This requires $\fPBH < 1$ for $M \gtrsim 10\.\Msun$, with the precise limit depending on model assumptions and the possible effect of an intermediate-mass black hole (IMBH) at their centre. However, it was suggested in Ref.~\cite{Clesse:2017bsw} that this argument can be reversed, with the observed mass and radius of UFDGs indicating that a large fraction of dark matter comprises stellar-mass compact objects. Although Ref.~\cite{2020MNRAS.492.5247S} claims that this is excluded, the presence of Poisson-induced clusters was not allowed for.

Figure~\ref{fig:rcvsmc} uses the tables of Ref.~\cite{2019ARA&A..57..375S} to show the stellar half-light radius $R_{1\mspace{-1mu}/\mspace{-0.5mu}2}$ and the virial mass $M_{1\mspace{-1mu}/\mspace{-0.5mu}2}$ enclosed within that radius, assuming $M_{1\mspace{-1mu}/\mspace{-0.5mu}2} = 4\.\sigma_{\text{los}}^{2} R_{1\mspace{-1mu}/\mspace{-0.5mu}2}/G$, where $\sigma_{\text{los}}$ is the line-of-sight velocity dispersion. These quantities are not identical to the total cluster mass and radius, even though they are compared in Fig.~\ref{fig:rcvsmc}. As indicated by Fig.~\ref{fig:relax}, we need $\fPBH\mspace{1mu}\langle\mspace{1mu}m\mspace{1mu}f \rangle \sim 10^{3}\.\Msun$ and $\beta \sim 1000$ to explain UFDGs. Even if most of the dark matter is in PBHs of $1\.\Msun$, this is possible if we have another population with $M \sim 10^{6}\.\Msun$ and $f \sim 10^{-3}$, so this requires the existence of SMBHs at the centres of UFDGs. As discussed in Sec.~\ref{sec:Thermal-History-Model-and-Multimodal-Mass-Functions}, this is expected in the {\it Thermal History} model. A comparison of the $\fPBH\mspace{1mu}\langle\mspace{1mu}m\mspace{1mu}f \rangle = 3\.\Msun$ and $\fPBH\mspace{1mu}\langle\mspace{1mu}m\mspace{1mu}f \rangle = 30\.\Msun$ cases with $\beta = 10$ and $\beta = 1000$ is shown in Fig.~\ref{fig:relax}. The characteristic minimum mass and radius decrease with $\beta$, so any inferred correspondence with UFDGs must account for both this and the distinction between stellar and halo radii. The qualitative correspondence with UFDGs was first noted in Ref.~\cite{Clesse:2017bsw}, which also proposed that PBH accretion could contribute to the large mass-to-light ratios of UFDGs. Figure~\ref{fig:rcvsmc} shows that the dynamical limits at the Galactocentric distances of the UFDGs are only weak.

In order to determine how realistic it is to obtain ultra-faint clusters with similar properties to UFDGs (i.e.~without a dominant stellar component), it is interesting to study the accretion of baryons and possible CDM particles by PBHs inside a Poisson cluster. The current explanation for the faintness of UFDGs and the existence of a very old stellar population formed over a few hundred million years invokes the reionisation of the Universe or reheating of the gas by supernova feedback. Together with the shallowness of the gravitational potential well, this would have prevented the gas from recollapsing. This only occurs under stringent conditions and it has long been a challenge to explain UFDG properties. If dark matter is composed even partially of PBHs, then this provides an alternative mechanism not only for UFDG formation through Poisson fluctuations, through the seed effect, or through the growth of enhanced primordial fluctuations, but also for their faintness, despite the accretion of matter in such a dense black hole environment. 

We use a simple spherical accretion model to estimate the surviving baryon fraction as a function of time and cluster mass and radius. For order-of-magnitude estimates, we assume that the PBH and baryon densities in such a cluster are homogeneous. In reality, the stellar and dark matter profiles, as well as the total dark matter mass, are not easy to estimate. We extend the calculation of Ref.~\cite{Clesse:2017bsw}, where this mechanism was proposed, to any broad PBH mass function and velocity distribution. The accretion process explored in this section therefore complements studies of PBH accretion in the early Universe and its impact on the CMB anisotropies, as well as local probes using radio, X-ray and gamma-ray observations towards the Galactic centre.

For a single PBH, the accreted mass is obtained using the Bondi--Hoyle accretion rate,
\begin{equation}
	\dot m_{\rm BH}
		=
			\frac{ 4\mspace{1mu}\pi \lambda\.\rho_{\infty}\.G^{2}\.m^{2} }
			{ v^{3} } 
			\, ,
\end{equation}
where $v$ is the relative velocity between the PBH and the gas, which is of order the virial velocity of the cluster, $v_{\text{vir}} = \sqrt{G M_{\crm} / 2\.r_{\crm}}$, and $\rho_{\infty}$ is the gas density (or particle density for dark matter accretion) far from the PBH. The parameter $\lambda$ is the accretion efficiency, whose value depends on the accretion physics and can be $\Ocal( 1 )$ for spherical accretion or $\Ocal\big( 10^{-2}$\,--\,$10^{-1} \big)$ for disc accretion. We adopt $\lambda = 1$ as a reference value and discuss below how our conclusions depend on this choice. The gas density is $\rho_{\infty} = f_{\brm}\.\rho_{\text{cl}} = 3\.f_{\brm}\.M_{\crm} / ( 4\mspace{1mu}\pi\.r_{\crm}^{3} )$, where $\rho_{\text{cl}}$ is the cluster density and $f_{\brm}$ is the unaccreted baryon fraction. The mass accretion rate for the PBH population in a single cluster is then 
\begin{equation}
	\dot m^{\text{tot}}_{\rm BH}
		=
			\frac{ 4 }{ 3 }\.\pi\.r_{\crm}^{3}
			\int\d\mspace{-1mu}\ln m\;\frac{ \dot m_{\rm BH} }{ m }\.
			\rho_{\text{cl}}\.\fPBH\.f( m )
			\, .
\end{equation}
We assume a Maxwellian distribution, which boosts the accretion rate by a factor
\begin{equation}
	b_{v}
		=
			v_{\text{vir}}^{3}
			\int_{0}^{\infty}
			\d\mspace{1mu}v\;
			\frac{ f( v ) }
			{ \big( v^{2} + c_{\srm}^{2} \big)^{3/2} }
		=
			\frac{ 4 }{ \sqrt{\pi} }
			\int_{0}^{\infty}
			\d\mspace{1mu}x\;
			\frac{
				x^{2}\,\erm^{-x^{2}}
			}{
				\big( x^{2} + s^{2} \big)^{3/2}
			}
		=
			U\big( 3/2,\mspace{1mu}1,\mspace{1mu}s^{2} \big)
			\, ,
\end{equation}
where $c_{\srm}$ is the sound speed, $U$ is the confluent hypergeometric function and we have put $x \equiv v / v_{\text{vir}}$ and $s \equiv c_{\srm} / v_{\text{vir}}$ in the last integral. Turbulence or coherent gas streaming can be included in the effective velocity. $b_{v} = 10$ corresponds to $s = 0.007$, representing an exceptionally cold gas and enhanced accretion rather than a typical value. Numerically, $b_{v} = 9.9,\.3.9,$ and $0.36$ for $s = 0.007,\.0.1$ and $1$, respectively. 

\begin{figure*} 
	\centering
	\vs{-1mm}
	\includegraphics[width=0.8\linewidth]{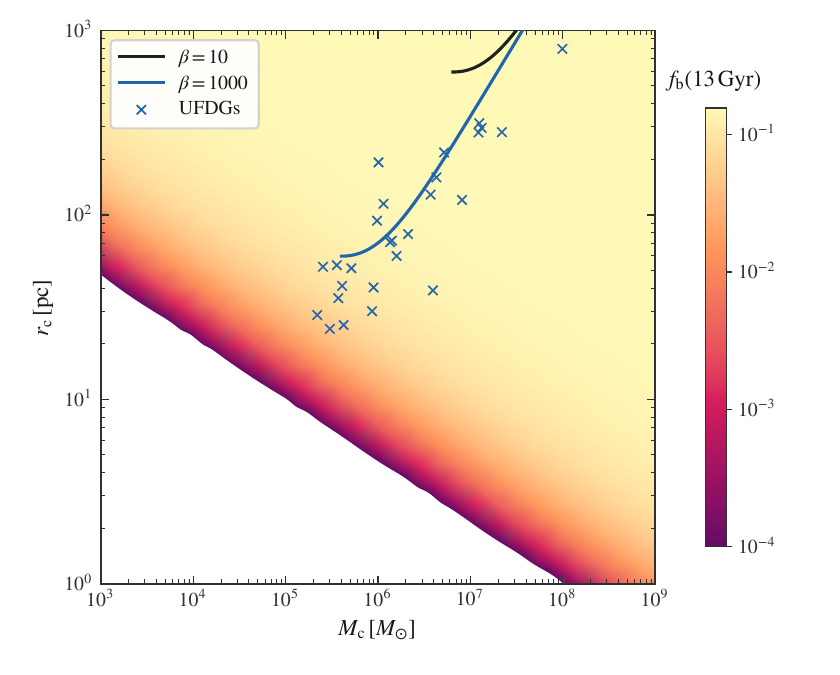}
	\caption{Remaining baryon fraction $f_{\brm}$ after $13\,$Gyr, due to Bondi--Hoyle accretion onto PBHs in Poisson clusters of mass $M_{\crm}$ and radius $r_{\crm}$, assuming $\lambda = 1$, $\fPBH = 1$ and $\langle\mspace{1mu}m\mspace{1mu}f \rangle = 3\.\Msun$. The solid black and blue curves show the predicted cluster relations for $\beta = 10$ and $\beta = 1000$, respectively. The baryon-fraction colour scale assumes a fixed cluster mass and radius and $b_{v} = 10$. It is unchanged when only $\beta$ is varied. The crosses indicate the stellar size and total mass for UFDGs but the latter does not include the gas fraction. 
		} 
	\label{fig:accretion} 
\end{figure*}

Combining the above equations gives a baryon accretion rate 
\begin{equation}
	\dot f_{\brm}
		=
			-\.6\,\sqrt{2}\.\lambda\.b_{v}\.f_{\brm}\.\fPBH\.
			\big\langle
				\mspace{1mu}m\mspace{1mu}f
			\big\rangle\,
			\sqrt{\frac{ G }{ M_{\crm}\.r_{\crm}^{3} }} 
			\; .
\end{equation}
Assuming that the PBH masses do not change significantly, which is true if the PBHs constitute a large part of the dark matter, the expected time evolution of the baryon fraction is
\begin{equation}
	f_{\brm}( t )
		\approx 
			f_{\brm}^{\irm}\.
			\exp\!
			\left[
				-\.5.6\,\lambda\.\fPBH\.
				\frac{ \langle\mspace{1mu}m\mspace{1mu}f \rangle }{\.\Msun }
				\left(
					\frac{ M_{\crm}\,r_{\crm}^{3} }
					{ 10^{6}\.\Msun\.{\rm pc}^{3} }
				\right)^{\!-1/2 }
				\left( \frac{ t }{ {\rm Gyr} }\right)
			\right]
			,
			\label{eq:baryonfraction} 
\end{equation}
where $f_{\brm}^{\irm} = \Omega_{\brm} / (\Omega_{\crm} + \Omega_{\brm}) \approx 0.15$ is the initial baryon fraction and we assume $b_{v} \approx 10$ in the last expression. In fact, the initial fraction is probably lower than this because the Poisson clusters form at high redshift and not all the gas will fall into the potential wells. $b_{v} \approx 10$ is also likely an overestimate but the two effects may compensate each other. Our calculation therefore describes an extreme case, but it can easily be adapted to other initial and environmental conditions. The surviving baryon fraction after $13\,$Gyr is shown in Fig.~\ref{fig:accretion} for different cluster sizes and masses. It indicates a clear transition between clusters where accretion is inefficient and more compact clusters in which the gas is accreted entirely. Figure~\ref{fig:accretion} identifies clusters in which the present model predicts efficient gas depletion. However, it does not establish that accretion removes the gas before the onset of star formation since that requires a redshift-dependent comparison of the accretion, free-fall, cooling, star-formation, reionisation and feedback timescales. 

Changing $\lambda$, $b_{v}$, $\fPBH$ or $\langle\mspace{1mu}m\mspace{1mu}f \rangle$ modifies the argument of the exponential in Eq.~\eqref{eq:baryonfraction} and therefore moves the transition line up or down in the $(M_{\crm},\mspace{1mu}r_{\crm})$ plane. The smaller radius obtained with $\beta = 1000$ increases the depletion exponent, which scales as $r_{\crm}^{-3/2}$. For an evolving cluster, the $t\,r_{\crm}^{-3/2}$ dependence in Eq.~\eqref{eq:baryonfraction} must be replaced with $\int_{0}^{t}\!\d\mspace{1mu}t'\,r_{\crm}(t')^{-3/2}$. For mass distributions with a peak at around a solar mass, the transition occurs at the scale of the most compact UFDGs. A more precise analysis, allowing for the mass evolution that could be important for intermediate-mass PBHs, as well as a more realistic PBH density profile that would enhance accretion towards the cluster centre, might lead to a smoother transition from bright dwarf galaxies to UFDGs in the form of PBH clusters.

When only a small fraction of the dark matter is PBHs, with the rest being CDM particles, direct capture of collisionless particles or interactions between them and the PBHs may modify the cluster structure through dynamical friction and heating. However, Bondi--Hoyle accretion does not apply in this case, so we do not try to quantify the analysis in this situation.

\subsection{Little Red Dots}

The discovery of LRDs with {\it JWST} has revealed a population of compact, very red sources at $z \sim 4$\;--\,$10$ that frequently exhibit broad Balmer emission lines and unusually large inferred ratios of black hole mass to stellar mass. These objects are abundant and often unresolved, with spectral energy distributions consistent with heavily enshrouded or compact active galactic nuclei (AGN)~\cite{2025ApJ...986..126K}. Follow-up X-ray analyses show that many LRDs are X-ray weak, suggesting significant obscuration or atypical accretion states~\cite{Ananna:2024jug}. High-redshift spectroscopic observations, including {\it CAPERS-LRD-z9} at $z \sim 9$, demonstrate the existence of a population of rapidly accreting massive black holes very early in cosmic history~\cite{2025ApJ...989L...7T}. The implied black hole masses, in some cases exceeding $10^{7}$\,--\,$10^{8}\.\Msun$ at $z > 6$, intensify the tension in early SMBH assembly scenarios.

PBHs offer a natural pathway in which the seed is of primordial origin. Supermassive PBHs accreting at sub-Eddington rates can reproduce LRD luminosities and number densities without requiring extreme baryonic growth histories~\cite{Hai-LongHuang:2024gtx}. For example, \citet{Matteri:2025vnv} suggest that a population of $10^{4}$\,--\,$10^{5}\.\Msun$ PBHs in low-mass haloes, accreting at sub-Eddington rates, can provide the very early UV luminosity density observed by JWST at $z \gtrsim 17$\,--\,$25$, thereby explaining the apparent overabundance of high-redshift galaxies without invoking standard stellar or direct-collapse seeds. At the higher-mass end, \citet{Huang:2024aog} argue that a population of initially clustered supermassive PBHs with $\sim 10^{9}\.\Msun$ may naturally explain the JWST high-redshift galaxies. An alternative framework invokes enhanced small-scale clustering of lighter PBHs, leading to hierarchical mergers that assemble $10^{5}$\,--\,$10^{8}\.\Msun$ seeds by $z \sim 10$; such scenarios may also generate a stochastic gravitational-wave background~\citep{Zhang:2025tgm}. 

Extending this framework, \citet{Zhang:2026uum} derive the SMBH mass function from the PBH cluster mass distribution induced by long--short mode coupling, finding that runaway mergers followed by accretion with duty cycles of $1$\,--\,$10\%$ can reproduce the shape inferred for LRDs. The seed effect and low halo spin parameters can account for the large black-hole-to-stellar mass ratios and compact sizes, respectively, while the high-mass tail above $10^{8}\.\Msun$ can enhance the cumulative stellar mass density at $z \sim 10$ and alleviate the tension with JWST observations. 

Recently, Ref.~\cite{Zhang:2026fhn} has provided numerical support for this environment-dependent picture through cosmological hydrodynamical simulations of individual PBH seeds with initial masses $10^{5}\.\Msun$ and $10^{7}\.\Msun$, bracketing the $10^{6}\.\Msun$ scale associated with $e^{+}e^{-}$ annihilation. Relatively isolated seeds produce faint, metal-poor systems strongly dominated by their central black holes, whereas sustained inflow and hierarchical assembly in overdense regions naturally build compact stellar hosts. Most strikingly, one $10^{7}\.\Msun$ seed passes through a compact LRD-like stage at $z \simeq 8.4$ with $R_{1/2} \simeq 110\.{\rm pc}$ and $M_{\bullet} / M_{\star} \simeq 10$, before its host develops into a more extended galaxy by $z \simeq 4.1$. Here $M_{\bullet}$ and $M_{\star}$ are the black hole mass and the total stellar mass, respectively. These results provide a concrete evolutionary route by which massive PBH seeds can naturally generate LRDs as a distinctive phase of early black hole--galaxy co-evolution.

More generally, for a wide spectrum of PBH seed masses, it has been shown that even a small PBH dark matter fraction can catalyse early structure formation and accelerate SMBH assembly~\citep{Yuan:2023bvh}. Related work demonstrates that clustered PBHs, undergoing dynamical friction, binary formation and runaway mergers, can yield $10^{4}$\,--\,$10^{5}\.\Msun$ seeds at $z \gtrsim 10$; these could subsequently power an LRD phase with modest accretion~\citep{Ziparo:2024nwh}. Cosmological simulations with massive PBH seeds and self-consistent accretion and feedback find that systems with low stellar mass arise naturally and can resemble even the most extreme LRD-like systems~\citep{Zhang:2025oyl, 2025ApJ...987..185Z}. The {\it Thermal History} model~\cite{Carr:2019kxo}, despite its simplicity, naturally provides PBHs with the required abundance and mass, the horizon mass at the epoch of $e^{+}e^{-}$ annihilation being $10^{6}\.\Msun$. 

Note that quantum diffusion during inflation produces a significant non-perturbative primordial non-Gaussianity~\cite{Ezquiaga:2019ftu}, which enhances both the probability of collapse to PBHs and the formation of $10^{6}$\,--\,$10^{9}\.\Msun$ clusters with a range of PBH masses. Some of these clusters inhabit dense gas regions, so the PBHs accrete gas and re-radiate at longer wavelengths. As first pointed out in Ref.~\cite{Siles:2024yym}, this may explain the observed clumpiness~\cite{Tanaka:2024, Rinaldi:2024fgn} of the LRDs detected by the Near-Infrared Camera (NIRCam) aboard JWST~\cite{Matthee:2023utn, Greene:2023}. As the gas heats up around the PBHs, they start merging and perhaps all of them end up as a single black hole of $10^{9}$\,--\,$10^{10}\.\Msun$ at the centre of the cluster, thus explaining the SMBHs seen by JWST at high redshift in all galaxies. Although LRDs have been discovered up to $z\sim9$, it is likely that they can be traced further back in the JWST Deep Field. The abundance and spatial distribution of LRDs are still uncertain, but this may eventually provide a clue as to their possible connection with PBH clusters, whether of Poisson or non-Gaussian origin.

According to this scenario, the astrophysical environment (mostly hydrogen gas) which PBH clusters inhabit determines their subsequent evolution. Isolated PBH clusters (i.e.~those far from large concentrations of gas) will be rather inert and survive until today with minor changes in size and composition~\cite{Siles:2024yym}. These could populate outer galactic haloes, as illustrated in Fig.~\ref{fig:clusters}. On the other hand, PBH clusters in clouds with a large column density of pristine hydrogen will accrete the gas and heat the cluster environment. This will induce diffusion and energy losses that will cause most of the PBHs to merge into a single black hole at the centre, giving rise to the observed SMBHs at the centres of galaxies and the bright LRDs seen by JWST at high redshift. Between these two extremes are PBH clusters within gas clouds of moderate density, whose dynamics will prompt star formation \cite{Cappelluti:2021usg} and create the observed cross-correlation between the source-subtracted infrared (IR) and X-ray backgrounds seen by Kashlinsky {\it et al.} \cite{Kashlinsky:2018mnu}. This last environment likely leads to the UFDGs with large mass-to-light ratios observed in the Milky Way.

Recent observations of the LRD Abell 2744-QSO1 give some insights into its origin~\citep{Juod_balis_2026, Maiolino:2025tih}. The gas velocity has been measured and is compatible with a Keplerian value. This indicates that most of the LRD mass is concentrated in a central SMBH rather than in a Poisson cluster of PBHs with a size comparable to that of the LRD itself. Furthermore, the SMBH mass is comparable to the stellar mass, the gas has a very low metallicity and there is no nearby UV emission that might be associated with SMBH formation through direct gas collapse. The LRD properties are therefore hard to explain with low-mass seeds and would only be compatible with quite specific direct collapse scenarios, suggesting a possible PBH origin. In this case, the LRD would be produced by the seed effect rather than the Poisson effect. 

The seed-assisted scenario proposed by \citet{Garcia-Bellido:2026LRD} offers a concrete route to early black hole growth: a pre-existing PBH nucleus of $10^{3}$\,--\,$10^{5}\.\Msun$ captures lighter PBHs in a dense, gas-rich core at $z \sim 12$. Gas-assisted contraction and repeated captures can permit assembly within $1$\,--\,$50\.\mathrm{Myr}$, with smaller cores evolving faster. The unequal masses reduce merger kicks, favouring retention, although the lightest nuclei remain sensitive to their environment. The resulting nuclei could seed the most massive LRD black holes. Mergers between nuclei of similar mass may be individually detectable, while the associated GW background could approach LISA sensitivity under optimistic assumptions. Assessing the cosmological contribution of this channel requires a self-consistent prediction of the seed abundance and an account of how dynamically heated Poisson clusters develop the much more compact, subparsec cores required by the mechanism.

The evidence for three massive accreting black holes in a galaxy at $z \approx 5$, reported by \citet{2026A&A...712A..80U}, is particularly exciting for clustered-PBH scenarios. The two central objects are separated by only $190 \pm 40\.{\rm pc}$ in projection, while the third lies $1.7\.{\rm kpc}$ away. In particle-CDM models, this early multiplicity is challenging, whereas clustered PBHs offer a natural explanation: an extended mass function containing massive seeds can provide several black holes within a common primordial overdensity before star formation begins. Spatial correlations beyond Poisson statistics make their initial association especially plausible~\cite{Choi:2025eqn}, while subsequent accretion and dynamical friction can produce a growing central concentration with surviving off-nuclear companions.

\subsection{Blue Monsters}

Another class of high-redshift objects uncovered by {\it JWST}, termed {\it Blue Monsters}, comprises luminous, compact and deep blue galaxies, distinct from the (red) LRDs with their broad emission lines. The term originally referred to spectroscopically confirmed systems at $z \gtrsim 10$, with inferred stellar masses $M_{\star} \sim 10^{8}$\,--\,$10^{9}\.\Msun$, bright rest-frame UV emission and very little dust attenuation~\citep{2023MNRAS.520.2445Z, 2023MNRAS.522.3986F, 2025A&A...694A.286F}. The challenge is to explain their rapid early assembly, substantial metal enrichment and intense star formation, together with almost unobscured UV emission. Current explanations therefore allow dust to be present, while reducing its obscuration for various reasons: radiation pressure may drive it to larger radii; dust and young stars may occupy different regions; larger grains may reduce the ultraviolet opacity; or clustered supernov{\ae} may clear gas and dust from star-forming clouds~\citep{2023MNRAS.520.2445Z, 2025A&A...694A.215F, 2026OJAp....959986N, 2026A&A...705L..10M}. These possibilities predict different dust and metal-line signals for the {\it Atacama Large Millimeter/submillimeter Array} (ALMA). Non-detections can constrain the intrinsic line luminosities, while extended cold dust would support a picture in which outflows displace dust from the central star-forming regions~\citep{2023MNRAS.520.2445Z, 2025A&A...694A.215F}.

The strongly lensed $z = 9.625$ Cosmic Gems system~\citep{2026A&A...705A.171V} illustrates that an extreme, clustered starburst can briefly enter the Blue Monster regime, with $M_{\rm UV} < -\.20$, $R_{\text{eff}} \sim 100\,{\rm pc}$ and a very large star-formation rate, suggesting that at least some Blue Monsters may represent short-lived burst phases of highly clustered early star formation. More recent work has interpreted Blue Monsters and dustier ``Red Monsters'' as different evolutionary phases, in which radiation-driven outflows clear an initially obscured compact system. However, the most compact $r_{\erm} \lesssim 150\,{\rm pc}$ sources may require an additional AGN-like component~\citep{2026arXiv260522914F}.

A potential PBH origin is consistent with the Poisson-clustering scenario. The most conservative possibility is that PBHs do not produce the Blue Monster spectrum on their own, but increase the abundance of compact, early haloes and provide additional accretion luminosity. Massive seed PBHs can accelerate collapse, reducing the required star-formation efficiency in $z \gtrsim 10$ massive galaxies; in existing models this can involve PBHs with $m \gtrsim 10^{9}\.\Msun$ or radiating intermediate-mass PBHs with $m \sim 10^{4}$\,--\,$10^{5}\.\Msun$ accreting at sub-Eddington rates, while a super-Eddington PBH component with $m \sim 10^{3.7}\.\Msun$ and tiny $\fPBH$ can reproduce bright UV luminosity functions in some models~\citep{2022ApJ...937L..30L, Matteri:2025klg, Matteri:2025vnv}.

The discreteness of PBHs generates small-scale isocurvature power, so compact Poisson clusters can form early and deepen the potential wells in which gas cools, forms stars and possibly expels dust. Indeed, a PBH-cluster interpretation of the JWST early-galaxy tension has been shown to require compact structures with characteristic mass $M_{\crm} \sim 4 \times 10^{6}\.\Msun\,( \fPBH / 0.005 )^{-1}$ for $\fPBH \gtrsim 0.005$, although with a high star-formation efficiency~\citep{2023PhRvD.107d3502H}. For the solar-mass PBH scenario, increasing $\fPBH\mspace{1mu}\langle\mspace{1mu}m\mspace{1mu}f \rangle$ makes early PBH clusters and their baryonic counterparts less rare, while subsequent dynamical evolution sets the surviving $M_{\crm}( r_{\crm} )$ relation. One should therefore not identify all Blue Monsters with PBH clusters, since their blue colours still require a low effective dust optical depth, but PBH clustering could provide the dense, compact high-redshift environments in which short-lived Blue Monster phases occur. This would predict a mixed population: starburst-dominated systems resembling the attenuation-free Blue Monster models, together with PBH-accretion-dominated systems with AGN-like compactness and extreme black-hole-to-stellar-mass ratios.

\section{PBH Clustering from Non-Gaussian Effects}
\label{sec:PBH-Clustering-from-Non--Gaussian-Effects}

\noindent Non-Gaussian corrections to the matter density distribution may provide an important additional component to the initial PBH clustering during the radiation-dominated era and this may modify Figs.~\ref{fig:deltas} and \ref{fig:relax}. Such non-Gaussianity arises during inflation, for instance, if quantum diffusion of either the inflaton field itself or a spectator curvaton field creates high exponential tails or even bimodality in the probability distribution function for the density contrast~\cite{Ezquiaga:2019ftu, Carr:2019hud, Stamou:2023vft, Stamou:2023vwz, Chen:2024pge, Kuroda:2025coa, Allegrini:2026jqt}. These non-Gaussian effects increase the probability of finding another PBH near a given one, with successive waves of enhanced curvature fluctuations inevitably generating dense clusters. Their size and mass depend very strongly on the form of the non-Gaussian tails. In particular, plateau-type potentials like those of Critical Higgs Inflation~\cite{Ezquiaga:2017fvi, Garcia-Bellido:2017mdw} naturally generate such tails. As a result, significant clustering beyond Poisson fluctuations can arise in the {\it Thermal History} model~\cite{Carr:2019kxo}. The non-Gaussian exponential tails arising from quantum diffusion during inflation cannot be described in terms of the local perturbative parameters, like $f_{\rm NL}$ or $g_{\rm NL}$~\cite{Ezquiaga:2022qpw}, which characterise primordial non-Gaussianity at CMB scales~\cite{Byrnes:2024vjt}. Indeed, they are always positively skewed, with a large number of significant higher-order cumulants, enhancing the probability of collapse to PBHs and the formation of PBH clusters. This may explain~\cite{Siles:2024yym} the observed clumpiness~\cite{Tanaka:2024, Rinaldi:2024fgn} of the Little Red Dots detected by NIRCam aboard JWST~\cite{Matthee:2023utn, Greene:2023}. 

These non-Gaussian effects can evade PBH constraints from CMB $\mu$-distortions and scalar-induced GWs from pulsar-timing arrays, allowing extended mass functions with some interme-diate-mass or supermassive PBHs \cite{Dave:2026gsd}. \citet{Ishijima:2026npf} derive the initial PBH correlation function for arbitrary local non-Gaussianity in the high-peak, weak-modulation limit, relating clustering beyond Poisson statistics to the primordial trispectrum in the collapsed limit. Their treatment emphasises that spatial correlations, rather than the non-Gaussian one-point tail alone, determine the initial clustering.

These non-Gaussian exponential tails can enhance the early formation of rare, massive structures~\cite{Ezquiaga:2022qpw}. JWST has spectroscopically established luminous galaxies at $z \geq 14$, notably JADES-GS-z14-0 at $z \simeq 14.18$~\cite{Carniani:2024zaf}, and MoM-z14 at $z = 14.44$~\cite{Naidu:2025xfo}, observed only about $280$\,--\,$300$~Myr after the Big Bang. Photometric searches have also identified a growing number of candidates at $z \simeq 15$\,--\,$17$~\cite{Castellano:2025vkt}, with more tentative samples currently extending to $z \sim 25$~\cite{Perez-Gonzalez:2025bqr}; these require spectroscopic confirmation, since lower-redshift galaxies and Galactic contaminants can mimic the relevant spectral breaks, but if confirmed would provide a definite conflict with the standard (particle) $\Lambda$CDM. The confirmed galaxies already demonstrate rapid assembly and chemical enrichment, while their abundance and the status of the more distant candidates remain under discussion~\cite{Carniani:2024zaf, Naidu:2025xfo, Castellano:2025vkt}. A contribution from black hole accretion has been proposed for MoM-z14~\cite{Fabian:2025get}; this provides supporting evidence for a high-mass bump, corresponding to the seed effect. 

The exponential tails also affect the evolution of PBH clusters in the matter era if they comprise an appreciable fraction of the dark matter. This has been explored using $N$-body simulations of large-scale structure that include gas and other components~\cite{Tkachev:2020uin}. Reference~\cite{Ali-Haimoud:2018dau} claims that one does not expect clustering on very small scales beyond what is expected from a Poisson distribution but Ref.~\cite{2024JCAP...08..026A} disagrees with this. The usual analysis of binary formation during the radiation-dominated era~\cite{Nakamura:1997sm} assumes Poisson statistics, with binaries forming due to (rare) three-body encounters. However, the additional clustering effect discussed here modifies both the formation of binaries and their disruption due to energetic close encounters, preventing the generation of a gravitational-wave background. This issue has been studied in Ref.~\citep{Siles:2024yym}, which finds that dense PBH clusters are rather robust, their mass being reduced due to ``evaporation'' (including slingshots) by no more than 20\%, while their cores remain compact and almost unchanged for the age of the Universe. Even if the clusters do not contain binaries initially, binaries form by dynamical capture and three-body encounters. However, these in-cluster binaries do not last long due to disruption by gravitational kicks in a dense environment, thus preventing them from finally merging within the cluster. Only those tight binaries that have been expelled from the cluster finally merge within the age of the Universe. Therefore, the GW emission by PBH binaries in sufficiently dense clusters is significantly suppressed compared with that from PBH binaries in the field. This important result completely alleviates the constraints on PBHs from GW measurements. 

Our analysis of the fate of PBH clusters and PBH binaries presented here is essentially independent of the cluster origin. One could try to apply our results to determine the size and mass of more compact clusters that would be predicted in scenarios with additional clustering from non-Gaussian fluctuations. However, it is difficult to predict the initial size and mass of clusters in the non-Gaussian case and numerical simulations are typically needed for an accurate treatment. \citet{Choi:2025eqn} have recently used such simulations to study the influence of spatial correlations on clustering properties of PBHs with a non-Gaussian exponential tail. Their analysis includes the non-sphericities of the emerging clusters, as well as the PBH mass function. These effects arise because any PBH-producing primordial power spectrum implies, for a range of scales, an increased likelihood for the occurrence of peaks near black holes. The associated dynamics take a finite time, possibly several e-folds (cf.~Ref.~\citep{EscrivaManas:2021mqb}). Therefore, a Hubble patch at the formation of an apparent horizon contains numerous initial Hubble patches (connected to the time of horizon entry) and therefore numerous other peaks which will merge and yield increased black hole masses.

\begin{figure*} 
	\centering
	\includegraphics[width=0.98\linewidth]{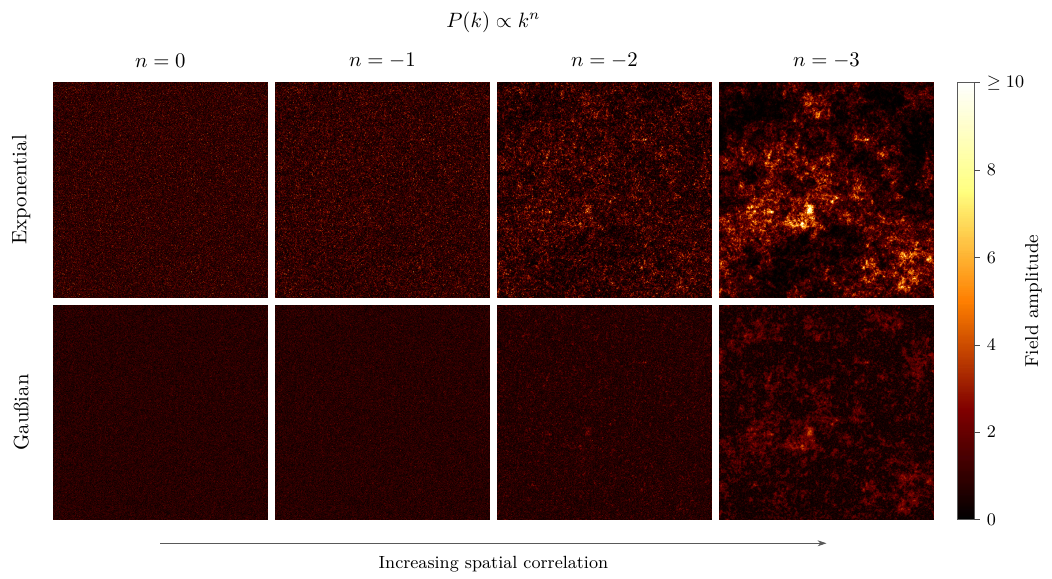}
	\caption{
		Two-dimensional slices through three-dimensional random fields with unit variance and power spectra $P( k ) \propto k^{n}$. The exponential (upper row) and Gaussian (lower row) fields are shown on a common amplitude scale for $n = 0,\.-\mspace{1mu}1,\.-\mspace{1mu}2,\.-\mspace{1mu}3$. Spatial correlations become increasingly apparent towards the right; the exponential fields display more pronounced high-amplitude concentrations. The colour scale saturates at amplitudes of $10$ and above.
		}
	\label{fig:intuitive} 
\end{figure*}

Concretely, Ref.~\cite{Choi:2025eqn} studied power spectra $P( k ) \propto k^{n}$, with $n = -\mspace{1mu}3,\.-\mspace{1mu}2,\.\ldots,\.3$, as the two-point function for the statistical distribution of the overdensities, while the one-point function was taken to be an exponential distribution. As discussed above, this describes the tail behaviour which generically emerges from quantum-diffusion calculations and typically leads to vastly enhanced spatial concentrations (see Fig.~\ref{fig:intuitive}). These are also measurable in finite neighbourhoods (representing the Hubble patch when the PBH actually forms) around an initial Hubble patch whose energy density exceeds the PBH formation threshold.

In this context, Ref.~\cite{Escriva:2023qnq} has performed a detailed numerical study on how the presence of additional overdensities alters the black hole formation process. The authors find a lowering of the PBH formation threshold, together with a mass increase of the final black hole, and a prolongation of the time until mass saturation is reached. For these effects to operate, the two overdensities must be close to each other; at larger separations, the collapses proceed independently. The associated changes in the PBH mass function are more pronounced the more steeply the power spectrum increases. As \citet{Choi:2025eqn} showed, this leads to a shift and broadening of the PBH mass spectrum even by more than an order of magnitude. It should be stressed that this effect is {\it generic for any primordial power spectrum with a peak}.

If above-threshold peaks are close to each other, one might expect them to merge. Concretely, for six-sigma fluctuations, the associated probabilities are $10^{-9}$ for $n = 0$, $10^{-7}$ for $n = -\.1$, $10^{-4}$ for $n = -\.2$ and $10^{-2}$ for $n = -\.3$. The corresponding (directly neighbouring) overdensities will clearly lead to the formation of a single black hole instead of two separate ones. Slightly more distant overdensities can likewise be expected to coalesce, thereby further altering the PBH mass distribution. Furthermore, as shown in Ref.~\cite{Choi:2025eqn}, non-trivial power spectra influence both the spatial concentrations and their structure, and these cluster-induced non-sphericities can be expected to substantially alter the PBH spin distribution. The associated effects are currently under investigation.\\

\section{Effect of Clusters on Microlensing}
\label{sec:Effect-of-Clusters-on-Microlensing}

\noindent The idea that stellar-mass PBHs or other compact bodies might make up the dark matter was first put to the test on the basis of an idea by Paczy{\'n}ski~\citep{Paczynski:1985jf}. His proposal was to monitor several million stars in the Magellanic Clouds to look for evidence of rare microlensing events in their light curves. This idea was followed up by the MACHO collaboration, which undertook a monitoring programme of 12 million stars in the Large Magellanic Cloud (LMC) covering four years to detect microlensing events. The project was a great success~\citep{MACHO:2000qbb}, detecting some $13$ to $17$ microlensing events, depending on selection criteria. These detections far exceeded the expected number of $4$ to $5$ from known stellar populations, and implied the discovery of a previously unknown population of compact stellar-mass objects, which were seen as possible dark matter candidates. However, to establish the MACHO content of the Galactic halo a mass model was needed. At that time, little was known about the rotation curve of the Galaxy, and the MACHO collaboration's favoured model was based on radial velocity measurements of nearby galaxies, which implied a flat rotation curve and a heavy halo. This implied a large optical depth to microlensing for compact bodies making up the dark matter, and on the basis of the number of MACHO microlensing detections set a limit of $\sim\!20$\% on the dark matter fraction of the halo in compact bodies. The MACHO collaboration saw this constraint as a disappointment, despite having discovered a previously unsuspected population of stellar-mass compact bodies, which they tentatively identified as PBHs~\citep{MACHO:1996qam}. 

Subsequently, microlensing surveys of the Magellanic Clouds have usually been invoked as constraints on compact objects, despite their being most notable for the detection of such a population. This point has been emphasised by~\citet{MACHO:2000qbb}, who showed that a detailed modelling of the stellar populations in the Milky Way halo and disc implies that the observed microlensing cannot be accounted for by any known population of stars. The idea that the lenses might be PBHs formed during the QCD phase transition was soon suggested by~\citet{Jedamzik:1998hc}, and a detailed analysis of other candidates~\citep{Hawkins:2020zie} confirmed that PBHs remain one possible interpretation. The remaining candidates were ruled out because they were baryonic, outside the observed mass range, insufficiently compact or very rare, or because of a combination of these factors.

Although it has been claimed that scenarios in which most of the dark matter is in solar-mass PBHs are excluded, it seems illogical to insist on theoretical constraints excluding their existence when compact bodies have actually been detected. The only way to challenge the detections is to show that they are not microlensing events; despite extensive efforts, this has so far proved unsuccessful. Much effort has therefore gone into challenging the robustness of the constraints. In particular, various factors have called this conclusion into question. 
	(1) The efficiency with which surveys detect microlensing, and hence the inferred constraint on the number of compact objects, is subject to numerous uncertainties. 
	(2) The MACHO constraints relied on the assumption of a flat Galactic rotation curve, implying a heavy halo, and this is now known not to be the case. 
	(3) The constraints only apply for a very narrow mass function, peaking in the solar-mass range, and they are not valid for a very broad mass function, as expected in the {\it Thermal History} model~\cite{Carr:2019kxo}. 
	(4) The clustering of the microlenses, as discussed in this paper, might reduce the probability of detecting lensing events. We now discuss these points in turn.

\subsection{Efficiency Analysis}

The constraints derived from the microlensing surveys rely crucially on estimates of the detection efficiency for flagging microlensing events. Given the crowded nature of the Magellanic Cloud star fields, this is a challenging problem, and comparison of event detections in overlapping fields for the different surveys has shown little consistency~\citep{Hawkins:2015uja}, raising doubts over the reliability of the adopted detection efficiencies. Of particular concern is the surprising result that when two surveys overlapped in time and space there were no common detections, despite estimated detection efficiencies of over $50$\%.

A recent paper~\citep{Mroz:2024wag}, reporting the latest results from the {\it Optical Gravitational Lensing Experiment} (OGLE) microlensing survey, illustrates the problem with the reliability of constraints on the MACHO content of the halo. The OGLE collaboration started their microlensing survey of the LMC in 1996 and have developed it through successive phases up to the present. The original idea was to replicate or confirm the MACHO collaboration's microlensing detections, but the results did not turn out as expected. From the outset, the OGLE detection rate of microlensing events was very low, which they claimed to be consistent with known stellar populations in the Milky Way and self-lensing from stars in the LMC. The combined OGLE-III/OGLE-IV analyses covered 300\.{\text{deg}}$^{2}$ and included some 62 million stars. The final result was the detection of 13 microlensing candidates, which is less than the number expected from known stellar populations. The SMC results have also been reported~\citep{Mroz:2025xbl} but are notoriously difficult to analyse as one is looking at a small tube-like structure end on, where the expected number of Galactic halo events is very small and typically dominated by Poisson statistics. The probability of self-lensing is then very high and hard to quantify due to the orientation of the SMC relative to the Galaxy. For this reason the SMC has generally been excluded from the microlensing analysis.

Another recent OGLE paper~\citep{Mroz:2024mse}, entitled {\it No massive black holes in the Milky Way halo}, received a lot of attention and prompted a response~\citep{Hawkins-2025} that challenged the basis of this claim. The inferred limits depend on a proper understanding of a number of issues. These include their choice of only a single photometric passband to select microlensing events, the quality and cadence of the light curves compared with those from the MACHO survey, the reliability of their event selection techniques, their estimates of detection efficiency, the contribution of self-lensing in the LMC and lensing by Galactic disc stars and other known stellar populations, and the choice of halo models in the context of the observed Galactic rotation curve. It is argued in Ref.~\cite{Hawkins-2025} that all of these issues remain unsettled and are consistent with a significant PBH component of dark matter in the Galactic halo. 

Perhaps the most convincing reason for distrusting the OGLE results is the status of their recent 13 detections of LMC microlensing events. Quite apart from the question of whether these events can be attributed to PBHs, their number falls far short of the roughly 40 detections expected from known stellar populations. In earlier surveys~\citep{2009MNRAS397}, with very small numbers of detections, it could be claimed that they were consistent with known stellar populations. However, in the new much larger sample, the number of OGLE detections~\cite{Mroz:2024wag} is inconsistent with that expected for known stellar populations at a high significance level. This failure to detect expected microlensing events motivates a sensitivity analysis of the results from the OGLE survey, so any proposed constraints are questionable and should not be considered as firm. It should also be noted that \citet{Sugiyama:2026kpv} claim to have found evidence for microlensing towards M31 by planetary mass black holes.

\subsection{Effect of Falling Rotation Curve}

The conclusion that the observed microlensing rate in the Galactic halo is inconsistent with a dark matter halo composed of stellar-mass compact bodies rests fundamentally on the assumption in the analysis of the surveys~\citep{MACHO:2000qbb, Tisserand:2006zx, 2011MNRAS.416.2949W} of a heavy Milky Way halo with a flat rotation curve. The results of the MACHO microlensing survey were published some 20 years ago, when very little was known of the shape of the rotation curve of the Milky Way, as direct measurement was challenging. At that time it was generally believed that the rotation curve was flat, as observed for a number of nearby spiral galaxies. This implied a massive halo, and a large optical depth to microlensing to the Magellanic Clouds, inconsistent with the observations if the halo was made up of compact bodies. When it became possible to make direct measurements of the Galactic rotation curve, it was clear that it declined out to the distance of the Magellanic Clouds~\citep{Battaglia:2005rj, SDSS:2008nmx, 2013PASJ...65..118S}. This decline was confirmed by subsequent measurements~\cite{Huang:2016tam} and put beyond reasonable doubt by data from {\it Gaia} Data Release 3~\citep{Garcia-Bellido:2024yaz}. This implies a halo mass which is smaller by an order of magnitude than assumed in the earlier analyses, with a corresponding reduction in the optical depth to microlensing from MACHO dark matter. This would be consistent with the results of \citet{MACHO:2000qbb} for an all-MACHO halo~\citep{Hawkins:2015uja}.

\subsection{Effect of Extended Mass Function}

The extended mass spectrum of the {\it Thermal History} model requires the microlensing signal to be calculated from the full distribution, including its QCD peak and the populations at other mass scales~\citep{Carr:2019kxo}. We first consider independent point lenses with the same spatial and velocity distributions. In the geometric-optics and point-source limits, the Einstein radius scales as $R_{\Erm} \propto m^{1/2}$, so the crossing time is $t_{\Erm} = R_{\Erm}/v_{\perp}$, where $v_{\perp}$ is the transverse speed of the lens relative to the line joining the observer and source. This time increases as $m^{1/2}$ for fixed geometry and transverse velocity. The optical depth at fixed PBH mass density is independent of the mass and hence the mass spectrum, but the event rate is mass-dependent, with heavier lenses producing fewer but longer events than lighter ones. Moreover, each lens mass generates a distribution of durations because the distances and velocities vary. Whether the resulting events are detected therefore depends on the combination of the PBH distribution and survey efficiency, as well as the mass function. 

The expected number of detected PBH events can be written as~\citep{Green:2016xgy, Calcino:2018mwh}
\begin{equation}
	N_{\text{exp}}
		=
			\fPBH \int \d\mspace{-1mu}\ln m\;f( m )\,\Ncal( m )
			\, ,
			\qquad 
	\Ncal( m )
		=
			\Ecal\int_{0}^{\infty}\d\mspace{1mu}t_{\Erm}\;\epsilon( t_{\Erm} )\,
			\frac{ \d\Gamma_{m} }{ \d\mspace{1mu}t_{\Erm} }
			\, ,
\label{eq:microlensing-extended-count}
\end{equation}
where $\d\Gamma_{m}/\d\mspace{1mu}t_{\Erm}$ is the differential event rate per monitored source for a monochromatic population with $\fPBH = 1$, integrated over the lens positions and velocities. $\Ecal$ is the exposure in star-years and $\epsilon$ is the detection efficiency for the corresponding source sample and duration convention. If a monochromatic abundance limit $f_{\text{max}}^{\text{mono}}( m )$ is obtained by imposing the same upper bound on the expected signal count at every mass, the corresponding constraint for an extended mass function is~\citep{2017PhRvD..96b3514C, Calcino:2018mwh}
\vs{-2mm}
\begin{equation}
	\fPBH
		\leq
			\Bigg[
				\int \d\mspace{-1mu}\ln m\;\frac{f( m )}
				{f_{\text{max}}^{\text{mono}}( m )}
			\Bigg]^{-1}
			\, .
			\label{eq:microlensing-extended-bound}
\end{equation}
The halo, efficiency and background assumptions must be held fixed in this conversion. A likelihood using individual event durations requires the corresponding mass-integrated differential prediction, and different surveys must be assessed through their own responses or a joint likelihood.

This mass weighting explains why the moment $\langle\mspace{1mu}m\mspace{1mu}f \rangle$ cannot be used as an effective lens mass. Before detection cuts, the rate under the above assumptions scales as $\Gamma \propto \int \d\mspace{-1mu}\ln m\;f( m )\.m^{-1/2}$, whereas the ensemble Poisson variance is controlled by the moment $\int \d\mspace{-1mu}\ln m\;m\.f( m )$. The distinction is particularly clear for the red distribution in Fig.~\ref{fig:thermal-mf-moment}: PBHs above $10^{6}\.\Msun$ supply approximately $92\%$ of the Poisson moment but only $9 \times 10^{-5}$ of the PBH mass density. Such a tail can strongly enhance the ensemble variance without appreciably reducing the microlensing signal of the lighter population. Indeed, transferring a mass fraction $\eta$ from the lighter to heavier component can reduce the expected event count by at most the fraction $\eta$, with equality when the added component produces no detectable events. The abundance limit can therefore weaken by at most a factor $1/(1 - \eta)$ and, for the assumed tail, the maximum relative weakening is less than $0.01\%$. Changes to the shape or position of the main mass distribution can have a much larger effect, so those changes must be distinguished from increasing the high-mass moment. The magnitude of the effect is consequently specific to the adopted spectrum and survey model. 

\citet{Calcino:2018mwh} show how extended distributions shift the microlensing bounds; their comparisons also illustrate why the mass scale must be specified when varying the width. For a Galactic model fitted to a declining rotation curve and a chosen {\it Thermal History} distribution, \citet{Garcia-Bellido:2024yaz} change the abundance upper limits by factors of $1.9$ for MACHO and $1.0$ for EROS-2, with the latter analysis adopting a factor of two reduction in the published detection efficiency. A limit above unity imposes no restriction. These results demonstrate the importance of the mass-function, Galactic and survey assumptions, but they do not give a bound for every {\it Thermal History} spectrum. In particular, the distributions used here require separate calculations for the event counts and durations, including comparison with the more recent OGLE analysis~\citep{Mroz:2024mse}. The consequences of spatial clustering are considered in the following subsection.

\subsection{Effect of Clustering on Microlensing Constraints}
\label{sec:Effect-of-Clustering-on-Microlensing-Constraints}

Another challenge to the constraints, pointed out by~\citet{Calcino:2018mwh} and~\citet{Garcia-Bellido:2017xvr}, is that clustering of the microlenses could reduce the probability of detecting lensing events. This is clearly relevant to our claim that a population of solar-mass PBHs could provide the dark matter in our halo. On the other hand, \citet{Petac:2022rio} and~\citet{Gorton:2022fyb} have shown, on the basis of detailed computations of PBH clustering, that the limits for clustered PBHs are still stringent enough to exclude this scenario. Poisson clusters are generally too numerous and extended to change the mean microlensing rate, so clustering merely increases the event-to-event variance. However, various different effects are involved, so we now disentangle them.

One reason PBH clustering might be expected to weaken the microlensing limits is the lower probability of finding a PBH cluster along the line of sight than of finding an individual PBH within a homogeneous distribution. The covering factor of clusters out to the distance of the LMC ($50$\.kpc) can be obtained by integrating the projected area of the clusters over halo radii between $R_{\orm} \approx$ 8\.kpc and $R_{\Hrm} \approx 50$\.kpc and then dividing by the area of the halo at that radius:
\begin{equation}
	{\rm CF}
		=
			\int^{R_{\Hrm}}_{R_{\orm}}\d R\;
			4 \pi f \rho_{\Hrm}( R )\.
			\frac{ R^{2}\mspace{1mu}\pi\.r_{\crm}^{2} }{ 4 \pi R_{\Hrm}^{2}\mspace{1mu}M_{\crm} }
		\approx
			\left(
				\frac{ f M_{\Hrm} }{ 4\mspace{1mu}M_{\crm} }
			\right)\!
			\left(
				\frac{r_{\crm} }{ R_{\Hrm} }
			\right)^{\!2}
			,
\label{CF}
\end{equation}
where $M_{\Hrm}$ is the halo mass within $R_{\Hrm}$ and we assume $\rho_{\Hrm} \propto R^{-2}$ from Eq.~\eqref{eq:rhohalo}. This expression does not allow for the anisotropy in the cluster distribution for an observer at $R_{\orm}$. However, the integral is dominated by the upper limit, $R_{\Hrm}$, where the anisotropy is small. The lower limit may also be increased if clusters at $8$\.kpc are disrupted by collisions but again this is only a minor effect if collisions are ineffective at $50$\.kpc. Therefore a necessary but not sufficient condition to avoid lensing of LMC stars is 
\vs{-2mm}
\begin{equation}
	{\rm CF}
		<
			1
	\quad \Rightarrow \quad r_{\crm}
		<
			180\.f^{-1/2}
			\left(
				\frac{ M_{\crm} }{ 10^{6}\.\Msun }
			\right)^{\!1/2} \left(
				\frac{ R_{\Hrm} }{ 50 {\rm kpc} }
			\right)^{\!1/2} 
			{\rm pc}
		\, ,
\label{eq:CFradius}
\end{equation}
where we have assumed $M_{\Hrm} \approx 3 \times 10^{11}\.\Msun$ at $R_{\Hrm} \approx 50$~kpc. This condition is shown in Fig.~\ref{fig:rcvsmc} for different values of $f$ and is necessarily consistent with the dynamical constraints at 50\.kpc. However, even if ${\rm CF} < 1$, some stars in the LMC will still be lensed if the number of clusters in front of the LMC ($N_{\text{LMC}}$) exceeds 1 or, equivalently, if the LMC diameter $d_{\text{LMC}}$ exceeds the projected separation $d_{\text{sep}}$ between the clusters at $R_{\Hrm}$. The condition for this is
\begin{equation}
	d_{\text{sep}}
		\sim
			{\rm CF}^{-1/2} r_{\crm}
		\sim
			160\.f^{-1/2}
			\left(
				\frac{ M_{\crm} }{ 10^{6}\.\Msun }
			\right)^{\!1/2}
			{\rm pc}
		<
			d_{\text{LMC}}
		\sim
			10\.{\rm kpc}
			\, ,
\end{equation}
which implies $M_{\crm} < 10^{9}\.f\.\Msun$. Since this upper limit can be expressed as $f\mspace{1mu}\rho_{\Hrm}\mspace{1mu}R_{\Hrm}\.d_{\text{LMC}}^{2}$, it is just the mass of dark matter in a cone with apex at Earth and base at the LMC, as also pointed out in Ref.~\cite{Gorton:2022fyb}. So only clusters larger than this (i.e.~on the right-hand side of Fig.~\ref{fig:rcvsmc}) can avoid lensing altogether. In this case, $N_{\text{LMC}}$ may be larger or smaller than 1. When there is a cluster along the line of sight, the number of expected microlensing events is the product of CF and the covering factor of all the PBH Einstein radii within the cluster:
\begin{equation}
	N_{\rm ev}
		=
			\frac{ N_{*}\.r_{\rm ER}^{2} }
			{ N_{\text{PBH}}\.r_{\text{LMC}}^{2} }
			\, ,
			\label{eq:lensrate}
\end{equation}
where $r_{\rm ER}$ is the PBH Einstein radius and $N_{*}$ is the number of stars in the LMC. However, this is precisely the number of microlensing events for a uniformly distributed PBH population, so the relative rate ($R$) for microlensing at $50$~kpc is unaffected. Therefore, unless $M_{\crm}$ is very large, only rare or unusually compact clusters can suppress microlensing. The clusters expected to exist are neither massive enough nor small enough to remove the mean signal, as also concluded in Ref.~\cite{Gorton:2022fyb}.

Another effect comes in if the cluster is very small. If its projected physical size is less than the Einstein radius associated with its total mass, it acts as a single lens and the condition for this is 
\begin{equation}
	r_{\crm}
		<
			\big(
				4\mspace{1mu}G M_{\crm}\.D_{\Lrm}\.D_{\rm LS} / 
				D_{\Srm}\.c^{2}\mspace{1mu} 
			\big)^{1/2}
		\approx
			0.07\.
			\big( M_{\crm}/10^{6}\.\Msun \big)^{1/2}\,
			\big( D_{\Lrm} / 50\.{\rm kpc} \big)^{1/2}\;
			{\rm pc}
			\, ,
	\label{eq:singlelens}
\end{equation}
where $D_{\Lrm}$ is the distance of the PBH cluster and $D_{\Srm}$ and $D_{\rm LS}$ are the observer-source and lens-source distances, respectively, all of these being comparable. The resulting event may be missed if its timescale lies outside the survey window or if finite-source effects are important. Even if $r_{\crm}$ exceeds the value given by Eq.~\eqref{eq:singlelens}, one no longer detects standard microlensing by individual PBHs if the Einstein radius $r_{\Erm}$ of the PBH lens is more than the projected separation $r_{\text{sep}}$ of PBHs in the cluster, i.e.~if
\begin{equation}
	r_{\Erm}
		\approx
			10^{-5}\.{\rm pc}
			\left(
				\frac{ m }{ \Msun }
				\frac{ D_{\Lrm} }{ {\rm kpc} }
				\frac{ D_{\rm LS} }{ D_{\Srm} }
			\right)^{\!1/2}
		>
			r_{\text{sep}}
		\approx
			2 \times 10^{-3}\,r_{\crm}
			\left(
				\frac{ m }{\.\Msun }
				\frac{ 10^{6}\.\Msun }{ M_{\crm} }
			\right)^{\!1/2}
			.
\end{equation}
This implies
\begin{equation}
	r_{\crm}
		<
			0.03\.
			\Bigg(
				\frac{ D_{\Lrm} }{ 50\.{\rm kpc} }
				\frac{ D_{\rm LS} }{ D_{\Srm} }
				\frac{ M_{\crm} }{ 10^{6}\.\Msun }
			\Bigg)^{\!1/2}\;{\rm pc}
			\, ,
			\label{eq:ML}
\end{equation}
which is the same as condition \eqref{eq:singlelens}. In this case, instead of well-defined light curves, the PBH cluster population will generate complex caustics that would not be detected by standard pipelines. Conditions \eqref{eq:singlelens} and \eqref{eq:ML} do not depend on the PBH mass for a monochromatic mass function and can only be satisfied for very large $M_{\crm}$ or very small $r_{\crm}$. As can be seen from Fig.~\ref{fig:rcvsmc}, neither of these conditions is likely to apply. However, the conditions must be modified for an extended mass distribution. This issue was recently investigated in Ref.~\citep{Toshchenko:2025esh}, using simulated caustic maps and OGLE-like criteria for the detection efficiency. It was found that 93\% of microlensing events would remain undetected by OGLE for a cluster radius of $0.01\.$pc with a core Plummer profile and a mass of $4 \times 10^{4}\.\Msun$. 

\begin{figure*} 
	\centering
	\includegraphics[width=0.75\linewidth]{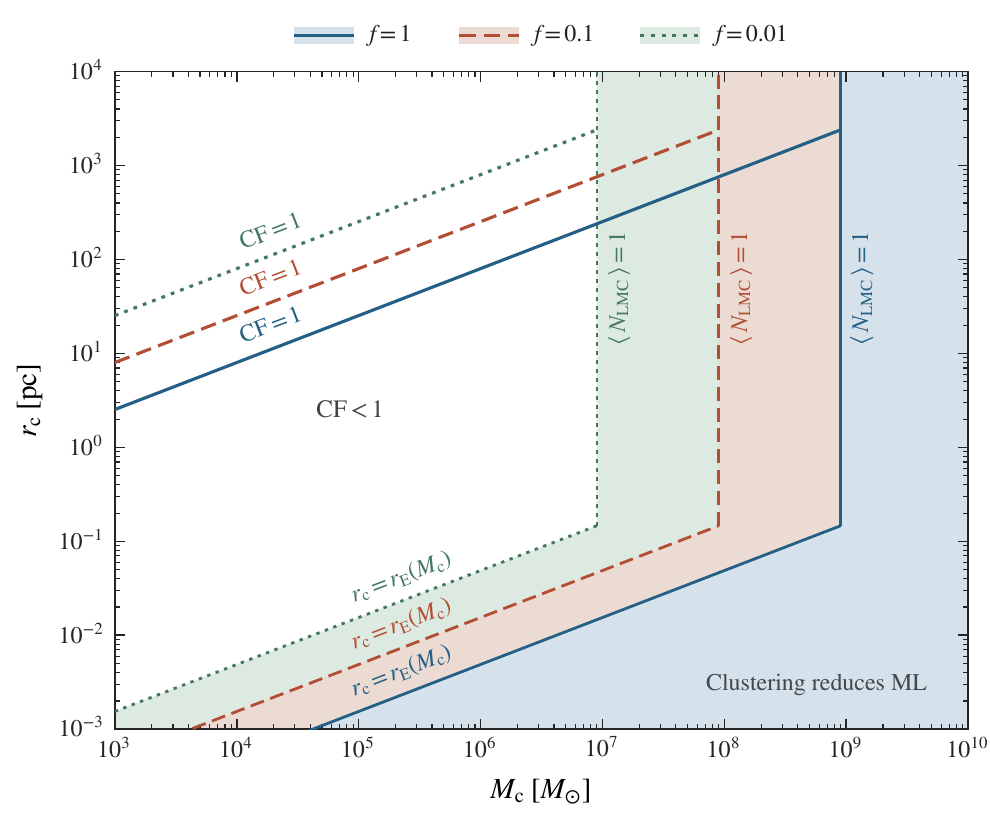}
	\caption{
		($r_{\crm},\.M_{\crm}$) regions in which LMC microlensing is modified by various effects for different values of the dark matter fraction $f$: the covering factor (CF) of the clusters being less than 1; the LMC containing more than one cluster; the cluster itself inducing microlensing.
		}
	\label{fig:lensing} 
\end{figure*}

The lines corresponding to Eqs.~\eqref{eq:CFradius} and \eqref{eq:singlelens} and the condition $N_{\text{LMC}} = 1$ (just one cluster in front of the LMC) are shown in Fig.~\ref{fig:lensing}, where the ranges of both axes are much larger than those for the dynamical limits shown in Fig.~\ref{fig:rcvsmc}. The grey region indicates that the microlensing limits can be evaded only if the clusters are very massive or very compact, which confirms the conclusions of previous studies. It is therefore implausible that the PBH Poisson clustering significantly alters the microlensing limits. Furthermore, the very compact option may be implausible since the clusters would have expanded due to dynamical heating on a timescale much shorter than the age of the Universe. This can only be avoided for scenarios where highly compact clusters can survive. For example, this might apply if there is a massive black hole at the centre or if a dense core remains due to gravothermal collapse. There remain, however, uncertain processes that may significantly alter microlensing constraints. For instance, the gravitational slingshot mechanism may result in PBHs with the relevant masses having a larger velocity than expected. Besides providing constraints on the PBH abundance in simple scenarios, microlensing surveys could therefore help to select scenarios with a large fraction of the dark matter if there is PBH clustering.

\section{Effect of Clusters on Gravitational Waves from PBH Mergers}
\label{Effect-of-Clusters-on-Gravitational-Waves-from-PBH-Mergers}

\noindent The size, mass and history of PBH clusters seeded by Poisson fluctuations strongly affect the formation rate of PBH binaries by two-body or three-body tidal capture and their subsequent merger rates. Another important effect is the perturbation by other PBHs of binaries formed before the clusters, which leads to a suppression of this early channel that also depends on the cluster mass and size. In this section, we use the results of the previous sections to re-estimate the effects of Poisson clusters on the expected PBH merger rates and compare our estimates with previous results. Finally, we discuss the implications for a PBH interpretation of GW observations of compact binary coalescences by LVK and the associated constraints on the fraction of dark matter in PBHs in the stellar-mass range.

\subsection{Perturbation of Early Binaries due to Light PBHs}

Strong limits on the PBH abundances down to $\fPBH \lesssim \Ocal( 10^{-3} )$ in the stellar-mass range, including the effects of an extended mass function, have been set by \citet{Andres-Carcasona:2024wqk, Andres-Carcasona:2026avd} and \citet{Franciolini:2022tfm}. However, these limits are very sensitive to the assumed mass distribution. More precisely, they depend on the presence of numerous light perturbers in the tail of the distribution. The current merger rate of early binaries (i.e.~those that form before the clusters due to two PBHs being born sufficiently close to each other) is exponentially dependent on the average number of nearby perturbing PBHs, as pointed out by \citet{Raidal:2018bbj} and \citet{Hutsi:2020sol}. Based on the standard merger-rate prescriptions, it therefore suffices to assume a mass distribution for which the light PBHs outnumber those in the stellar-mass range; this is the case for our adopted QCD mass distribution, as discussed in Section~\ref{sec:Thermal-History-Model-and-Multimodal-Mass-Functions}. \citet{Andres-Carcasona:2024wqk} show the limits as a function of some average mass $\langle m \rangle$ for such extended mass distributions. (Note that the averaging notation here is different from that used in Sec.~\ref{sec:Formation-of-PBH-Clusters-from-Poisson-Fluctuations}.) Since $\langle m \rangle$ can be arbitrarily small and much less than a solar mass, there is no firm and model-independent GW limit on the PBH abundance. The mass cut-offs introduced in Ref.~\cite{Franciolini:2022tfm}, together with the correction of a slight inconsistency with the rates of Refs.~\cite{Raidal:2018bbj, Hutsi:2020sol} identified by \citet{Escriva:2022bwe}, allow stringent but model-dependent limits to be set on PBHs.

There are two counter-arguments. First, sufficiently light PBHs should not perturb an early solar-mass binary, leading Ref.~\cite{Escriva:2022bwe} to use the QCD-peaked-distribution prescription. However, the PBH mass below which perturbations do not significantly alter the merger time is still unclear. Preliminary results indicate that merger rates can be heavily suppressed for realistic mass distributions~\cite{Biot:2026xxx}, at least in the mass range between $10\.\Msun$ and $100\.\Msun$. Second, perturbed early binaries have their own merger channel, which dominates for peaked distributions when $\fPBH \gtrsim 0.1$~\cite{Raidal:2024bmm}. However, these merger rates scale as $\big[ \langle m \rangle / ( m_{1} + m_{2} ) \big]^{\alpha}$ with $\alpha \gtrsim 8$; here $m_{1}$ and $m_{2}$ are the component masses and $m$ is the total mass. Therefore one can always consider a mass distribution extending down so far that $\langle m \rangle \ll ( m_{1} + m_{2} )$, heavily suppressing this channel.

Altogether, for sufficiently extended mass functions and large $\fPBH$, the merger rate of early binaries is heavily suppressed by interactions with light perturbers, so there is no significant constraint on the PBH abundance from GW observations for this channel. As discussed below, additional suppression of these rates comes from perturbations inside the Poisson clusters, which we re-evaluate in light of our results for the dynamics of these clusters. The merger rate is therefore dominated by binaries formed inside Poisson clusters and depends sensitively on the cluster properties.

\subsection{Perturbation of Early Binaries due to Poisson Clusters}

A large fraction of the early binaries end up in PBH clusters, in which they are quickly perturbed, at least for a peaked mass distribution and binaries with modest mass ratios. For an extended mass function, the effect of light PBHs on these binaries is studied below. The merger rate of early binaries is therefore suppressed by a factor $S = ( 1 - F_{\crm} )$, the probability that a Poisson fluctuation does not collapse to form a PBH cluster, where $F_{\crm}$ is given by Eq.~\eqref{eq:F}. For a peaked mass distribution, based on the probability of cluster formation at redshift $z_{\crm}$, this factor was estimated by~\citet{Hutsi:2020sol} to be
\begin{align}
\label{eq:supp-factor-poisson}
	S( z )
		&=
			1 - \sum_{N\mspace{1mu}=\mspace{2mu}3}^{N_{\crm}( z )}
			\left[
				p^{}_{N}( z_{\crm} ) 
				+ \sum_{N' > N_{\crm}( z )}\.
				p^{}_{N,\mspace{1mu}N'}( z_{\crm} )
			\right]
			\displaybreak[1]
			\notag
			\\[3mm]
		&\approx
			\min\!
			\left\{
				1,\mspace{1mu}0.01
				\left[
					\left(
						\frac{ t }{ t_{0} }
					\right)^{\!0.44}
					\fPBH
				\right]^{-0.65}
				\exp\!
				\left(
					0.03 \ln^{2}\!
					\left[
						\left(
							\frac{ t }{ t_{0} }
						\right)^{\!0.44}
						\fPBH
					\right]
				\right)
			\right\}
			.
\end{align}
Here, $p_{N}$ is the probability of forming a cluster with $N$ PBHs at redshift $z_{\crm}$ and $p_{N,\mspace{1mu}N'}$ is the probability that such a cluster is inside a larger cluster of $N'$ PBHs. In this expression, $N_{\crm} \approx 5.2 \times 10^{3}\.\fPBH^{10/7}$ is the maximum number of PBHs in a cluster for any primordial binary to be perturbed. A good approximation for this suppression factor today is then $S \approx 0.01 \times \fPBH^{-0.65}$ when $\fPBH > 0.03$. For smaller values of $\fPBH$, there is no suppression. We identify below four possible caveats with this approach, associated with the evolution of Poisson fluctuations, the dynamics of PBH clusters and the possibility of extended mass distributions. 

First, the suppression factor is obtained by considering the probability that a binary ends up in a Poisson cluster at redshift $z_{\crm} \approx \fPBH\.z_{\text{eq}}/ \sqrt{N_{\crm}}$. This assumes the linear growth of a Poisson fluctuation with scale factor but the number of black holes in a cluster does not uniquely specify the fluctuation. The relation $\delta_{\mspace{1mu}\irm} = ( \fPBH\.m /M )^{1/2}$ from Eq.~\eqref{eq:initial} only defines the average Poisson fluctuation but one expects variations around this. Fluctuations of mass $M$ with a smaller number of black holes could still collapse later than $z_{\crm} $, in which case the probability that a binary is in a cluster at redshift $z$ becomes
\begin{equation}
	p_{N}( z )
		=
			\frac{ N^{-1/2}\.{\erm}^{-N/N_{*}( z )} }
			{ \sum_{N\mspace{1mu}>\mspace{1mu}2}
			N^{-1/2}\.{\erm}^{-N/N_{*}( z )} }
			\, ,
\label{eq:PN}
\end{equation}
instead of $p_{N}( z_{\crm} )$, where 
\be
	N_{*}( z )
		=
			\Big[
				\ln\big( 1+ \delta_{*} \big) - \delta_{*}/
				\big( 1 + \delta_{*} \big)
			\Big]^{-1}
			\, ,
\ee
with $\delta_{*} \approx 1.69 / \big[ \fPBH\,g( z ) \big]$ and $g( z ) \approx ( 1 + z_{\text{eq}} ) / ( 1 + z )$. Changing $z_{\crm}$ to $z$ strongly reduces the threshold $\delta_{*}$, which pushes $p_{N}( z )$ towards unity. 

Second, the probability that a binary is in a cluster of $N_{\text{PBH}}$ black holes comes from the Press--Schechter formalism, which does not distinguish between haloes and haloes-in-haloes. Since $p_{N}( z )$ already allows for a cluster being inside a larger cluster, we should not include the second term with $p_{N,\mspace{1mu}N'}$ in Eq.~\eqref{eq:supp-factor-poisson}. 

Third, two effects are involved in determining the value of $N_{\crm}$: 
	(1) encounters of the binary with other PBHs in the cluster with impact parameters sufficiently small to perturb it; 
	(2) the expected gravothermal collapse of the cluster before today. 
But this does not consider the dynamical heating of clusters that should lead to their expansion if they are sufficiently compact. One should therefore check that encounters can perturb binaries before dilution of the cluster renders such encounters ineffective. Additionally, the claim that gravothermal collapse contributes to the perturbation of binaries only applies when PBHs are a subdominant fraction of dark matter and in this case the collapse would stop at the same time the dark matter halo is heated dynamically anyway. The impact on binary perturbations is therefore complex and model-dependent and requires further investigation. 

Fourth, the rate suppression factor in Eq.~\eqref{eq:supp-factor-poisson} should depend on $\fPBH\mspace{1mu}\langle\mspace{1mu}m\mspace{1mu}f \rangle$, as well as on the mass--radius relation of the PBH cluster. Only the $\fPBH$ dependence remains for a monochromatic mass function. However, for an extended mass function, one should retain the explicit dependence on $\langle\mspace{1mu}m\mspace{1mu}f \rangle$ in the suppression formula.

Given these considerations, and using Eq.~\eqref{eq:F}, we propose the following formula for the suppression factor for any PBH mass function which considers the product integral of the probability of not being included in a cluster $[1 - F_{\crm}( M,\mspace{1mu}z )]$ at each mass $M_{\crm}$ in the relevant range,
\begin{equation}
	S( z )
		=
			\prod_{M_{\text{min}}}^{M_{\text{max}}}
			\big[
				1 - F_{\crm}( M_{\crm},\.z)
			\big]^{\drm\mspace{-1mu}\ln M_{\crm}}
		=
			\exp\mspace{-2mu}
			\left\{
				\int_{M_{\text{min}}}^{M_{\text{max}}}
				\ln\!
				\left[
					1 - F_{\crm}( M_{\crm},\.z )
				\right]
				{\drm \ln M_{\crm}}
			\right\}
			.
\end{equation}
Here, $\prod$ denotes a product integral over logarithmic cluster mass. It is defined by dividing the interval from $M_{\text{min}}$ to $M_{\text{max}}$ into bins with representative masses $M_{i}$ and logarithmic widths $\Delta \ln M_{i}$. Taking its logarithm gives a sum weighted by $\Delta\mspace{-1mu}\ln M_{i}$, which approaches the integral in the exponent as the largest bin width tends to zero. The exponent $\d\mspace{-1mu}\ln M_{\crm}$ therefore specifies the weight assigned to each mass interval and ensures that refining the numerical partition does not itself increase the suppression. Physically, this prescription estimates the fraction of early binaries that remain unperturbed by the relevant cluster environments, with $M_{\text{min}}$ and $M_{\text{max}}$ defining the range of cluster masses for which encounters can appreciably alter their merger times~\cite{LIGOScientific:2025vwc}. It treats $1 - F_{\crm}$ as an effective survival factor per unit interval in $\ln M_{\crm}$ and neglects correlations between the contributions from different mass scales. This probabilistic interpretation is an additional modelling assumption: a binary may belong to a small cluster embedded in a larger one, so the corresponding cluster-membership probabilities need not be independent. The expression should therefore be regarded as a phenomenological estimate.

Using the approximation ${\text{erfc}}( x ) \approx \exp\big( -x^{2} \big) / \big( \sqrt{\pi}\.x \big)$ for $x \gg 1$ gives Eq.~\eqref{eq:PN}. By changing variable to $N = M / m$ and putting $M_{\text{min}} = 2\.m$ and $z = z_{\crm}$, one can recover the first term of Eq.~\eqref{eq:supp-factor-poisson} in the monochromatic case. The dependence on $\fPBH\mspace{1mu}\langle\mspace{1mu}m\mspace{1mu}f \rangle$ arises from $P_{\Prm}$. From Ref.~\cite{Raidal:2024bmm}, binaries are perturbed by encounters with cluster PBHs of mass $m_{3}$ when the impact parameter satisfies 
\vs{-2mm}
\begin{equation}
	b
		\lesssim
			\frac{
				\big[
					m_{3}\.( m_{1} + m_{2} + m_{3} )
				\big]^{1/3}\.r_{\arm}^{1/2} }
			{ j_{\tau}^{1/3}\.( m_{1} + m_{2} )^{1/6}\.v_{\text{rel}} }
			\, ,
\end{equation}
where $r_{\arm}$ is the semi-major axis, $j_{\tau}$ is the reduced (i.e.~dimensionless) angular momentum of a binary merging today and we put $G = c = 1$. The estimates provided in Ref.~\cite{Raidal:2024bmm} are
\begin{equation}
	j_{\tau}
		=
			\frac{ L/\mu }{ \sqrt{( m_{1} + m_{2} )\.r_{\arm}} }
		=
			1.7 \times 10^{-2} 
			\left(
				\frac{ m_{1} + m_{2} }{\.\Msun }
			\right)^{\!5/37}
			\frac{ ( 4\.m_{1}\.m_{2} )^{3/37} }
			{ ( m_{1} + m_{2} )^{6/37} }\;
			\fPBH^{16/37}
\end{equation}
and
\vs{-1mm}
\begin{equation}
	r_{\arm}
		\simeq
			6.02 \times 10^{-8}
			\left[
				\frac{ m_{1}\.m_{2}\.( m_{1} + m_{2} ) }
				{ \Msun^{3} }
			\right]^{1/4}
			j_{\tau}^{-7/4}
			\;{\rm pc}
			\, .
\end{equation}
The numerical coefficient assumes an inspiral time of $13.8\,{\rm Gyr}$ and applies to a highly eccentric binary evolving through GW emission alone. For the encounter estimate, we take the relative speed between the binary and the perturbing PBH to be the cluster virial velocity. The corresponding interaction cross-section is $\sigma = \pi\.b^{2}$ and one can then estimate, for each PBH mass $m_{3}$ with an abundance $f( m_{3} )$, the time taken to perturb the binary, $t_{\text{pert}} = 1/[ n( m_{3} )\.\sigma\.v_{\text{rel}} ]$, where $n( m_{3} )$ is the number density of perturbers of mass $m_{3}$. The time to perturb binaries with masses $m_{1}$ and $m_{2}$ is then
\vs{-2mm}
\begin{align}
\begin{split}
	t_{\text{pert}}
		&\approx
			2.7 \times 10^{10}\.{\rm yr} 
			\times \fPBH^{-79/111}\.f( m_{3} )^{-1}
			\\[2mm]
		&\phantom{\approx}
			\times
			\left(
				\frac{ m_{1}\.m_{2} }{\.\Msun^{2} }
			\right)^{\!-29 / 148}
			\left(
				\frac{ m_{3} }{\.\Msun }
			\right)^{\!1/3} 
			\left(
				\frac{ m_{1} + m_{2} }{\.\Msun }
			\right)^{\!29 / 444} 
			\\[2mm]
		&\phantom{\approx}
			\times
			\left(
				\frac{ m_{1} + m_{2} + m_{3} }{\.\Msun }
			\right)^{\!-2 / 3} 
			\left(
				\frac{ M_{\crm} }{\.\Msun }
			\right)^{\!- 1 / 2} 
			\left(
				\frac{ r_{\crm} }{ \rm pc }
			\right)^{\!5 / 2}
			\, .
\end{split}
\end{align}
At fixed binary and perturber properties, the perturbation time scales as $t_{\text{pert}} \propto M_{\crm}^{-1/2}\.r_{\crm}^{5/2}$.

The above formula is useful because it can be applied to clusters of arbitrary mass and size. In order to examine whether a binary with masses $m_{1}$ and $m_{2}$ is perturbed, one must integrate over the full mass distribution of the PBH perturbers. The perturbation time scales as $m_{3}^{1/3}$ for light perturbers, so with an extended mass distribution (e.g.~with a peak in the asteroid range) even very light PBHs are able to perturb solar-mass binaries. Motivated by the mass distributions imprinted by the QCD epoch, with a peak at $m_{3} \approx 2\.\Msun$, we have indicated the typical perturbation time for $m_{1} = m_{2} = 30\.\Msun$, corresponding to the black hole population observed by LVK. One sees that more massive clusters will be able to perturb these binaries, which suppresses the rates further. 

\begin{figure}[t!]
	\includegraphics[width=0.85\textwidth]{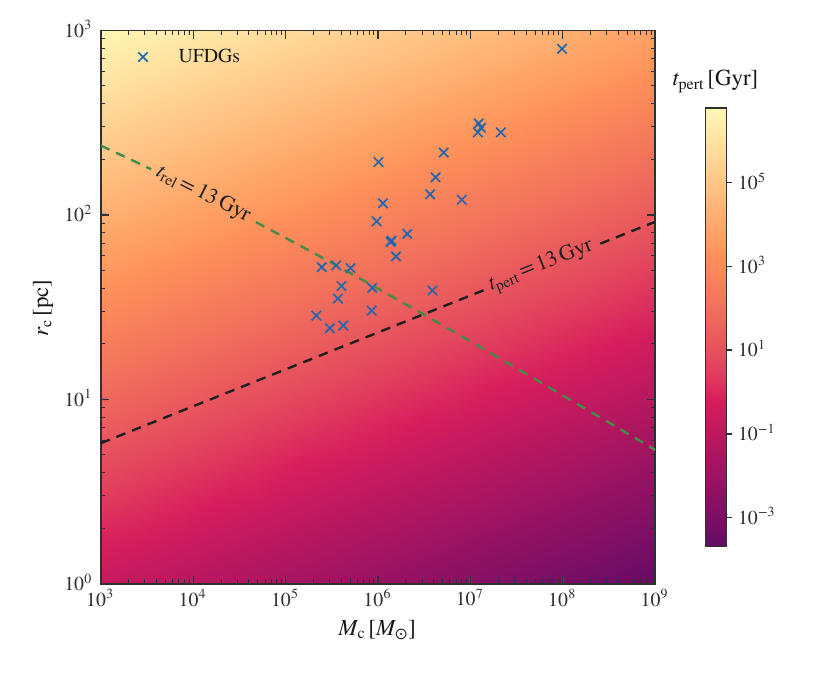}
	\caption{
		Time required to perturb a PBH binary of equal-mass components $m = 3\.\Msun$ in a Poisson cluster of mass and radius $( M_{\crm},\mspace{1mu}r_{\crm} )$. The contours for $t_{\text{pert}} = 13\,$Gyr and for a similar relaxation time are also represented.
		}
	\label{fig:LVK-fpbhX}
\end{figure}

\subsection{Two-Body PBH Binary Formation in Clusters}

PBH binary formation can take place in Poisson clusters via two-body capture. Consider two PBHs with masses $m_{1}$ and $m_{2}$ moving with a relative velocity $v$ in a cluster of mass $M_{\crm}$ and radius $r_{\crm}$ and with a PBH mass fraction $\fPBH$. The two PBHs form a binary if they get sufficiently close that the energy loss due to gravitational waves (GWs) exceeds their orbital kinetic energy. The merger rate in a single cluster and the total merger rates for the expected halo mass function are calculated in Ref.~\cite{Clesse:2024epo}. Here we provide the total rates and discuss their dependence on the Poisson cluster properties. 

The differential merger rate per unit logarithmic mass, as a function of redshift, is given by
\begin{align}
\label{eq:Merger-Rate}
	\frac{ \d\mspace{1mu}R_{\text{late}}( z ) }
	 {\d\mspace{-1mu}\ln m_{1}\.\d\mspace{-1mu}\ln m_{2} }
		&=
			\fpbh^{2}\.R_{\text{cl}}( z )\.
			f( m_{1} )\.f( m_{2} )\,
			\frac{ ( m_{1} + m_{2} )^{10/7} }
			{ ( m_{1}\.m_{2} )^{5/7} }
			\, ,
\end{align}
where $R_{\text{cl}}( z )$ is given by the following approximation:
\begin{align}
\label{eq:integral-Rcl}
	R_{\text{cl}}( z )
		&\approx
			\sqrt{\pi}\,
			\bigg(
				\frac{ 85\mspace{1mu}\pi }{ 6 \sqrt{2} }
			\bigg)^{\!2/7}\,
			\frac{ \big( G\,\Omega_{\text{DM}}\.
			\rho_{\text{crit}} \big)^{2} }
			{ c^{10/7} }
			\notag
			\displaybreak[1]
			\\[1mm]
		&\phantom{=\;}
			\times \int_{M_{\text{min}}( z )}^{\infty} \d\mspace{1mu}M_{\crm}\;
			v_{\text{vir}}^{-11/7}\.\Fcal_{\text{av}}\,
			\Gcal_{\text{prof}}\;\delta_{\crm}\,
			\sqrt{\frac{ M_{\crm} }{ M_{*}( z ) }}\;
			\frac{ e^{-M_{\crm} / M_{*}} }{ M_{\crm} }
 			\, ,
\end{align}
with $\Omega_{\text{DM}}$ being the dark matter density parameter in units of the critical density $\rho_{\text{crit}}$. The factor $\Fcal_{\text{av}}$ is of order one (we adopt $\Fcal_{\text{av}} = 3$, which provides a good approximation~\cite{Clesse:2024epo}) and is obtained by averaging the $-11/7$ power of the ratio of PBH relative velocities to the cluster virial velocity over the Maxwellian distribution of relative velocities. The factor $\Gcal_{\text{prof}}$ describes the radial density distribution. In the approximation that the velocity average is independent of radius, it is given by
\vs{-2mm}
\begin{equation}
	\Gcal_{\text{prof}}
		=
			\frac{ V_{\crm} }{ M_{\crm}^{2} }
			\int_{0}^{r_{\crm}}\d\mspace{1mu}r\;
			4\mspace{1mu}\pi\mspace{1mu}r^{2}\.\rho^{2}( r )
			\, ,
	\qquad
	V_{\crm}
		\equiv
			\frac{ 4\mspace{1mu}\pi }{ 3 }\.r_{\crm}^{3}
			\, .
\end{equation}
Here $\rho( r )$ is the cluster density profile, normalised to $M_{\crm}$ inside $r_{\crm}$, and the PBHs are assumed to trace this profile without mass segregation. A uniform sphere has $\Gcal_{\text{prof}} = 1$, the minimum value at fixed mass and volume within this approximation. For an NFW profile,
\begin{equation}
	\Gcal_{\rm NFW}
		=
			\frac{ C^{3}\big[ 1 - (1 + C)^{-3} \big] }
			{ 9\mspace{1.5mu}\big[ \ln(1 + C) - C/(1 + C) \big]^{2} }
			\, ,
\end{equation}
where $C \equiv r_{\crm}/r_{\srm}$ and $r_{\srm}$ is the scale radius. This gives $\Gcal_{\rm NFW} \approx 50$ for $C = 10$. At fixed cluster population and velocity prescription, the two-body capture rate is proportional to $\Gcal_{\text{prof}}$; the density-profile uncertainty therefore need not be a small correction. The factor $\delta_{\crm}$ parametrises the overdensity of the PBH cluster compared to the cosmological background density,
\begin{equation}
	\delta_{\crm}
		=
			\frac{ 3 M_{\crm} }{ 4\mspace{1mu}\pi\,r_{\crm}^{3}\,
			\bar{\rho}_{\text{DM}} }
			\, ,
			\label{eq:overdensity}
\end{equation}
which depends on the mass and size of Poisson clusters.

Reference~\cite{Clesse:2024epo} illustrates the dependence of $R_{\text{cl}}$ on $\fPBH$ for several primordial spectra, using the NFW profile and cluster-evolution prescription adopted there. Its results are sensitive to the lightest surviving haloes. In applying Eq.~\eqref{eq:integral-Rcl} here, the lower limit $M_{\text{min}}( z )$ must therefore be calculated consistently with the assumed expansion and disruption history; the NFW results of that reference should not be identified with the uniform-density illustration below.

The Poisson clustering has a very strong impact on the PBH merger rates of late binaries. The first effect is that it fixes the value of $M_{\text{min}}( z )$ and we have provided theoretical and observational arguments that this is $10^{5}$\,--\,$10^{6}\.\Msun$. The second effect is through the cluster virial velocity $v_{\text{vir}}$ since this depends on the cluster mass and size. In order to examine the impact of the cluster mass and radius, we have computed the merger rates for such clusters by approximating the clustering factor by its integrand at $M_{\crm} = M_{\text{min}}$, i.e.~by using ${\d\mspace{1mu}R_{\text{cl}}( z = 0 ) / \d\mspace{-1mu}\ln M_{\crm}}$, without integrating over the cluster mass function. We assume $f( m = 2\.\Msun ) \approx 1$, so that most of the PBH density comes from the QCD-induced peak. For this illustrative estimate, we take $\Fcal_{\text{av}} = 3$ and $\Gcal_{\text{prof}} = 1$, corresponding to a uniform-density cluster. This is an explicit approximation, rather than an NFW profile with $C = 10$. It does not assume that dynamical heating necessarily removes a central cusp. An NFW profile with $C = 10$ would increase the rate by a factor $\approx 50$.

One can now compare these merger rates with the local binary neutron star merger rate inferred by LVK for component masses $1\.\Msun \leq m_{1},m_{2} \leq 2.5\.\Msun$. The GWTC-5.0 population analysis gives $5.1$\,--\,$154.7\,{\rm Gpc}^{-3}\,{\rm yr}^{-1}$~\cite{LIGOScientific:2026ctl}. Although no new neutron star binaries enter the population sample, the additional observing exposure updates the rate inference and lowers its upper bound relative to GWTC-4.0.

\begin{figure}[t!]
	\hs{-2mm}\includegraphics[width=0.95\textwidth]{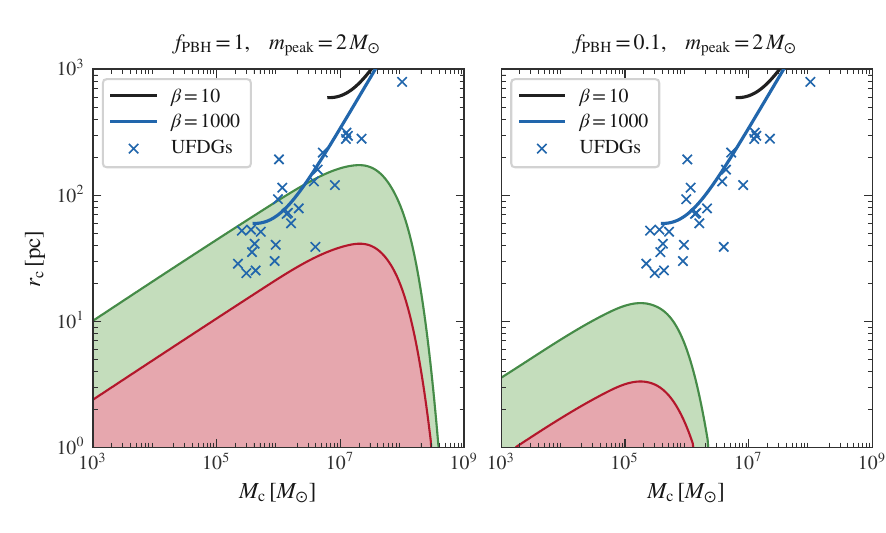}
	\caption{
		Constraints in the $( M_{\crm},\mspace{1mu}r_{\crm} )$-plane from the limits on the merger rates of compact binary coalescences observed by LVK after the observing run O4a, and attributed to binary neutron stars, assuming $\fPBH = 1$ (left panel) and $\fPBH = 0.1$ (right panel) and a high peak at $m = 2\.\Msun$, as expected for a mass distribution impacted by the thermal history of the Universe. The predicted cluster mass--radius relations for $\beta = 10$ and $\beta = 1000$ are overlaid for $\fPBH\,\langle\mspace{1mu}m\mspace{1mu}f \rangle = 3\.\Msun$, together with the UFDG estimates. The overlap with the allowed or favoured region must be assessed separately for each value of $\beta$. {\it Red}: predicted PBH rate above the LVK upper limit; {\it green}: predicted rate within the LVK binary-neutron-star interval, conditional on a PBH interpretation; {\it unshaded}: predicted rate below that interval; {\it labelled curves}: predicted Poisson-cluster mass--radius relations for $\beta = 10$ and $\beta = 1000$.
		}
	\label{fig:LVK-fpbh}
\end{figure}

Figure~\ref{fig:LVK-fpbh} compares the local rate estimates with the adopted LVK interval for $\fPBH = 1$ and $\fPBH = 0.1$, together with $\beta = 10$ and $\beta = 1000$. At fixed cluster mass, radius and other parameters, the rate boundaries are independent of $\beta$. The smaller radii attainable with $\beta = 1000$ can move the predicted relation towards higher rates. A predicted rate below the observed binary-neutron-star interval is not an exclusion, since the events need not be PBH mergers. Within this halo estimate, points above the adopted upper limit indicate potential tension; quantitative exclusion requires integration over the surviving cluster population. Whether $\fPBH < 0.1$ can reproduce the observed rate therefore depends on the adopted expansion history.

Finally, we stress that this approach still contains some uncertainties and, as discussed in Ref.~\cite{Carr:2023tpt}, several physical effects may modify the tidal and collisional disruption of clusters in the haloes of massive galaxies and their hierarchical mergers. More massive PBHs may also explain the observed mergers between $10\.\Msun$ and $100\.\Msun$ black holes. Na{\"i}vely, their merger rate should be suppressed compared to that at the solar-mass scale, leading to much lower merger rates than the ones of order $\Ocal( 10 )\;{\rm yr}^{-1}\.{\rm Gpc}^{-3}$ observed by LVK. However, this neglects the possible mass segregation of the more massive but subdominant PBH population towards the centres of the clusters, which may boost their merger rates and possibly explain the observations in this range. Alternatively, in scenarios where PBHs are not formed by Gaussian primordial fluctuations, such as those with an ultra-slow-roll phase or with a light curvaton field during inflation, it is expected that the peaks in the mass distribution due to the thermal history of the Universe are less pronounced, because a transient reduction of the density threshold will have a lower impact on the initial PBH abundance. Further analysis is underway to determine the origin of these black hole merger events.

\subsection{PBH Binary Formation Through Three-Body Channel}

When a third body is close to a binary forming in a PBH Poisson cluster, it can modify the binary coalescence time and subsequently the merger rate of binaries there. Two ingredients are needed to estimate these rates: an estimate of the probability of finding a nearby third body and the impact on the coalescence time. The mass and size of clusters have a strong impact on the first factor. A detailed calculation of merger rates for the late three-body channel can be found in Ref.~\citep{Raidal:2024bmm}. This channel is claimed to dominate the late two-body channel for PBHs with masses larger than $\Ocal( 10 )\.\Msun$. The rate is then given by
\begin{align}
\begin{split}
	\frac{ \d\mspace{1mu}R }{ \d\mspace{-1mu}\ln m_{1}\,\d\mspace{-1mu}\ln m_{2} }
		&\approx
			1.3 \times 10^{-16}\;{\rm Gpc}^{-3}\.{\rm yr}^{-1}\,
			\erm^{-6\mspace{1mu}( \gamma - 1 )}\.\fPBH^{3}\;
			\delta_{\text{eff}}^{2}\mspace{-2mu}
			\left(
				\frac{ \sigma_{v} }{ \rm km/s }
			\right)^{\!-9 + 8\mspace{1mu}\gamma / 7}
			\\[3mm]
		&\phantom{\approx\;}\times
			\eta^{-1 + \gamma / 7} 
			\left(
				\frac{ m_{1} + m_{2} }{\.\Msun }
			\right)^{\!3 - \gamma / 7}\!
			\left(
				\frac{ t }{ t_{0} }
			\right)^{\!\gamma / 7}
			\Fcal\mspace{-2mu}
			\left(
				\frac{ \langle\mspace{1mu}m\mspace{1mu}\rangle\.\kappa_{\text{min}} }
				{ 2\.\eta\.( m_{1} + m_{2} ) }
			\right)\.f( m_{1} )\.f( m_{2} )
			\, ,
\end{split}
\end{align}
where $\gamma$ parametrises the angular momentum distribution; it is typically between 1 and 2, with $\gamma = 2$ corresponding to a thermal distribution. The function $\Fcal$ is given by Eq.~(100) of Ref.~\cite{Raidal:2024bmm} and is $\Ocal( 1 )$ for the mass functions of interest here.

The relative importance of the three-body channel depends on the density and velocity dispersion, both of which change when the cluster expansion is modified. For fixed cluster mass and profile, $\delta_{\text{eff}} \propto r_{\crm}^{-3}$ and $\sigma_{v} \propto r_{\crm}^{-1/2}$, so the rate scales as $R_{3} \propto r_{\crm}^{-3/2 - 4\mspace{1mu}\gamma/7}$. Thus, although increasing the velocity dispersion suppresses three-body binary formation at fixed density, the associated density increase can raise the rate in a more compact virialised cluster.

The simulations of \citet{Barcelo-Sastre:2026nbs} illustrate the strong dependence of PBH mergers on cluster compactness: their initially non-spinning, binary-free cores with radii of order $10^{-5}\,{\rm pc}$ undergo rapid mergers and runaway IMBH growth, whereas their $10\,{\rm pc}$ clusters produce very few in-cluster mergers over a Hubble time. In the latter case, the observable mergers mainly arise from ejected binaries. These independently simulated regimes motivate distinguishing a dense core from an extended halo, but do not establish that such an initial configuration follows from Poisson clustering.

\section{Thermal History Model and Multimodal Mass Functions}
\label{sec:Thermal-History-Model-and-Multimodal-Mass-Functions}

\noindent The previous discussion motivates scenarios in which a large fraction of the dark matter consists of PBHs with an extended mass distribution. In this context, one must distinguish between a simple power-law function and a multimodal function~\cite{Carr:2018poi}. The power-law case was discussed in Ref.~\cite{Carr:2018rid}. Of particular interest in the multimodal case until now are distributions with a dominant solar-mass population and a subdominant massive tail. This can yield distributions in which the density peaks at a solar mass, while $\langle\mspace{1mu}m\mspace{1mu}f \rangle$ could lie in a much broader range of masses.

\subsection{Mass Function in Thermal History Model}

\begin{figure}
	\centering
	\vs{-3mm}\includegraphics[width=0.85\columnwidth]{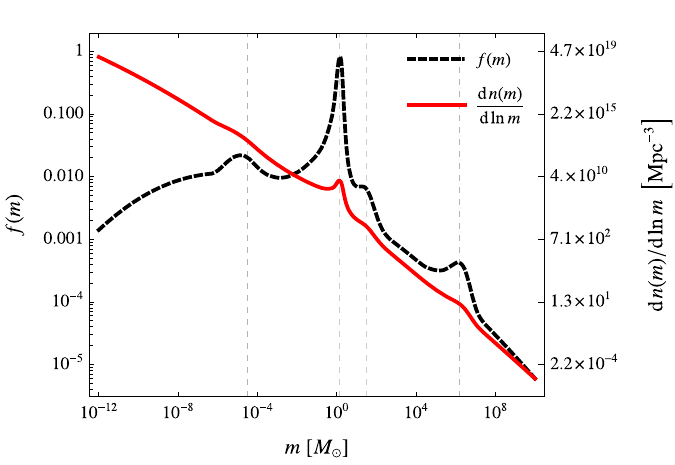}
	\caption{
		Mass and number-density spectra of the {\it Thermal History} model of Ref.~\cite{Carr:2019kxo}, using a spectral index $n_{\srm} = 0.973$, a spectral running $\alpha_{\srm} = -\.0.0011$ and an amplitude $A = 0.021946855$, with the latter chosen to yield $\fPBH = 1$.
		}
	\label{fig:fPBH-dndlnM} 
\end{figure}

A particular realisation of this is the {\it Thermal History} model of Ref.~\cite{Carr:2019kxo}, in which changes in the cosmic equation of state enhance the probability of collapse at several characteristic epochs. The resulting mass function can exhibit features at planetary masses around the electroweak epoch, at solar masses around the QCD transition, and around $10^{6}\.\Msun$ at electron--positron annihilation. Their relative amplitudes depend on the primordial fluctuation spectrum, allowing the dominant PBH population and the massive tail to contribute differently to the clustering and seed effects discussed above. In particular, a simple primordial spectrum of the form
\vs{-0.5mm}
\begin{align}
	\delta_{\text{rms}}( M )
		=
			A
			\left(
				\frac{ M }{\.\Msun }
			\right)^{\!(1 - n_{\srm}) / 
			4\.+\. [\alpha_{\srm} / 16] \ln( M/\Msun )}
			\label{eq:delta-power-law}
\end{align}
can resolve numerous cosmic conundra \cite{Carr:2023tpt, Carr:2019kxo}. Rather surprisingly, the spectral tilt $n_{\srm}$ and running $\alpha_{\srm} \equiv \d\mspace{1mu}n_{\srm}( k ) / \d\mspace{-1mu}\ln k$ are within the range allowed for equivalent parameters on the vastly larger Planck pivot scale~\citep{Planck:2018vyg}. For Fig.~\ref{fig:fPBH-dndlnM}, which depicts the {\it Thermal History} mass function, as well as the corresponding number-density spectrum, we adopt $n_{\srm} = 0.973$ and $\alpha_{\srm} = -\.0.0011$, as phenomenological parameters of the small-scale spectrum. However, these should not be interpreted as the Planck best-fit parameters. They may be compared with Planck constraints only within an explicitly specified model allowing running, at a stated pivot scale and using the joint ($n_{\srm},\mspace{1mu}\alpha_{\srm}$) confidence region. The amplitude $A$ is treated as a free phenomenological parameter, chosen to yield $\fPBH = 1$ for the indicated values of $n_{\srm}$ and $\alpha_{\srm}$. 

Such a spectral shape is natural in single-field models of inflation with quasi-inflection points \citep{Garcia-Bellido:2017mdw}, where a plateau-like feature in the potential enhances the amplitude of the power spectrum without affecting CMB scales. A concrete realisation is Critical Higgs Inflation~\cite{Ezquiaga:2017fvi}, where the renormalisation-group running of the Higgs self-coupling ensures this feature. This is particularly appealing, given the recent determination of the top quark and Higgs masses \cite{ATLASCMS:2024TopMass, ATLAS:2023HiggsMass}. Nevertheless, one expects less pronounced thermal features in this model, due to the non-Gaussian statistics, and further investigations are needed to accurately calculate the expected mass function for this model.

The peaks in Fig.~\ref{fig:fPBH-dndlnM} relate to specific observational effects:
	(1) microlensing events towards the Galactic bulge generated by planet-mass objects with about 1\% of the cold dark matter density~\cite{Niikura:2019kqi}, well above most expectations for free-floating planets; 
	(2) microlensing of LMC stars, as discussed in Sec.~\ref{sec:Effect-of-Clustering-on-Microlensing-Constraints};
	(3) the unexpectedly high number of microlensing events towards the Galactic bulge by dark objects in the mass gap between $2$ and $5\.\Msun$~\cite{Wyrzykowski:2019jyg}, where stellar evolution models fail to form black holes~\cite{Brown:1999ax}; 
	(4) unexplained correlations in the source-subtracted X-ray and cosmic infrared background fluctuations~\cite{2005Natur.438...45K}; 
	(5) the non-observation of UFDGs below the critical radius for which dynamical heating by PBHs is important, as discussed in Sec.~\ref{sec:Ultra--Faint-Dwarf-Galaxies-and-Accretion-in-PBH-Clusters}A;
	(6) the masses, spins and coalescence rates for the black holes found by LVK~\cite{LIGOScientific:2018mvr}; 
	(7) the relationship between the mass of a galaxy and that of its central black hole.

\subsection{Higher Moments in Thermal History Model}

\begin{figure}[t]
	\centering
	\hs{3mm}\includegraphics[width=0.85\columnwidth]{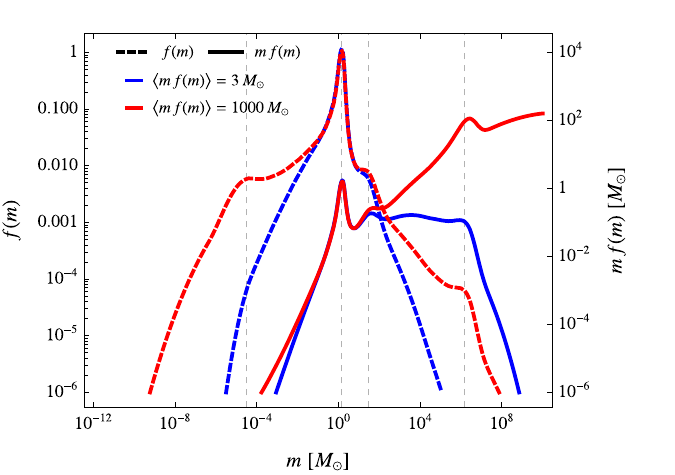}
	\caption{
		 PBH mass-density distributions $f( m )$ (dashed, left axis) and the corresponding contributions $m\mspace{1mu}f( m )$ (solid, right axis) for the {\it Thermal History} model. Both examples have $n_{\srm} = 0.973$. The blue and red curves have $\alpha_{\srm} = -\.0.01655$ and $-\.0.005374$, corresponding to $\langle\mspace{1mu}m\mspace{1mu}f \rangle = 3\.\Msun$ and $10^{3}\.\Msun$, respectively. The fluctuation amplitude is chosen to yield $\fPBH = 1$, evaluated over the mass range $10^{-12}\.\Msun \leq m \leq 10^{10}\.\Msun$.
		}
	\label{fig:thermal-mf-moment}
\end{figure}

For the clustering and seed effects, it is useful to display the mass moment, $m\mspace{1mu}f( m )$, which gives the contribution of each logarithmic mass interval to the moment entering the Poisson fluctuations. Figure~\ref{fig:thermal-mf-moment} compares two examples with $\langle\mspace{1mu}m\mspace{1mu}f \rangle = 3\.\Msun$ (blue curves) and $10^{3}\.\Msun$ (red curves). Both have the spectral index $n_{\srm} = 0.973$, while the running is $\alpha_{\srm} \simeq -\.0.016550$ and $-\.0.005374$, respectively. The equation of state and collapse prescriptions are the same. For each running, the fluctuation amplitude is adjusted to obtain $\fPBH = 1$.

In order to explain the significance of the mass moment, we first distinguish averages over PBH number from averages over PBH mass. If $\d\mspace{1mu}n_{\text{PBH}} / \d\mspace{-1mu}\ln m$ denotes the PBH number density per logarithmic mass interval, the number-weighted average of any function $g( m )$ corresponds to weighting by $f( m ) / m$ and is given by
\begin{equation}
	\big\langle\mspace{1mu}g( m ) \big\rangle^{}_{\rm N}
		\equiv
			\frac{ 1 }{ n_{\text{PBH}} }
			\int_{m_{\text{min}}}^{m_{\text{max}}}\d\mspace{-1mu}\ln m\;
			g( m )\.
			\frac{ \d\mspace{1mu}n_{\text{PBH}} }{ \d\mspace{-1mu}\ln m }
			\, ,
\end{equation}
where $n_{\text{PBH}}$ is the total number density. In particular, $\langle\mspace{1mu}m \rangle^{}_{\rm N}$ is the mean mass per PBH, so $\rho^{}_{\text{PBH}} = n_{\text{PBH}}\mspace{1mu}\langle\mspace{1mu}m \rangle^{}_{\rm N}$ and one can show
\begin{equation}
\label{eq:thermal-mf-number-moment}
	\langle\mspace{1mu}m\mspace{1mu}f \rangle
		=
			\int_{m_{\text{min}}}^{m_{\text{max}}}\d\mspace{-1mu}\ln m\;
			m f( m )
		=
			\frac{
				\big\langle\mspace{1mu}m^{2} \big\rangle^{}_{\rm N}
			}{
				\langle\mspace{1mu}m \rangle^{}_{\rm N}
			}
			\, .
\end{equation}
The relevance of this mass-weighted mean PBH mass to Poisson fluctuations follows because independent Poisson counts in a volume $V$ give a total PBH mass with mean $V n_{\text{PBH}}\mspace{1mu}\langle\mspace{1mu}m \rangle^{}_{\rm N}$ and variance $V n_{\text{PBH}}\mspace{1mu}\big\langle\mspace{1mu}m^{2} \big\rangle^{}_{\rm N}$. The ratio of the variance to the mean is precisely the mass moment in Eq.~\eqref{eq:thermal-mf-number-moment}. 

Massive PBHs contribute quadratically to the mass variance but only linearly to the mean mass, so a rare massive population can substantially enhance the Poisson fluctuations. The relevant moment can therefore differ substantially from both the number-weighted mean and the mass at which $f( m )$ peaks. In the blue and red examples shown in Fig.~\ref{fig:thermal-mf-moment}, with $\langle\mspace{1mu}m\mspace{1mu}f \rangle$ being $3\.\Msun$ and $10^{3}\.\Msun$, respectively, both distributions retain the QCD peak near one solar mass, with $98\%$ and $93\%$ of the PBH mass density lying between $10^{-2}\.\Msun$ and $10\.\Msun$, respectively. On the other hand, PBHs above $10^{6}\.\Msun$ contribute $4\%$ and $92\%$, respectively, of the clustering moment, while accounting for only $7 \times 10^{-8}$ and $9 \times 10^{-5}$ of the PBH mass density. The larger moment of the red example is therefore predominantly generated by its massive tail, even though most of its mass density is concentrated near a solar mass.

For the two distributions in Fig.~\ref{fig:thermal-mf-moment}, increasing $\langle\mspace{1mu}m\mspace{1mu}f \rangle$ from $3\.\Msun$ to $10^{3}\.\Msun$ increases the ensemble Poisson power by a factor $10^{3}/3 \simeq 333$ at fixed $\fPBH$ and redshift, assuming the same growth factor [Eq.~\eqref{eq:power}]. For the same smoothing window and scale, the Poisson variance increases by the same factor, whereas the rms Poisson density contrast increases by approximately $18.3$. The enhancement is dominated by the rare massive tail of the red distribution and depends on the adopted high-mass spectrum, here integrated up to $10^{10}\.\Msun$. On scales where most regions contain no PBH from the population dominating this moment, the enhanced variance primarily reflects rare regions containing massive seeds and does not characterise typical clusters. Where that population is sufficiently sampled, the larger moment favours earlier collapse and smaller initial radii at fixed cluster mass in the treatment leading to Eq.~\eqref{eq:rM}. The subsequent heating in Eqs.~\eqref{eq:growth} and \eqref{eq:dyntime} also depends on the mass spectrum, but its application requires the PBH population and spatial distribution within the cluster to be taken into account. Increasing $\beta$ reduces the subsequent expansion without changing the initial collapse relation.
\newpage

In the mass range $\Ocal( 10$\,--\,$10^{4} )\.\Msun$, another relevant constraint comes from accretion. Indeed, it has recently been claimed that this would exclude the QCD transition as the origin of the LVK events unless one invokes an {\it ad hoc} evolution of the PBH mass function and a cut-off in the power spectrum close to the QCD scale~\citep{Juan:2022mir}. In addressing this criticism, we first remark that Ref.~\cite{Juan:2022mir} uses values for the parameters $A$ and $n_{\srm}$ that differ from those adopted in our {\it Thermal History} model, leading to a drastically different mass spectrum. This spoils the compatibility with the data and leads to a less pronounced QCD peak, lowering $\fPBH$ in the LVK mass range.

\subsection{Modifications to Thermal History Model}

The multi-peaked PBH mass function is not uniquely specified in the {\it Thermal History} scenario. The amplitude and detailed shape depend on the small-scale primordial curvature spectrum{\,---\,}including its amplitude, tilt, running and any localised features{\,---\,}as well as the perturbation profile and collapse threshold~\cite{Carr:2019kxo, Escriva:2022bwe, Musco:2023dak}. Primordial lepton-flavour asymmetries also provide an important source of additional freedom. Through the associated chemical potentials, they modify the equation of state around the QCD epoch and thereby shift the thermal-history features in the PBH mass spectrum~\cite{Middeldorf-Wygas:2020glx, Bodeker:2020stj}. 

\citet{Bodeker:2020stj} have shown that mass functions obtained for non-zero flavour asymmetries can yield a closer correspondence with the LIGO--Virgo events than the standard, nearly lepton-symmetric case. More recently, Ref.~\cite{Gonin:2026tyu} numerically followed the self-consistent evolution of the primordial baryon and flavour-dependent lepton asymmetries from the QCD epoch through neutrino decoupling, including neutrino oscillations, and combined the resulting equation of state with collapse thresholds and peak theory. Within the merger prescription adopted there, several cases with non-zero asymmetries yield a better fit to the observed GWTC-5 component masses than the case with zero asymmetry, while the neutrino-oscillation history also modifies the high-mass feature associated with $e^{+}e^{-}$ annihilation. The comparison with GWTC-5 is merely indicative because of merger-model and population uncertainties. The spectrum in Fig.~\ref{fig:fPBH-dndlnM} is therefore not a unique prediction. Accretion bounds must also be evaluated for the particular mass function under consideration.

The lepton asymmetries have a pronounced impact on PBH formation because the collapse probability depends exponentially on the collapse threshold. Consequently, the position and amplitude of the QCD peak need not coincide with those obtained for negligible flavour asymmetries~\cite{Gonin:2026tyu}. In the asymmetric cases investigated, PBH formation around the QCD transition is suppressed at fixed primordial fluctuation amplitude. So the QCD peak becomes less dominant, with increased relative contributions from larger masses. Furthermore, sufficiently large tau-lepton asymmetries can shift the principal maximum to lower PBH masses by enhancing the annihilation of charged tau leptons and their antiparticles before the QCD transition~\cite{Gonin:2026tyu}. A lower peak mass may improve correspondence with the subsolar lens masses inferred by \citet{MACHO:2000qbb}.

\subsection{Astrophysical Consequences}

Recently, \citet{Magaraggia:2026ngp} found that the Poisson contribution of a thermal-history extended mass function moderately enhances the formation of small-scale structure, advancing star formation and reionisation while remaining consistent with the Planck optical-depth constraint. Their separate binary prescriptions also allow a PBH contribution to LVK mergers and predict LISA sources from the high-mass tail. This complements our analysis, but their analysis does not include the relaxation and dynamical evolution of bound PBH clusters.

There are two other issues with the accretion constraints. First, there are large uncertainties associated with feedback from the radiation emitted by the accreting gas; this could heat and ionise the intergalactic medium~\citep{Oh:2003pm}. Second, structure formation causes PBHs to fall into the gravitational potential wells of larger structures, increasing their peculiar velocities. Both effects could reduce PBH accretion after some redshift. In particular, Ref.~\cite{DeLuca:2020fpg} studies the effect of a redshift cut-off, arguing that $z_{\rm cut} = 10$ is realistic but also considering $z_{\rm cut} = 7$ and $15$. Even the (realistic) case with $z_{\rm cut} = 10$ is marginally consistent with the {\it Thermal History} mass function and the $z_{\rm cut} = 15$ case is even consistent to within an order of magnitude.

Redshift dependence is not the only source of uncertainty in these constraints, since it is still unknown how PBHs accrete throughout cosmic history. The suggested scenarios include spherical accretion, accretion through thick, thin or slim discs, and advection-dominated accretion (see Ref.~\citep{Abramowicz:2011xu} for a recent review). This yields an uncertainty of several orders of magnitude in the accretion luminosity. Therefore, one may conclude that PBH accretion neither significantly constrains nor excludes the QCD transition as the origin of the LVK events. Clustering also has a large effect on the accretion constraints.

\section{Conclusion}
\label{sec:Conclusion}

In this work, we have discussed the inevitable consequences of the seed and clustering effects of PBHs in any scenario for their formation. We have gone beyond the usual consideration of these effects for a monochromatic PBH mass function and investigated a rich phenomenology of physical processes, including dynamical heating, core collapse, collisions, tidal disruptions, dynamical friction, and dilution within larger clusters. Our analysis connects the survival and observable properties of PBH clusters to their expansion history. In particular, their final mass and radius depend crucially on the efficiency of dynamical heating (corresponding to the amount of expansion). Reduced expansion can also enhance gas depletion. However, all these effects are model-dependent, so further investigation through dedicated $N$-body and hydrodynamical simulations is required.

PBH cosmologies suggest a new and potentially coherent picture of the cosmic dawn. In this picture, Poisson clustering and massive-seed effects{\,---\,}together with enhanced primordial fluctuations{\,---\,}operate simultaneously, generating a Universe that becomes highly inhomogeneous at earlier times than in the standard scenario, with dwarf galaxies and LRD-like objects forming at high redshifts. Poisson clusters initially have their own dark matter haloes but are progressively disrupted by dynamical processes. This naturally leads to a characteristic dark matter distribution, with PBH Poisson clusters in the outskirts of galaxies, embedded within an increasingly smooth halo mass towards galactic centres. The clusters may play an important r{\^o}le in explaining the origin of LRDs, one of the most active areas of current research. PBHs as large as {$10^{7}\.\Msun$} would naturally seed the formation of objects composed of gas and smaller PBHs; the latter would rapidly sink through dynamical friction and merge with the central PBH at high redshifts. On the other hand, Poisson clusters without a massive central seed would generate a population of fainter high-redshift objects, potentially accounting for the observed correlations between the source-subtracted infrared and X-ray backgrounds. The clusters could potentially be identified with UFDGs but the agreement is merely qualitative and based on proxy ${( M_{1\mspace{-2mu}/\mspace{-1mu}2},\.2\mspace{1mu}R_{1\mspace{-2mu}/\mspace{-1mu}2} )}$ values rather than measured total virial masses and radii.

Understanding the evolution of clusters directly impacts the strength of the constraints on the PBH abundance. However, the compactness required to explain UFDGs does not suffice to reduce microlensing constraints, which can only be weakened by invoking a modification to the halo density profile. Changes to the variance of the microlensing rate should be distinguished from changes to its mean value. GW limits based on the primordial or early-binary channel are strongly dependent on the assumed PBH mass function and cluster evolution. But due to dynamical heating, we have argued that binaries should survive in Poisson clusters. We have generalised the calculation of the PBH binary fraction ending up in Poisson clusters to extended mass functions and argued that this fraction should be evaluated at the redshift of merging rather than cluster formation. For sufficiently broad mass functions, we have emphasised that there are no robust, model-independent GW constraints on the PBH abundance from the early-binary channel. In such cases, binaries formed inside the clusters can become the dominant merger channel. However, the corresponding rates depend on the cluster density profile, the surviving mass distribution and dynamical heating. The UFDG, accretion, microlensing and GW comparisons therefore all depend on these factors.

It should be stressed that we are not claiming that the various observations discussed in this paper can {\it only} be explained by PBHs. The more mainstream models, which have been the focus of most numerical simulations and in which the dark matter is assumed to be some form of elementary particle, may also explain some of them. Indeed, whether these models can explain UFDGs and LRDs is a crucial aspect of such numerical investigations. Our approach in this paper has been primarily analytical but clearly numerical work is required to study such issues more rigorously. Our main point is that the possibility of macroscopic dark matter (even if not PBHs) introduces a wide range of astrophysical signatures, especially if it is clustered, which do not arise with particle dark matter. Indeed, it is intriguing that specialised cosmological zoom-in simulations now employ dark matter ``particles'' with masses of order a solar mass or even substantially below it.

\acknowledgements

\noindent The authors thank Marc Barcel{\'o}-Sastre, Simon Biot, Julien Lavalle, Cedric Lacey, Ester Ruiz Morales, Mairi Sakellariadou, Dominik Schwarz, Jos{\'e} Francisco Nu{\~n}o Siles, Joe Silk, Sonali Verma, Sam Young, Saiyang Zhang for useful discussions and remarks. S.~C.~is supported by an {\it Interuniversity Institute for Nuclear Science} (IISN) grant of the Belgian Fund for Research FNRS-F.R.S.~and the {\it PROgramme for the Development of scientific EXperiments} PRODEX grant {\it Belgian Euclid Science Exploitation} (BESE) of the {\it European Space Agency} (ESA) [Euclid]. J.~G.-B.~acknowledges support from the Spanish Research Project PID2024-159420NB-C43 [MICIU-FEDER], and the framework of the R\&D\&I Severo Ochoa Project CEX2025-001574-S at IFT, funded by MICIU/AEI/10.13039/501100011033.

\section{List of Acronyms}
\label{sec:List-of-Acronyms}

\begin{description}
	\setlength{\itemsep}{0pt}
	\setlength{\parskip}{0pt}

	\item[AGN] Active galactic nucleus
	\item[ALMA] Atacama Large Millimeter/submillimeter Array
	\item[BBN] Big Bang nucleosynthesis
	\item[BH] Black hole
	\item[CDM] Cold dark matter
	\item[CMB] Cosmic microwave background
	\item[DESI] Dark Energy Spectroscopic Instrument
	\item[DM] Dark matter
	\item[GW] Gravitational wave
	\item[GWTC] Gravitational-Wave Transient Catalog
	\item[IMBH] Intermediate-mass black hole
	\item[IR] Infrared
	\item[ISCO] Innermost stable circular orbit
	\item[JWST] James Webb Space Telescope
	\item[KAGRA] Kamioka Gravitational Wave Detector
	\item[$\Lambda$CDM] Cosmological-constant-plus-cold-dark-matter model
	\item[LIGO] Laser Interferometer Gravitational-Wave Observatory
	\item[LISA] Laser Interferometer Space Antenna
	\item[LMC] Large Magellanic Cloud
	\item[LRD] Little Red Dot
	\item[LVK] LIGO--Virgo--KAGRA collaboration
	\item[MACHO] Massive Compact Halo Object
	\item[NFW] Navarro--Frenk--White
	\item[NIRCam] Near-Infrared Camera
	\item[OGLE] Optical Gravitational Lensing Experiment
	\item[PBH] Primordial black hole
	\item[QCD] Quantum chromodynamics
	\item[QSO] Quasi-stellar object
	\item[SMBH] Supermassive black hole
	\item[SMC] Small Magellanic Cloud
	\item[UFDG] Ultra-Faint Dwarf Galaxy
	\item[UV] Ultraviolet
\end{description}


\begingroup
\renewcommand{\theequation}{A.\arabic{equation}}
\setcounter{equation}{0}

\endgroup

\setlength{\bibsep}{6pt}

\let\originalbibinfo\bibinfo
\renewcommand{\bibinfo}[2]{%
	\ifstrequal{#1}{title}{\textit{#2}}{\originalbibinfo{#1}{#2}}%
}

\providecommand{\enquote}[1]{#1}
\renewcommand{\enquote}[1]{\textit{#1}}

\interlinepenalty = 10000
\bibliography{refs}

\end{document}

%% file: header.tex
\usepackage{amsmath}	
\usepackage{amssymb}	
\usepackage{graphicx}
\usepackage{bm}

\usepackage{color}

\usepackage[colorlinks=true, citecolor=midblue, linkcolor=midblue, urlcolor=midblue]{hyperref}

\usepackage{setspace}

\usepackage{datetime}

\usepackage{tgpagella}
\usepackage[T1]{fontenc}
\usepackage[protrusion=true,expansion=false]{microtype}

\usepackage{fancyhdr}

\usepackage{lastpage}

\usepackage{titlesec}
\usepackage[normalem]{ulem}     

\usepackage{letltxmacro}		

\usepackage{etoolbox}

\usepackage{eso-pic}

\numberwithin{equation}{section}

\titleformat{\section}
       {\centering\normalfont\fontsize{14}{17}\bfseries}{\thesection}{1em}{}

\makeatletter
	\renewcommand{\fnum@figure}{{\bf Figure \thefigure}}
	\renewcommand{\fnum@table}{{\bf Table \thetable}}
\makeatother

\makeatletter
\let\oldr@@t\r@@t
\def\r@@t#1#2{%
\setbox0=\hbox{$\oldr@@t#1{#2\,}$}\dimen0=\ht0
\advance\dimen0-0.2\ht0
\setbox2=\hbox{\vrule height\ht0 depth -\dimen0}%
{\box0\lower0.4pt\box2}}
\LetLtxMacro{\oldsqrt}{\sqrt}
\renewcommand*{\sqrt}[2][\ ]{\oldsqrt[#1]{#2}}
\makeatother

\definecolor{grey}{rgb}{0.4,0.4,0.4}
\definecolor{dullmagenta}{rgb}{0.4,0,0.4}
\definecolor{darkblue}{rgb}{0,0,0.4}
\definecolor{midblue}{rgb}{0,0,0.7}
\definecolor{midred}{rgb}{0.5,0,0}
\definecolor{orange}{rgb}{1,0.5,0}
\definecolor{lightbrown}{rgb}{0.75,0.5,0.25}
\definecolor{tan}{cmyk}{0.14,0.42,0.56,0}
\definecolor{djunglegreen}{cmyk}{0.99,0,0.52,0}
\definecolor{lightgreen}{rgb}{0,1,0}
\definecolor{olivegreen}{cmyk}{0.64,0,0.95,0.40}
\definecolor{midgreen}{rgb}{0.0,0.675,0.0}
\definecolor{darkgreen}{rgb}{0,0.5,0}
\definecolor{pink}{rgb}{1,0.078,0.57}
\definecolor{darkpink}{RGB}{191,0,96}

\newcommand{\vs}{\vspace}
\newcommand{\hs}{\hspace}
\renewcommand{\.}{\hspace{0.5mm}}

\newcommand{\la}{\ensuremath{\leftarrow}}

\newcommand{\Erm}{\ensuremath{\mathrm{E}}}

\newcommand{\Grm}{\ensuremath{\mathrm{G}}}
\newcommand{\Hrm}{\ensuremath{\mathrm{H}}}

\newcommand{\Lrm}{\ensuremath{\mathrm{L}}}

\newcommand{\Nrm}{\ensuremath{\mathrm{N}}}

\newcommand{\Prm}{\ensuremath{\mathrm{P}}}

\newcommand{\Srm}{\ensuremath{\mathrm{S}}}

\newcommand{\arm}{\ensuremath{\mathrm{a}}}
\newcommand{\brm}{\ensuremath{\mathrm{b}}}
\newcommand{\crm}{\ensuremath{\mathrm{c}}}
\newcommand{\drm}{\ensuremath{\mathrm{d}}}
\newcommand{\erm}{\ensuremath{\mathrm{e}}}

\newcommand{\hrm}{\ensuremath{\mathrm{h}}}
\newcommand{\irm}{\ensuremath{\mathrm{i}}}

\newcommand{\mrm}{\ensuremath{\mathrm{m}}}

\newcommand{\orm}{\ensuremath{\mathrm{o}}}

\newcommand{\srm}{\ensuremath{\mathrm{s}}}

\newcommand{\Ecal}{\ensuremath{\mathcal{E}}}
\newcommand{\Fcal}{\ensuremath{\mathcal{F}}}
\newcommand{\Gcal}{\ensuremath{\mathcal{G}}}

\newcommand{\Ncal}{\ensuremath{\mathcal{N}}}
\newcommand{\Ocal}{\ensuremath{\mathcal{O}}}

\renewcommand{\d}{\ensuremath{\mathrm{d}}}

\newcommand{\be}{\begin{equation}}
\newcommand{\ee}{\end{equation}}
\newcommand{\ba}{\begin{eqnarray}}
\newcommand{\ea}{\end{eqnarray}}

\settimeformat{ampmtime}

\def\ga{\mathrel{\raise.3ex\hbox{$>$\kern-.75em\lower1ex\hbox{$\sim$}}}}
\def\la{\mathrel{\raise.3ex\hbox{$<$\kern-.75em\lower1ex\hbox{$\sim$}}}}

\def\Msun{M_{\odot}}

\def\fpbh{f_{\text{PBH}}}
\def\fPBH{f_{\text{PBH}}}

\def\be{\begin{equation}}
\def\ee{\end{equation}}
\def\bea{\begin{eqnarray}}
\def\eea{\end{eqnarray}}

\usepackage[intoc]{nomencl}

\makenomenclature

\usepackage{array}
\usepackage{booktabs}
\usepackage{longtable}

\makeatletter

\protected\def\mathsymbolrow#1#2#3{#1 & #2 & p.~#3\\}
\newcommand{\DeclareMathSymbolEntry}[3]{%
	\expandafter\protected@xdef\csname mathsymbollist#1\endcsname{%
		\csname mathsymbollist#1\endcsname
		\noexpand\mathsymbolrow{\unexpanded{#2}}{\unexpanded{#3}}{\thepage}%
	}%
}
\makeatother

\makeatletter
\patchcmd{\l@f@section}{1.0em}{0.7em}{}{}
\makeatother
